\documentclass[twocolumn]{aa} 
\usepackage[varg]{txfonts}
\usepackage{placeins}
\usepackage{natbib}
\usepackage{url}
\PassOptionsToPackage{hyphens}{url}
\usepackage{graphicx}
\usepackage{subfigure}
\usepackage{comment}
\usepackage{amsmath}
\usepackage{txfonts}
\usepackage{xcolor,soul}
\sethlcolor{lightgray}
\usepackage{listings} 
\usepackage{times}
\usepackage{multirow}
\usepackage{booktabs}
\usepackage{float}
\usepackage{epstopdf}
\usepackage{longtable}
\usepackage{soul}
\usepackage{gensymb}
\usepackage[colorlinks=true,linkcolor=red,citecolor=blue, breaklinks=true]{hyperref}
\restylefloat{figure}  
\usepackage{lscape}
\usepackage{stfloats}
\usepackage[switch]{lineno}

\newcommand{\herschel}{{\it Herschel}}
\newcommand{\spitzer}{{\it Spitzer}}

\newcommand{\lsun}{\,L$_{\odot}$\,}
\newcommand{\msun}{\,M$_{\odot}$\,}
\def\NHUNIT{\ifmmode {\rm \,cm^{-2}} \else $\rm \,cm^{-2}$ \fi} 
\newcommand{\mum}{$\mu\rm m$\,}
\def\nhh{\ifmmode N_{\rm H_{2}}\else $N_{\rm H_{2}}$\fi} 
\def\nhhc{\ifmmode N_{\rm H_{2}}^0\else $N_{\rm H_{2}}^0$\fi} 
\def\nhhbg{\ifmmode N_{\rm H_{2}}^{\rm bg}\else $N_{\rm H_{2}}^{\rm bg}$\fi} 
\def\nh{\ifmmode N_{\rm H}\else $N_{\rm H}$\fi}
\def\ml{\ifmmode M_{\rm line}\else $M_{\rm line}$\fi}  
\def\sunpc{\ifmmode \rm \,M_\odot/\rm pc\,\else $\rm \,M_\odot/\rm pc\,$\fi}  
\def\bang{$\chi_{B_{\rm POS}}$}
\def\fang{$\theta_{\rm fil}$}
\def\fdiff{$\theta_{\rm diff}$}
\def\rflat{\ifmmode R_{\rm flat}\else $R_{\rm flat}$\fi}   
\def\kms{\ifmmode {km\,s$^{-1}$}\else km\,s$^{-1}$\fi}  
\def\arcm{\ifmmode {^{\scriptstyle\prime}}
          \else $^{\scriptstyle\prime}$\fi}
\newdimen\sa  \newdimen\sb
\def\parcs{\sa=.07em \sb=.03em
     \ifmmode \hbox{\rlap{.}}^{\scriptstyle\prime\kern -\sb\prime}\hbox{\kern -\sa}
     \else \rlap{.}$^{\scriptstyle\prime\kern -\sb\prime}$\kern -\sa\fi}
\def\parcm{\sa=.08em \sb=.03em
     \ifmmode \hbox{\rlap{.}\kern\sa}^{\scriptstyle\prime}\hbox{\kern-\sb}
     \else \rlap{.}\kern\sa$^{\scriptstyle\prime}$\kern-\sb\fi}
\def\parcd{\sa=.08em \sb=.03em
     \ifmmode \hbox{\rlap{.}\kern\sa}^{\scriptstyle\circ}\hbox{\kern-\sb}
     \else \rlap{.}\kern\sa$^{\scriptstyle\circ}$\kern-\sb\fi}
\begin{document} 

\title{Evolution of magnetized hub-filament systems}
\subtitle{Comparing the observed properties of W3(OH), W3\,Main, and S\,106}

 \titlerunning{Evolution of magnetized hub-filament systems}

   \author{M. S. N. Kumar\inst{1,5}
   \and
   D. Arzoumanian\inst{2,3,4}
      \and
      S. Inutsuka\inst{5}
            \and
        R. Furuya\inst{6}
    \and
    N. K. Bhadari\inst{7}
            }
   \institute{Instituto de Astrof\'isica e Ci{\^e}ncias do Espa\c{c}o, Universidade do Porto, CAUP, Rua das Estrelas, 4150-762 Porto, Portugal\goodbreak
      \and
         Institute for Advanced Study, Kyushu University, Japan\goodbreak
   \and
   Department of Earth and Planetary Sciences, Faculty of Science, Kyushu University, Nishi-ku, Fukuoka 819-0395, Japan\goodbreak
   \and Division of Science, National Astronomical Observatory of Japan, 2-21-1 Osawa, Mitaka, Tokyo 181-8588, Japan\goodbreak
        \and
        Department of Physics, Graduate School of Science, Nagoya University, Furo-cho, Chikusa-ku, Nagoya 464-8602, Japan\goodbreak
\and
Institute of Liberal Arts and Sciences Tokushima University, Minami Jousanajima-machi 1-1, Tokushima 770-8502, Japan\goodbreak
\and
Kavli Institute for Astronomy and Astrophysics, Peking University, Beijing, 100871, China
                  }
                  
     \date{}

 \abstract
   {Hub-filament systems (HFS) span a broad range of star-forming gas densities and are widely recognized as the progenitors of open star clusters. They serve as ideal targets for investigating the physical properties of star-forming gas and observing how dense gas is removed during the assembly of star clusters.}
   {In this study, we explore the characteristics of three cluster-forming HFS - W3(OH), W3 Main, and S 106 - representing evolutionary stages from early to evolved, with a particular focus on the structure of their magnetic fields (B-field) and filament line-mass distributions. The goal here is to identify indicators of the evolution of the HFS, in particular, their hubs, as star formation proceeds.}
   {Our analysis combines observations of dense star-forming gas and young stellar populations. We present new JCMT/POL-2 observations of 850\,\mum dust polarized emission to probe the dense gas and magnetic field structures. Additionally, we utilize archival infrared and radio data from WIRCAM, WFCAM, \spitzer, \herschel, and the VLA to identify markers of star formation. We derive radial column density profiles centered in the hubs and use them to define distinct filament and hub regions. Histograms of line mass ($M_{\rm line}$), polarization intensity ($\it PI$), polarization fraction ($\it PF$), and the relative orientation between the magnetic field and the filaments are analysed.}
   {Each hub contains two adjacent nodes or peaks of star formation, with one peak consistently more evolved than the other. The radial intensity profiles for all three targets fit well with two distinct power laws, with turning points between 0.6–0.8\,pc — defining the semi-major axes of the hub, which approximates an elliptical shape. The power-law indices for the hub regions (0.6–0.8\,pc) are -2.1, -1.7, and -0.9, and for the filament regions ($>$0.8\,pc) are -2.9, -4.2, and -11.7, corresponding to W3(OH), W3\,Main, and S\,106, respectively. These power-law slopes indicate different dynamical behaviors (where $\le$ -2 suggests global collapse) which is important to understand HFS evolution. The hubs contain the highest line masses across all targets. In the earliest stage W3(OH), the filament line-mass function (FLMF) smoothly includes both the hub and filament regions in a Salpeter-like slope. In the evolved S\,106, the hub FLMF slope is -0.85 and the filament region FLMF slope is -1.4.  The plane-of-sky (POS) magnetic field structures display two notable features: (a) at low densities, B-field lines are misaligned with filaments but gradually align with them as density increases toward the hub; (b) B-field lines trace the walls of bipolar cavities formed by massive outflows from stars in the hub. PF  and PI contour maps show disk and bipolar outflow-like patterns centered on the most luminous sources. 
   Additionally, we identify a foreground mini-spiral HFS in W3\,Main, previously recognized as the coldest clump in the region.}
   {As HFS evolve, discernible changes can be found in the FLMF, $PF$ and B-field-Filament angles, especially inside the hub which is also found to increase in size. Massive bipolar outflows and radiation bubbles significantly reshape POS magnetic fields, aligning them along cavity walls and shells, adding to the well-documented rearrangements around HII region cavities. We notice there is an intriguing similarity between hub sizes and young cluster radii. 
   The presence of `double-node' star formation within hubs — characterized by systematic evolutionary differences — appears to be a common feature of hub-filament systems (HFS). We present evidence for their widespread occurrence in several well-studied, nearby star-forming clouds.}

   \keywords{star: formation -- ISM: clouds, magnetic fields, polarization  -- submillimeter: ISM
               }
\maketitle

\section{Introduction}\label{intro}

Star formation predominantly occurs in a clustered mode \citep{Lada2003}, with hub-filament systems (HFS) \citep{Myers2009,Peretto2014} now widely recognized as the progenitors of stellar clusters \citep{Schneider2012,Kumar2020}. HFS naturally encompass a broad range of filament densities \citep{Hacar2023}, as they gather gas from the surrounding diffuse cloud via low-density filaments \citep{Kirk2013,Palmeirim2013,Schisano2014}, while simultaneously building high-density filament networks within the central hub \citep{Myers2009,Kumar2021,Hacar2023}.
It has been established that most massive stars form within hubs \citep{Kumar2020}, as the high-density filament network within the hub \citep{Kumar2021} creates the conditions necessary to achieve the highest critical densities required for massive stellar seeds to form. These seeds are further nourished to grow in mass \citep{Kumar2020,Kumar2021}, as the hub serves as the central convergence point for gas drawn from the surrounding cloud \citep{Peretto2013,Peretto2014}. 

\citet{Myers2009} defined a hub as the central region of a HFS where the column density exceeds 10$^{22}$ cm$^{-2}$, noting that in NGC\,1333, the hub had the morphology of a segmented ellipse with a major axis of approximately 0.6\,pc. In Mon\,R2, radial density profiles revealed a distinct turnover in slope at a radius of 0.8\,pc, which was used to define the hub radius \citep{Trevino-Morales2019, Kumar2021}. Morphological analyses of other nearby ($<$1\,kpc) HFS prompted \citet{Kumar2020} to describe hubs as elliptical structures, with their foci marking the centers of star formation activity. However, the general properties of hubs within the broader context of HFS remain in need of a more precise definition. 

Several observational indicators, such as; a) magnetic fields (B-fields) alignment with filaments and flows, b) suppression of fragmentation \citep{Chen2019}, c) hour-glass morphologies \citep{Beltran2019}, and, d) magnetic braking and angular momentum transfer \citep{hullzhang2019} indicate that B-field play a significant role in facilitating massive star formation \citep[see also][]{Girart2009,Pattle2023}.  Despite the challenges in constraining their influence, significant progress has been made over the past decade in understanding the role of B-fields in the star formation process \citep[cf.][for reviews]{Hennebelle2019,Pattle2023}. Polarimetric measurements in the far-infrared (FIR) to (sub)-millimeter (sub-mm) wavelengths have become routine with many single-dish and interferometric telescopes, enabling detailed mapping of B-fields in star-forming regions. These observations span spatial scales ranging from clouds ($\sim$10\,pc) \citep[e.g.][]{planck2016-XXXV} to dense cores ($\sim$0.1\,pc) \citep[e.g.][]{Kwon2018,Liu2019}. Sub-mm polarization measurements of thermal dust emission from cold, dense molecular filaments reveal specific trends \citep{Palmeirim2013,Cox2016,Soler2016,planck2016-XXXIII,Pillai2020} in the plane-of-the-sky (POS) B-fields:\\
a) Parallel to non-self-gravitating, low-density filaments,\\
b) Perpendicular to high-density, self-gravitating filaments,\\
c) Parallel within very high-density, self-gravitating filaments connected to massive hubs.\\
These results are based on observations conducted across diverse environments in both low- and high-mass star-forming regions. The relative orientation of the filament axis and B-fields at parsec scales generally follows the patterns described above but can be altered at core scales ($\sim$0.1\,pc) due to local phenomena such as outflows, ionizing sources, or the interplay of gravity, turbulence, and B-fields \citep{Pattle2023}.

Recent studies of B-fields in HFS provide new insights into their structure and dynamics. For instance, \citet{Arzoumanian2021} traced three orders of magnitude in polarization intensity across the 10\,pc-long NGC\,6334. This work highlights the influence of young massive stars on local B-fields and explores the variation in energy balance—gravity, turbulence, and magnetic forces—throughout the HFS. In other regions, such as Mon\,R2 \citep{Hwang2022} and IRAS 18089-1732 \citep{Sanhueza2021}, striking toroidal magnetic fields have been observed, revealing an intricate relationship between the morphology of the B-fields and the structure of the HFS.

Therefore, it is evident that understanding the properties of hubs and magnetic fields in HFS is key to unraveling the process of massive star formation. HFS are expected to remain longer than the timescales of local individual star formation events. This follows the scenario conjectured in filaments-to-clusters (F2C) paradigm \citep{Kumar2020}, and is consistent with the theoretical expectation that star formation continue and accelerate over many millions of years \citep{Inutsuka2015,Inutsuka2017}. Since the formation of massive stars generates feedback that can reshape the structure of both the HFS and the associated magnetic fields, it is equally important to examine how HFS evolve as star formation progresses. The goal of this study is to explore the global properties of HFS, particularly the influence of B-fields resulting from the evolutionary stage of the HFS and its star formation history. The evolution of an HFS is directly influenced by star formation within the hub, especially the development of the OB stellar content \citep{Kumar2020}. To achieve this, we have identified three targets within 2\,kpc distance that will allow us to map the full extent of the HFS while maintaining a spatial resolution of approximately 0.1\,pc, necessary to trace the filaments and resolve the hubs. These targets — W3(OH), W3\,Main, and S\,106 — are well-studied in the literature for their distinctive massive star formation activity and evolutionary sign-posts.

{\em Targets of study:} The W3 molecular complex is well known for its rich star formation activity \citep{Rivera-Ingraham2011}, with young massive stars located in W3-Main and the very young, OH and H$_2$O maser-rich region W3(OH). The entire W3 complex is systematically associated with a distance of approximately 2\,kpc. Gaia-DR2 analysis by \citet{Navarete2019} estimates a mean distance of 2.14\,kpc for the complex, with W3-Main and W3(OH) at distances of 2.3\,kpc and 2.14\,kpc, respectively.

W3(OH) is an ultra-compact HII region \citep[e.g.][]{Wilner1995}, famously associated with the embedded source to its east, an H$_2$O maser known as the Turner-Welch (TW) object \citep{TurnerWelch1984}, or W3(H$_2$O). These sources are located within a bright FIR/sub-mm peak, which represents the hub of a network of filaments. \citet{Rivera-Ingraham2013} estimate a luminosity of 2.3$\times$10$^4$\lsun and a mass of 1.6$\times$10$^3$\msun for this source. The estimated diameter of the hub is 0.42\,pc, and it contains at least two forming O-type stars as represented by W3(OH) and W3(H$_2$O). These sources drive multiple outflows \citep[e.g.][]{Zapata2011} and have been found to drive a rich chemistry in the hot cores \citep[e.g.][and references therein]{Giese2024}, with various molecular lines tracing the dense gas. Therefore, W3(OH) represents the youngest hub in our study.

W3\,Main hosts several hyper- and ultra-compact HII regions with diameters ranging from $<$1000\,au to $>$20000\,au \citep{Tieftrunk1997}, indicative of massive stars in various evolutionary states. This dense region, approximately 1\,pc in size, also hosts very luminous infrared sources, the brightest being IRS\,5 (L$\sim$3$\times$10$^5$\lsun) \citep{Campbell1995}. IRS\,5 is surrounded by a rich cluster of stars \citep{Megeath2005}, including many sub-stellar objects \citep{Ojha2009}. Several H$_2$O masers are found in correspondence with radio continuum sources around IRS\,5, exhibiting radial velocities indicative of powerful molecular outflows \citep{Claussen1994}. \herschel\, observations \citep{Rivera-Ingraham2013} reveal the HFS nature of W3-Main, with a spiral structure (at least 5 arms). The luminous infrared sources and compact HII regions are located within the dense hub. Early sub-mm observations by \citet{Ladd1993} only traced the elongated dense hub, where they identified the "two centers" as SMS1 and SMS2 peaks.

The entire W3 complex hosts clusters of stars representing a large age spread, as shown by \citet{FeigelsonTownsley2008}, who suggest that star formation has occurred in a prolonged fashion, with a time-dependent IMF in this region. Within W3-Main, \citet{Bik2014} show that the young low-mass star population is at least 2-3\,Myr old, while the massive star population is younger, confirming the well-known sequence of low and high-mass star formation \citep{Kumar2006,Ojha2010}.

The S\,106 molecular cloud is a nearby, 1.09-1.3\,kpc \citep{Xu2013,Zucker2020} star-forming region, known for its distinct bipolar radiation bubble/wind cavity, driven by a late O-type star. This is evident from optical and infrared images from the Hubble Space Telescope (HST) \citep{Bally1998,Bally2022}. The young massive star, with dust luminosity inferred from sub-mm emission $>$10$^4$\,L$_{\odot}$ \citep{Adams2015}, is located in a `cold dust bar' \citep{Mezger1987}, which is clearly a dense hub of an HFS, as evidenced by recent observations of cold dust and molecular gas \citep{Schneider2007,Adams2015}. The nearly symmetric bipolar radiation bubble implies that the HFS is at a low inclination angle to the plane-of-the-sky, with the hub viewed almost edge-on.

In summary, the three O-star forming targets— W3(OH), W3\,Main, and S\,106—represent an evolutionary sequence, from the youngest (W3(OH)) to the oldest (S\,106), respectively. While W3(OH) and S\,106 represent HFS with relatively isolated O stars in early and evolved stages, W3\,Main is more complex due to its total mass, multiple OB sources, and rich cluster. This study discusses
new sub-mm dust polarization observations of these targets. Centered on the most luminous peak (hub) of the HFS, the observations cover an area of $\sim$10\,pc$\times$10\,pc (W3(OH) and W3\,Main, d=2.0\,kpc) and $\sim$5\,pc$\times$5\,pc (S\,106, d=1.3\,kpc), with a spatial resolution of 0.14\,pc (W3(OH) and W3\,Main) and 0.09\,pc (S\,106). Archival data from the Very Large Array (VLA), and other near- and far-infrared facilities are analyzed to examine the star formation history in all targets.

This paper is organized as follows: in Sect.\,\ref{Sect-Obs}, we describe the new JCMT observations and the archival data towards the three targets analyzed in this work. In Sect.\,\ref{Int-Bfield-structure}, we present the intensity and B-field structures of our three targets. 
Sect.\,\ref{obsprop} presents the observed properties towards the hub and filament regions of the HFSs. In particular, in Sect.\,\ref{hubdef}, we propose a method to identify the hub, in  Sect.\,\ref{profiles}, we identify different sections in the radial column density profiles of the HFSs centered onto the hubs, and in  Sect.\,\ref{Sec.Mline} we identify an evolutionary pattern in the filament line mass histograms towards the three HFSs. We then present the observed polarization properties in Sect.\,\ref{Sec.PolProp}. We discuss our  results and their interpretations in Sect.\,\ref{Sec.disc}, in particular the relation of hubs to clusters and the identification of the two peaks of star formation within the hubs (Sect.\,\ref{Sec.2eyes}) and the observational signatures to trace  the evolution of magnetized hub-filament systems (Sect.\,\ref{sec.disc.evolution}). Sect.\,\ref{Summary} summarizes and concludes the paper. 

\begin{figure*}[ht!]
   \centering
  \resizebox{9.7cm}{!}{\includegraphics[angle=0]{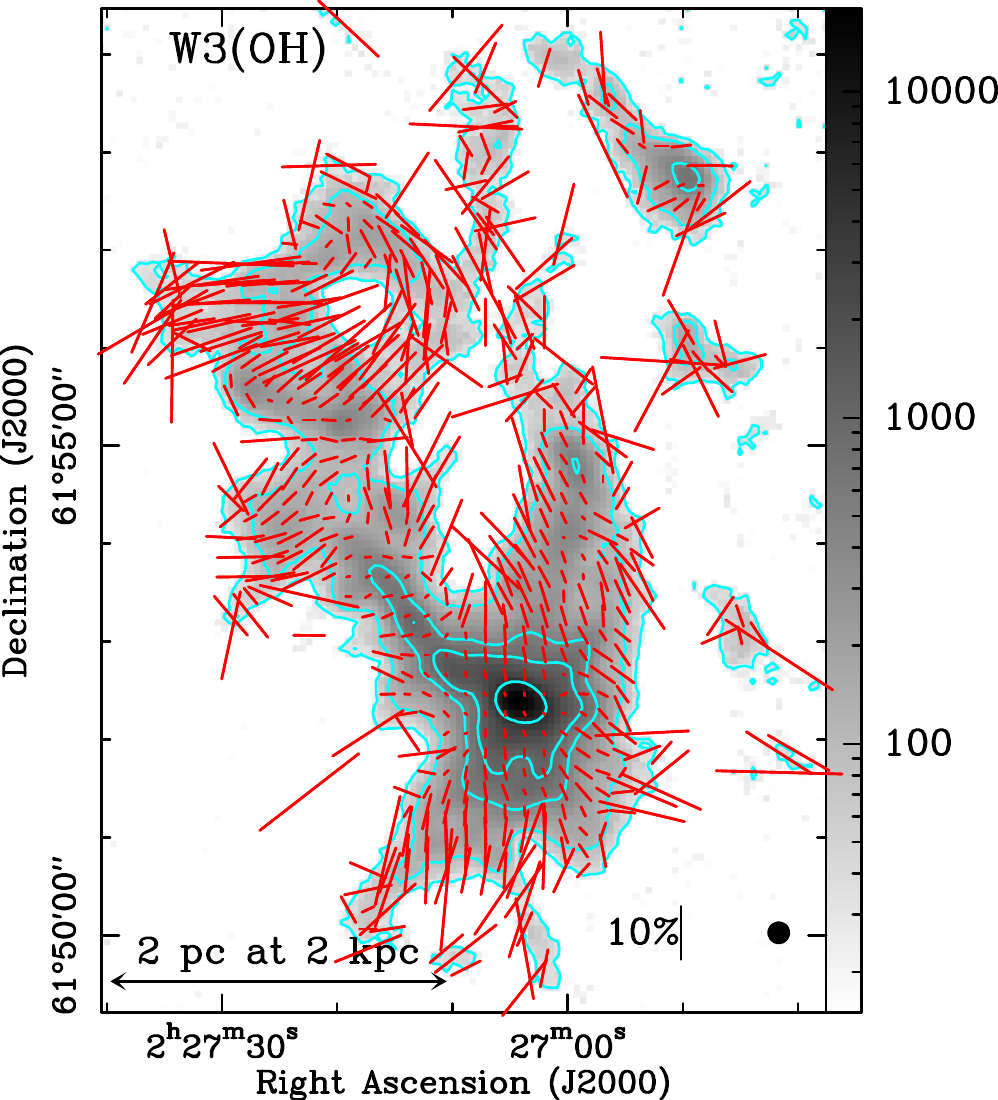}}
    \resizebox{8.0cm}{!}{\includegraphics[angle=0]{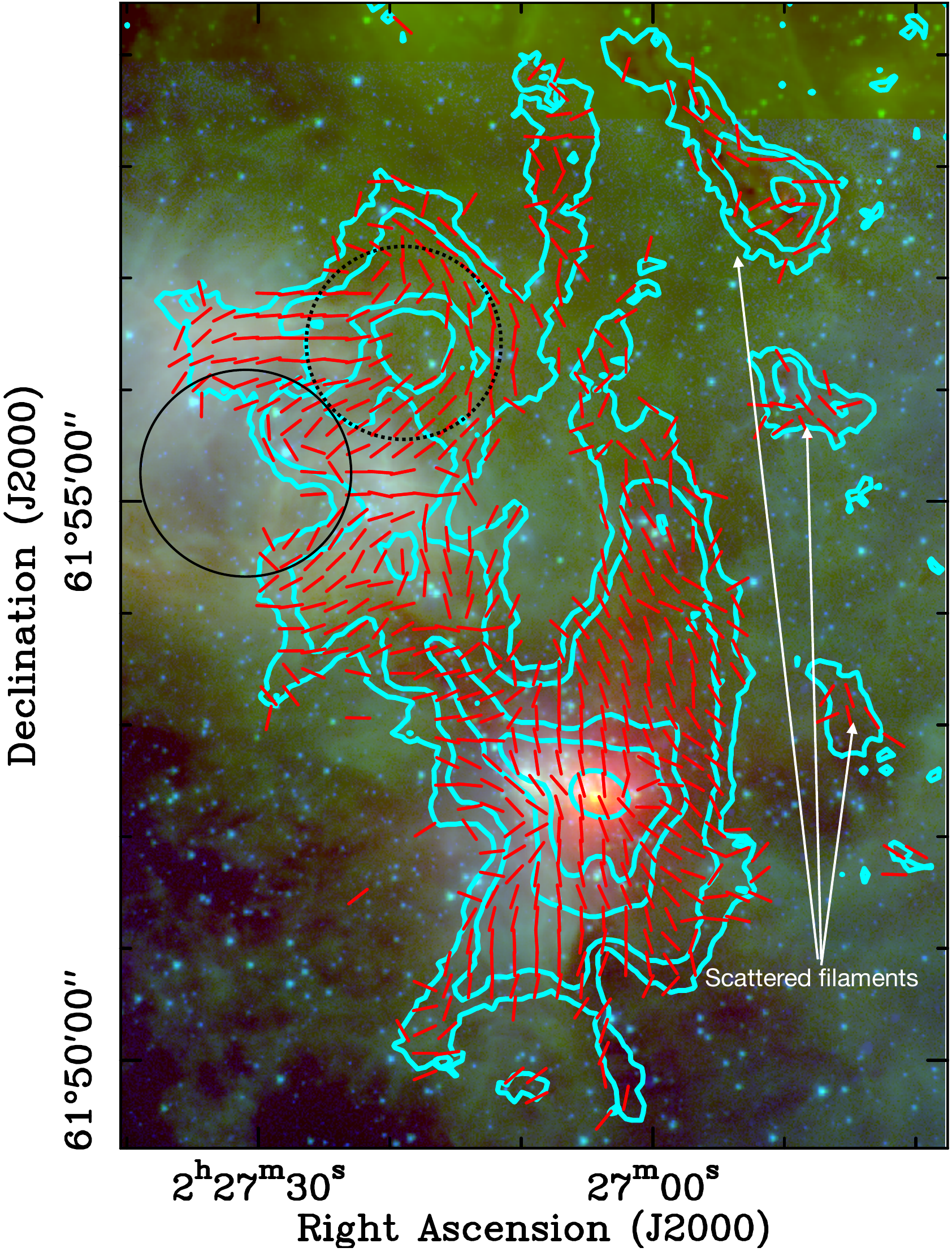}}
  \caption{
  {\it Left:} Total intensity Stokes $I$ map  at 850\,$\mu$m  observed with the JCMT SCUBA-2/POL-2  towards W3(OH) in unit of mJy\,beam$^{-1}$.  The contours are $I=30, 100, 500, 1000$, and $5000\,$mJy\,beam$^{-1}$. The lowest contour of $I=30\,$mJy\,beam$^{-1}$ is equivalent to 
  $I/\delta I\sim 13$.
 The  red  lines  show the orientation of the POS B-field angle ($\chi_{B_{\rm POS}}$) for $I/\delta I>5$, $PI/\delta PI>1$, and $PF<25\%$.
 The lengths of the red lines are proportional to $PF$. A line showing $PF=10\%$ is indicated on the  plot. The data are at an angular resolution of $14\arcsec$ or $\sim0.14$\,pc at the 2\,kpc distance of the source. 
{\it Right:} The contours and red lines are the same as in the left panel. The red lines are plotted with the same length to better show the B-field orientation. The background is a color image composed using {\em Spitzer} 3.6\,\mum, 4.5\,\mum, and 8.0\,\mum images for the  blue, green, and red, respectively. 
  }          
  \label{W3OH-I-Bfield-Spitzer}
\end{figure*}

\section{Observations}\label{Sect-Obs}

\subsection{JCMT SCUBA-2/POL-2 observations}\label{POL2-Obs}

We observed W3(OH), W3\,Main, and S\,106 at 850\,$\mu$m  using SCUBA-2/POL-2  \citep{Bastien2011,Holland2013,Friberg2016} installed on the JCMT.  The observations were carried out between August and November 2021 under dry (band 2) weather conditions with the standard SCUBA-2/POL-2  CV DAISY polarization mapping mode with a constant scanning speed of $8\arcsec$\,s$^{-1}$ and a data sampling rate of 2\,Hz. Single DAISY fields were observed for W3(OH) and W3\,Main, while 3 DAISY fields were mosaicked to map  S\,106.
The spatial distributions of the Stokes $I$, $Q$, and $U$  parameters are derived from their time-series measurements  using the {\it pol2map}  
data reduction pipeline (including {\it skyloop}) that is based on  the 
{\it Starlink} routine {\it makemap} \citep{Chapin2013,Currie2014} and   optimized for the SCUBA-2/POL-2 data. 

We adopted the flux calibration presented in  \citet{Mairs2021}.
The data are projected onto grid maps with pixel sizes of $4\arcsec$ 
at the $14\arcsec$ HPBW spatial resolution of JCMT  at 850\,$\mu$m \citep{Dempsey2013,Mairs2021}.
The mean uncertainties of the  Stokes  $I$ maps at 850\,$\mu$m are $\delta I\sim$ 2.4\,mJy/beam, 2\,mJy/beam, and  3.8\,mJy/beam, for W3(OH), W3\,Main, and  S\,106, respectively, for the 14\arcsec-beam (see also Appendix\,\ref{App1a}). 

The polarized intensity ($PI$) and polarization fraction ($PF$) are calculated  using the following relations: $PI= \sqrt{Q^2+U^2}$ and $PF=PI/I$. $PI$ and $PF$ are further debiased using the modified asymptotic estimator \citep{Plaszczynski2014}.
 The mean uncertainties of the  polarized emission maps  are $\delta PI\sim$ 1.5\,mJy/beam for W3(OH), 1.2\,mJy/beam for W3\,Main, and  1.7\,mJy/beam for S\,106, for the 14\arcsec-beam. 

 The polarization angle $\psi=0.5\,\arctan(U,Q)$ is   calculated in the IAU convention, i.e., North to East in the equatorial coordinate system. The POS magnetic field ($B_{\rm POS}$) orientation is obtained by adding $90^\circ$ to the polarization angle  $\chi_{B_{\rm POS}}= \psi+90^\circ$ \citep{Andersson2015ARAA}.

\subsection{Column density maps}\label{Coldens}

 We estimate the  column density (\nhh)  from the total intensity  Stokes $I$ values at 850\,$\mu$m ($I_{850}$) with the relation $\nhh=I_{850}/(B_{850}[T]\kappa_{850}\mu_{\rm H_2}m_{\rm H})$, where $B_{850}$ is the Planck function, $T=20\,$K is the mean dust temperature, 
 $\kappa_{850}=0.0182 \,{\rm cm}^2$/g is the dust opacity per unit mass of dust + gas at 850\,$\mu$m    \citep[e.g.,][]{Ossenkopf1994},  $\mu_{\rm H_2}=2.8$  is the mean molecular weight per hydrogen molecule \citep[e.g.,][]{Kauffmann2008}, and $m_{\rm H}$ is the mass of a hydrogen atom. We adopt  $T=20\,$K, which is the mean temperature of the dense gas (traced by the 850\,$\mu$m emission) derived from \herschel\ data towards  the W3  \citep{Rivera-Ingraham2013} and Cygnus \citep{Duarte-Cabral2013,Schneider2021} complexes. 
The column density maps are shown in Fig.\,\ref{coldensMaps_angle}.

\subsection{Archival data}\label{Spitzer-Obs}
Near-to-far-infrared archival data of all the three targets were processed to produce overlays comparing stellar and dust continuum distribution. Radio free-free emission in the centimeter wavelengths obtained by the VLA are used to trace the ionizing sources and HII regions.

Ks-band (2.2\,\mum) images of W3(OH) and W3\,Main were created by combining several exposures of the region obtained using WIRCAM on the 3.6\,m Canada France Hawaii Telescope (CFHT). The archival data was retrieved by searching the Canadian Astronomy Data Center (CADC). 
The K-band (2.2\,\mum) image for the S\,106 was obtained from the UKIDSS Galactic Plane Survey \citep{Lucas2008} which used the 3.8\,m United Kingdom Infrared Telescope (UKIRT) and the Wide-Field Camera (WFCAM). The mean full-width half maximum (FWHM) of the point sources are better than 1.0\arcsec\ in these images.

\spitzer\ data in the 3.6\,\mum, 4.5\,\mum, 5.8\,\mum, and 8\,\mum bands were obtained from the Spitzer Heritage Archive. Basic calibrated data (BCD) were retrieved and mosaicked using the  MOPEX software with standard mopex scripts for each band. The observations were made in the high-dynamic-range (HDR) mode, with the long exposures having 10.6\,seconds.

\herschel\ data of all the three targets were obtained by searching on the ESA Herschel science archive and processed using the Herschel interactive processing environment (hipe) pipelines specific to SPIRE and PACS data. We combined data from multiple programs to produce a deeper image while effectively recovering saturated pixels in the center of W3\,Main region. 

VLA data were retrieved from the NRAO data archive by searching for all past observations on the three targets and selecting high-sensitivity and high-spatial-resolutions observations. A large number of observations exist in the archive on these popular targets, observed in several centimeter bands and different interferometer configurations. To match the resolution of the infrared data above and maintain uniformity between targets we retrieved the 4.89 GHz continuum data observed in the B and C configurations. These data have a beam size of $\sim$0.7\arcsec.

\section{Total intensity images and POS B-Field angles}\label{Int-Bfield-structure}

\begin{figure*}[ht!]
   \centering
  \resizebox{9.5cm}{!}{\includegraphics[angle=0]{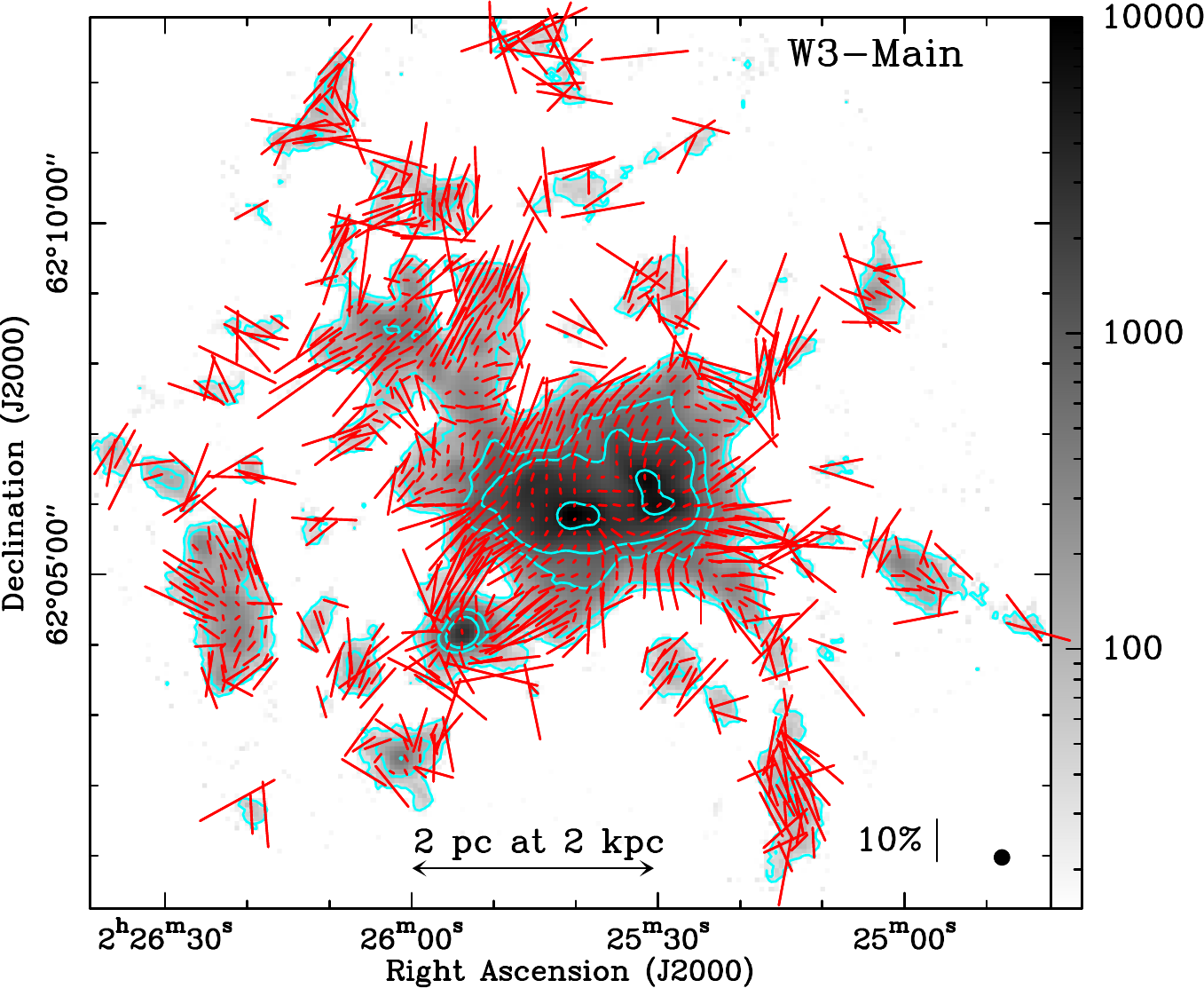}}
    \resizebox{8.5cm}{!}{\includegraphics[angle=0]{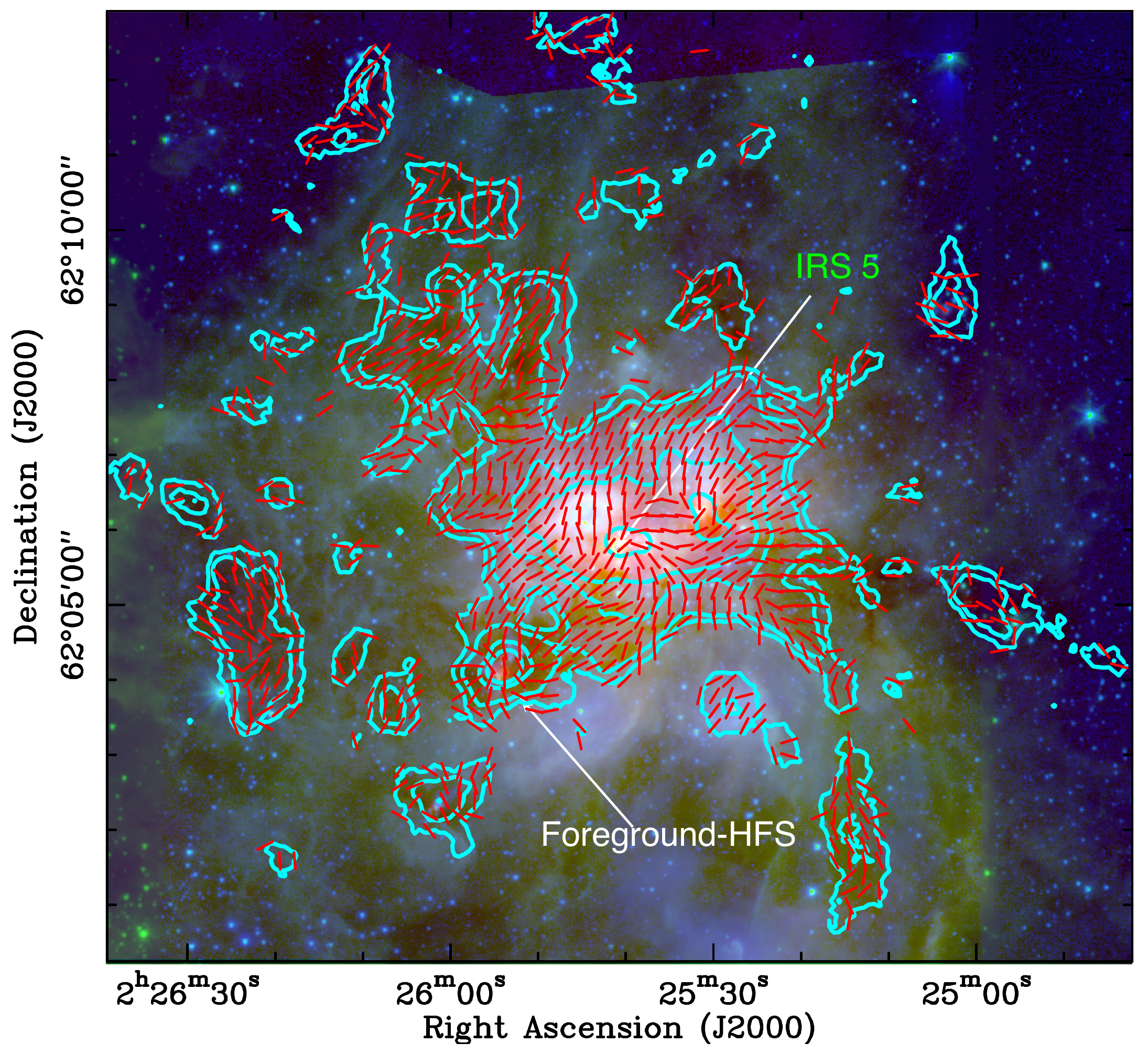}}
  \caption{Same as Fig.\,\ref{W3OH-I-Bfield-Spitzer} for W3-Main. 
 The data are at an angular resolution of $14\arcsec$ or $\sim0.14$\,pc at the 2\,kpc distance of the source. 
  }          
  \label{W3Main-I-Bfield-Spitzer}
\end{figure*}

\begin{figure*}[ht!]
   \centering
  \resizebox{9.5cm}{!}{\includegraphics[angle=0]{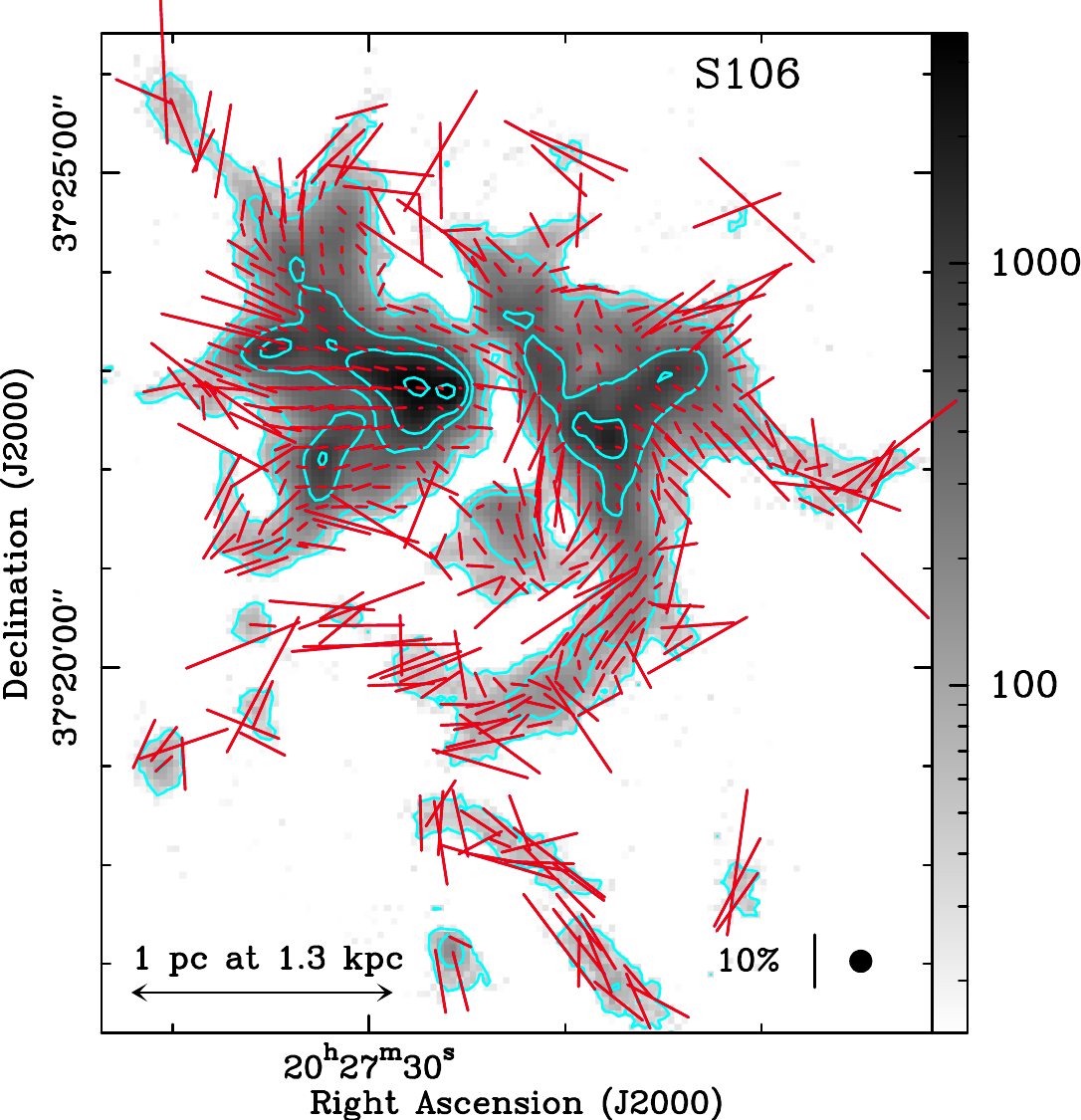}}
    \resizebox{8.4cm}{!}{\includegraphics[angle=0]{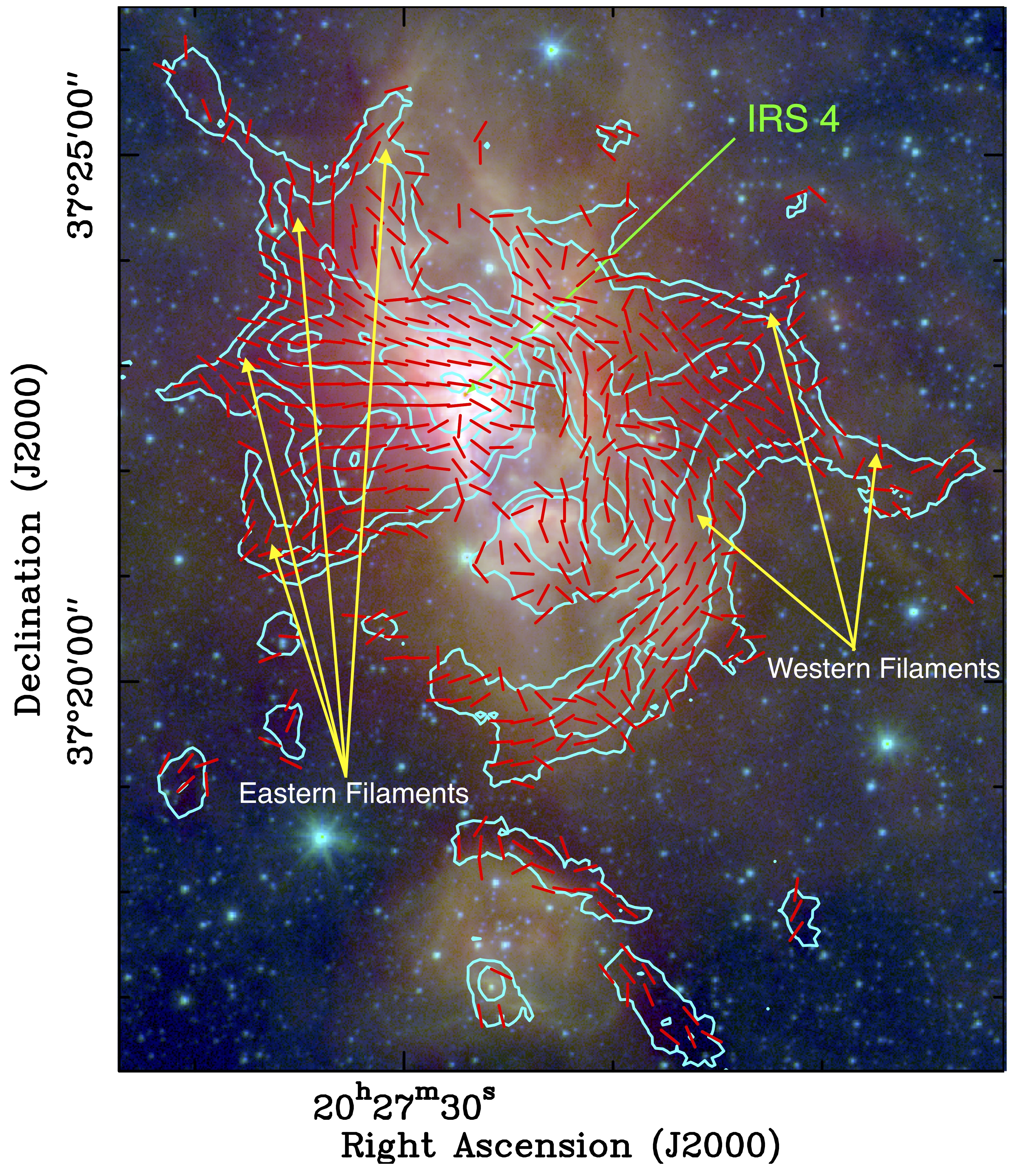}}
  \caption{
  Same as Fig.\,\ref{W3OH-I-Bfield-Spitzer} for S106. The data are at an angular resolution of $14\arcsec$ or $\sim0.09$\,pc at the 1.3\,kpc distance of the source. 
  }          
  \label{S106-I-Bfield-Spitzer}
\end{figure*}

The Stokes I images of W3(OH), W3\,Main and S\,106 at 850 \mum\, are displayed in the left panels of Figs.\,\ref{W3OH-I-Bfield-Spitzer}, \ref{W3Main-I-Bfield-Spitzer}, \& \ref{S106-I-Bfield-Spitzer}, respectively. The POS B-field is plotted using red lines showing its morphology in each of the three targets. Here, we show the POS B-field   lines for $I/\delta I>5$, $PI/\delta PI>1$, and $PF<25\%$.
From Figure\,\ref{coldensMaps_angle} in Appendix\,\ref{App1b}, one can see that the data points with  $PI/\delta PI>1$ are consistent and coherent with those with $PI/\delta PI>5$. We thus  have used  $PI/\delta PI>1$  for display purposes while a cut of $PI/\delta PI>5$ is used for the analysis (see Sec.\,\ref{Sec.PolProp}). In the right panels of these figures, the  normalized POS B-field lines and Stokes I contours are overlaid on {\em Spitzer} three colour composites of each target field where the 8.0\,\mum, 4.5\,\mum, and 3.6\,\mum band images are coded red, green and blue, respectively. These colour images roughly depict the general features of these star forming regions; a) W3(OH) appears as a deeply embedded red object, with a blue companion to its left (East), surrounded by infrared dark regions some of which trace the surrounding filaments. b) W3\,Main appears as bluish-white with embedded red sources, the entire region mostly illuminated by active star formation, with infrared dark patches at the outskirts of the region. The exception is a an object to the south east which appears compact and dark, with an embedded red source.  We name it foreground HFS because the filaments appear dark due to extinction of background diffuse emission from the W3\,Main HII region. c) S\,106 prominently reveals the bipolar shaped infrared nebula, which is largely surrounded by the 850\,\mum\ emission, the peak coinciding with the IRS\,4 massive protostar.

\begin{figure*}[!h]
   \centering
  \resizebox{9.5cm}{!}{\includegraphics[angle=0]{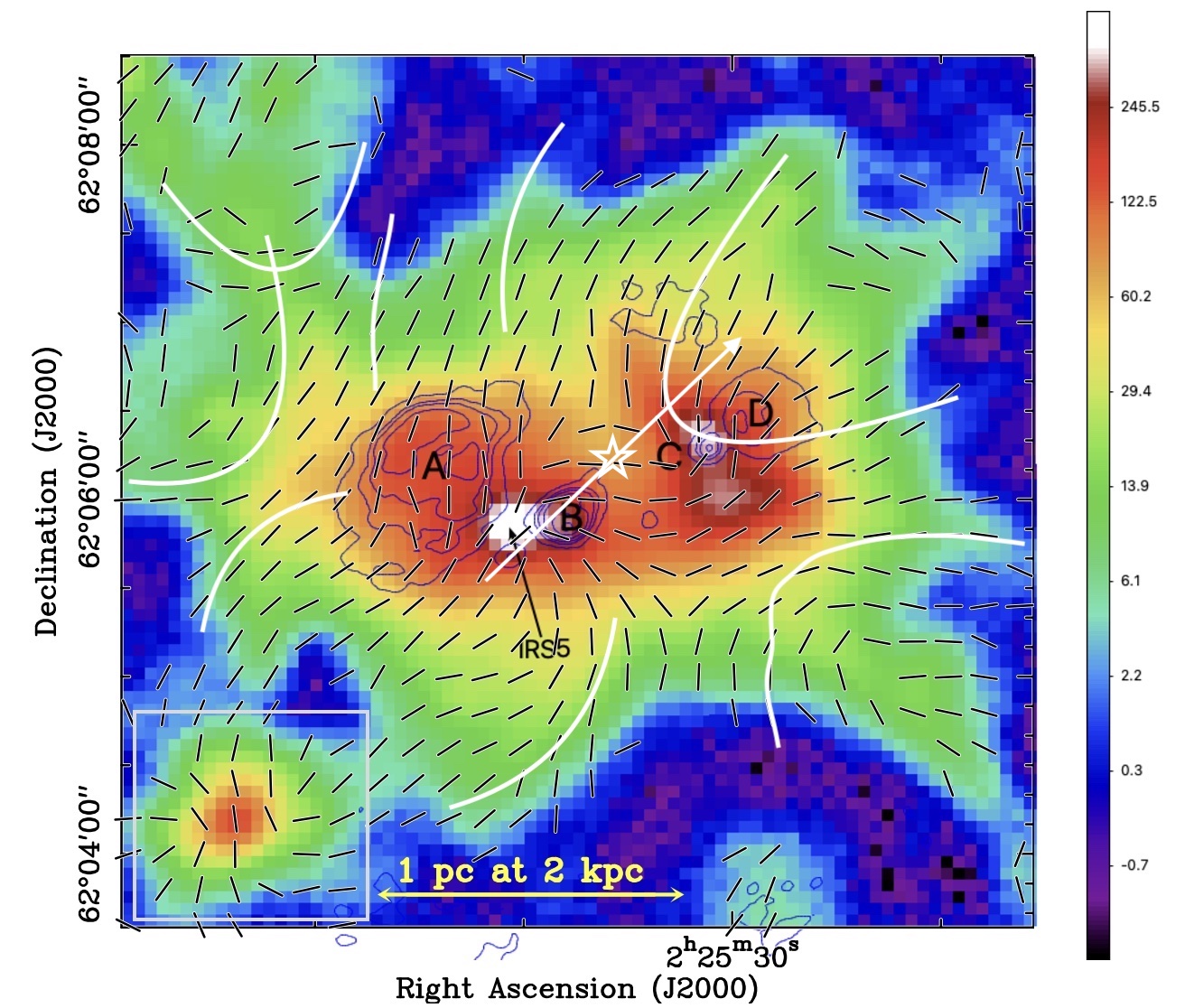}}
    \resizebox{8.cm}{!}{\includegraphics[angle=0]{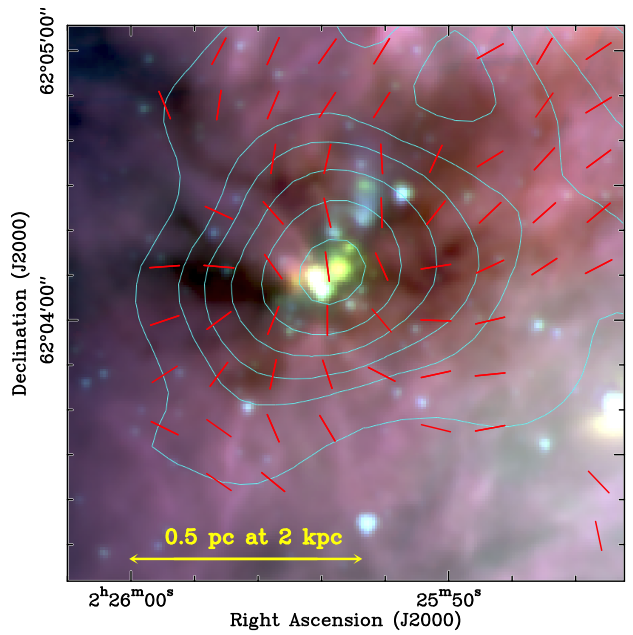}}
  \caption{Left: Blow-up image of the 850\,$\mu$m 
 Stokes $I$  emission
  towards the hub of W3\,Main (in units of Jy/beam), overlaid by normalized POS B-field lines. Grey contours display VLA 6\,cm emission at 0.1\arcsec resolution, the symbols A, B, C, and D mark the known compact HII regions. The youngest most luminous source IRS\,5 is also indicated. The white arcs aid to visualize the organized B-field pattern.  
  The white star and arrow indicate the center of a CO outflow and its direction, respectively (see text for details).   Right: W3 foreground mini-HFS, also referred to as W3\,Main\,SE indicated with the square on the left panel. The {\em Spitzer} three color image is overplotted with Stokes I contours and normalized POS B-field line.
       }          
  \label{W3main-zoom}
    \end{figure*}  

\subsection{W3(OH)}

W3(OH) (Fig.\,\ref{W3OH-I-Bfield-Spitzer}) appears as a centrally condensed peak in Stokes I, with a slight elongation to the east leaning towards the infrared peak that appear bluish in color (right panel of Fig.\,\ref{W3OH-I-Bfield-Spitzer}). The central peak or hub region is surrounded by three prominent filaments, two to the north and one to the south. To the north east, the Stokes I emission reveals a circular shell like structure (dotted circle in Fig.\,\ref{W3OH-I-Bfield-Spitzer} right panel), and adjacent to this another circular shell like structure (marked by black circle) that appears as a bright nebular emission in the {\em Spitzer} color image. Some scattered filaments (marked on the image) can also be seen to the west of the main HFS. The POS B-field lines appear as large (PF) and roughly orthogonal to the filament axis at the outer edges of the X-shaped HFS, smoothly streaming towards the central peak as they become small (PF). The northern shell like features are threaded by an organized POS B-field pattern that smoothly follows the curvature of the shells. Only half of the infrared bright nebular shell is traced by the sub-mm emission but the POS B-field traces the curvature of this shell remarkably well. The smaller filament fragments (marked on the image) display randomly scattered orientations for the B-field.

\subsection{W3\,Main}\label{W3main}

W3\,Main (Fig.\,\ref{W3Main-I-Bfield-Spitzer}) reveals itself as a hub region with two peaks of Stokes I emission, and it is surrounded by lower intensity emission filaments that stretches to much larger area compared to W3(OH). The POS B-field lines appear in groups of relatively scattered large (PF) lines spread all over the field, however appear as well organized streams of small (PF) lines in the hub region, that seem to focus on the two peaks. W3\,Main is known to be comprised of young stars from multiple episodes of star formation (cf. Sect.\,\ref{intro}), and the clumps of scattered large $PF$ values 
spread all over the observed field might be one of the signatures of multiple generation star formation. A small well condensed peak in  Stokes I  to the south east traces the foreground HFS. A rough spiral or radial shaped B-field pattern is centered on this object, especially evident from the normalized B-field lines (Fig.\,\ref{W3Main-I-Bfield-Spitzer} right panel).  As this object is newly discovered here, we list some of its estimated properties, in Tables.\,\ref{table1} \& \ref{table2}, however is omitted from the main analysis given its smaller angular extent and therefore less number of pixels to analyze. (See Sec.\,\ref{profiles})

To better display the details hidden in the B-field morphology, we zoom-in on the hub region in Fig.\,\ref{W3main-zoom} (left panel). Here the normalized POS B-field lines are overlaid on the Stokes I image. The thin grey contours represent the VLA 6\,cm continuum emission tracing the free-free emission in the compact HII regions in the hub. The UCHII regions A, B, C, D are marked following the notation from \citet{Tieftrunk1997}. The youngest most luminous source IRS\,5 is also marked. In this figure displaying the central 2\,pc region, the B-field structure appears relatively more organized compared to the larger (5\,pc) scales shown in Fig.\,\ref{W3Main-I-Bfield-Spitzer}. The white arcs aid to visualize this organized pattern. The B-field lines to the top left and bottom right of the image, represent the filaments from the larger HFS, where the B-field  appears to form arc shaped patterns pointing towards the hub. This is similar to the commonly observed pattern of the field dragged in by gravity \citep[e.g.][]{Sanhueza2021}. A similar symmetric arc shaped pattern aligns very well along a bipolar massive outflow detected by \citet{Li2019}. The B-field pattern in this direction is likely influenced by the powerful massive (4.92\,\msun) outflow. The driving source of this outflow is not identified, but the coordinates for the outflow center and its direction are represented by the white star symbol and arrow, respectively. Considering the low angular resolution of the CO outflow data, and absence of any sub-mm peak at this position, it is very likely that the actual driving source of the massive outflow is either IRS\,5 or the UCHII region B. The estimated dynamical age of the outflow is $\sim$1$\times$10$^5$ yrs \citep{Li2019}, representative of either of the sources IRS\,5 or UCHII region B. Besides the possible influence of the outflow, the B-field lines in the densest region appear to be parallel to the elongations of the total intensity emission around the UCHII regions A and D, which appear to be linked/connected through a horizontal magnetic stream. 

\begin{figure}[htpb]
   \centering
   \resizebox{7.5cm}{!}{\includegraphics{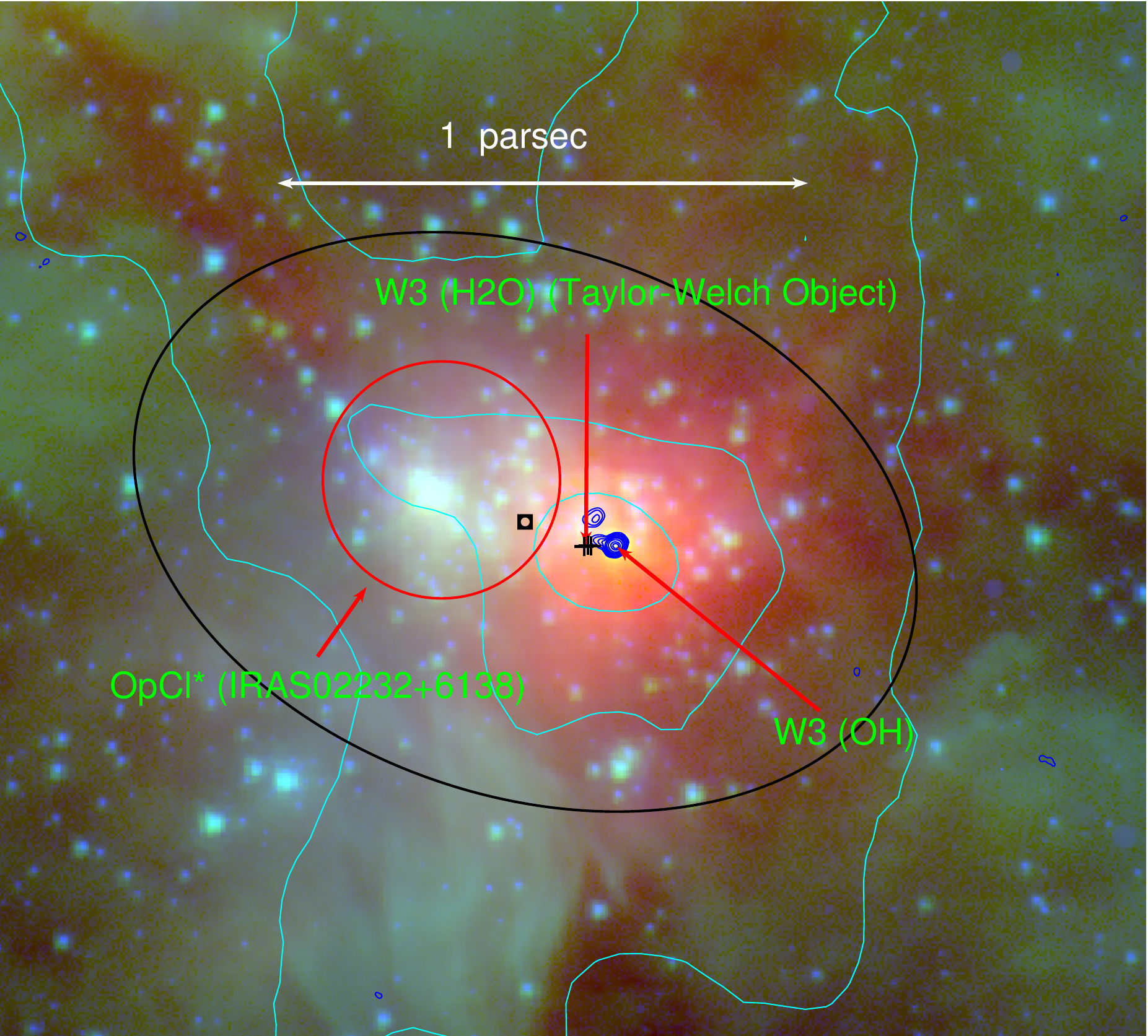}}
  \resizebox{7.5cm}{!}{\includegraphics{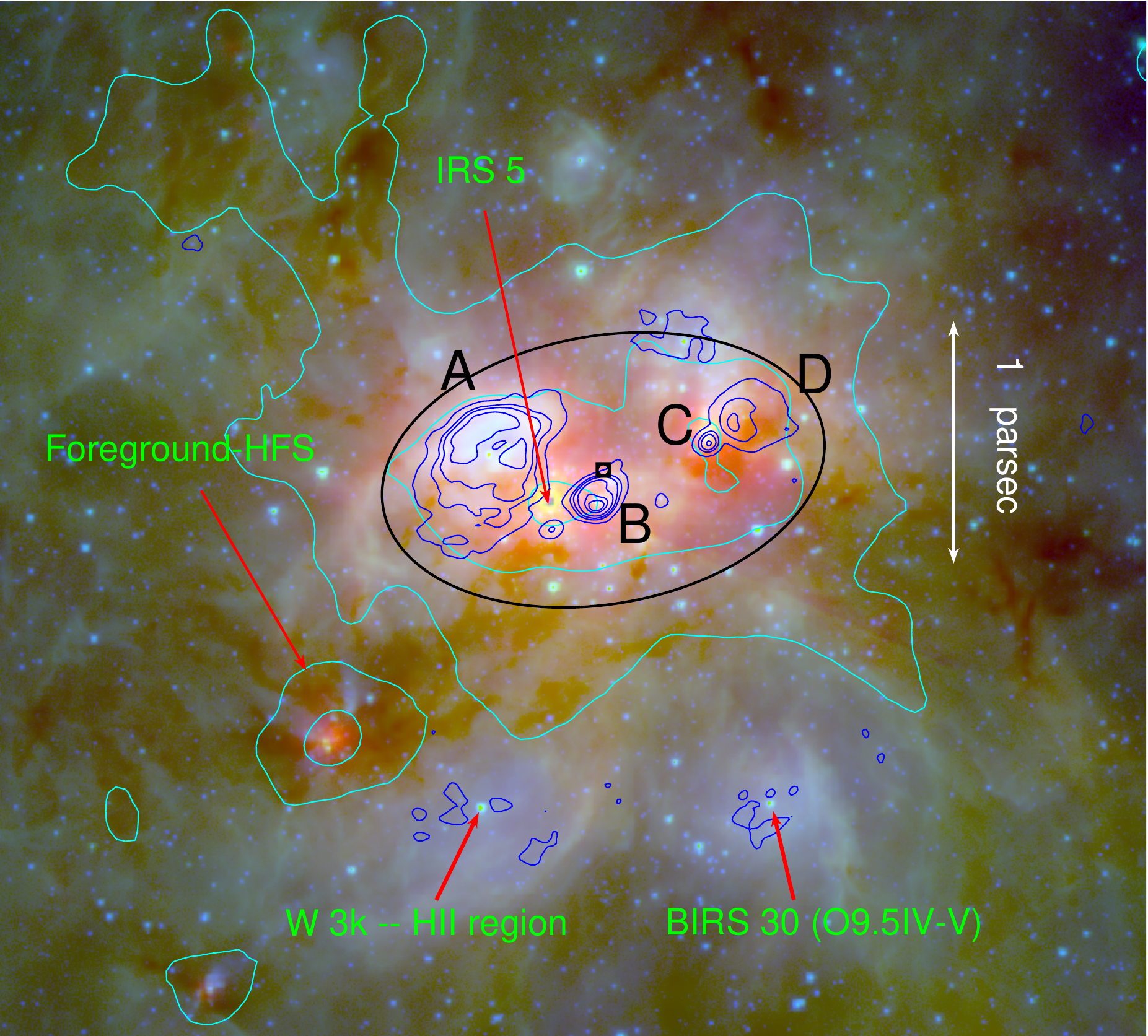}}
  \resizebox{7.5cm}{!}{\includegraphics{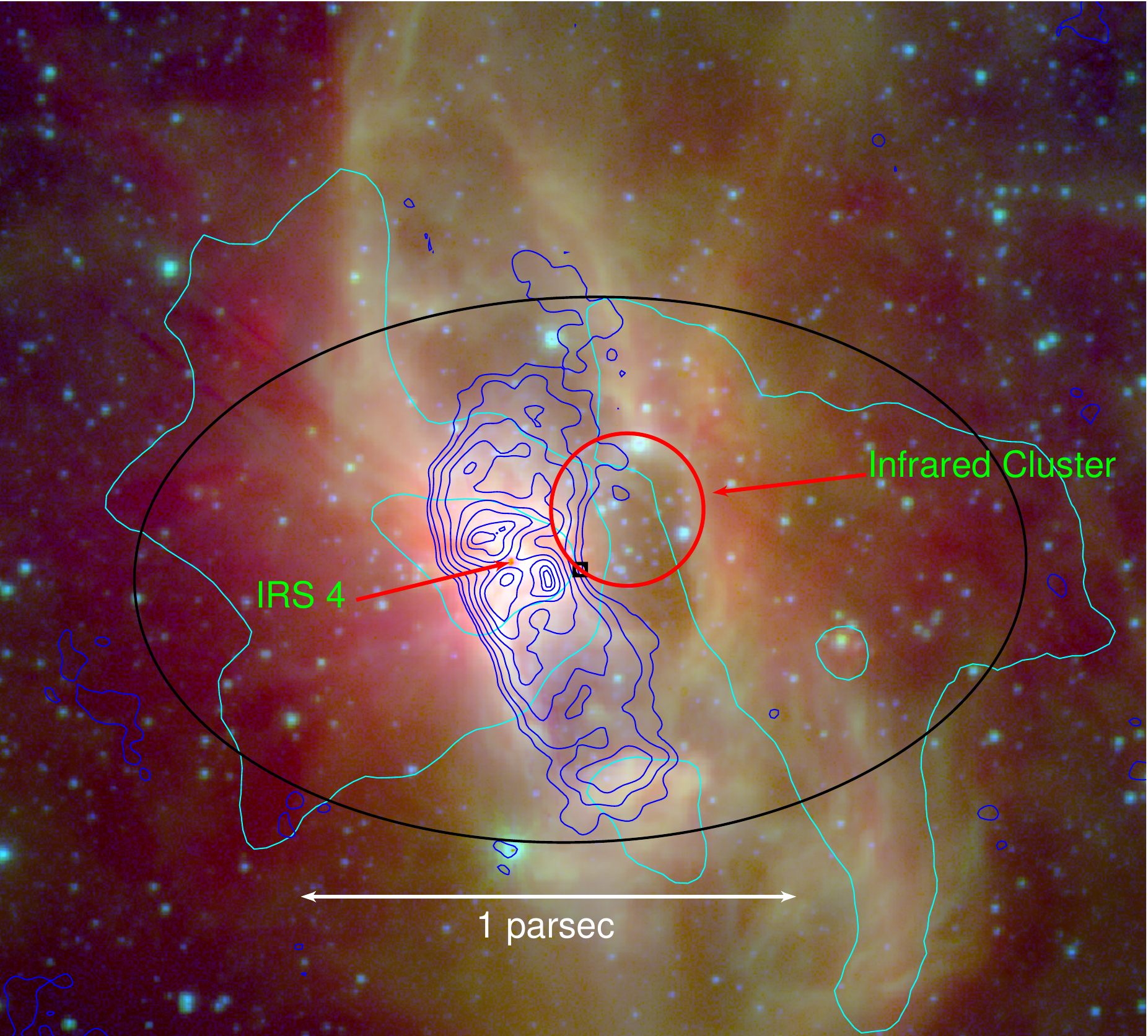}} 
  \caption{Hub regions of W3(OH), W3-Main, and S106 (clockwise from top left). The color compositions are the same as the right panels of Figs.\,\ref{W3OH-I-Bfield-Spitzer},\,\ref{W3Main-I-Bfield-Spitzer}, and \ref{S106-I-Bfield-Spitzer}. Here the cyan contours trace column densities of 10$^{22}$, 10$^{23}$ and 5$\times$10$^{23}$ cm$^{-2}$. The black ellipses represent the hubs as described in Table.\,\ref{table1}, and their centers are marked with a black box-circle symbol. The blue contours represent VLA 4.89\,GHz continuum emission obtained in B/C configurations. This emission is shown with a beam size of 0.7\arcsec\, for W3(OH), however, it is compact at the scale of 0.08\arcsec. In the "two node" system, the OpCl* \citep{Carpenter2000} represents the more evolved node compared to the younger node composed of W3(H$_2$O) and W3(OH) \citep[e.g.][]{Qin2016,Wyrowski1999}.  In W3-Main,  IRS\,5 surrounded by compact HII regions \citep{Tieftrunk1997} A \& B represent the younger node, while the evolved node corresponds to the compact regions C \& D. In S\,106, the younger and older nodes correspond to IRS\,4 and the infrared cluster, respectively \citep{Saito2009}.
}          
  \label{hubzoom}
    \end{figure}
    
The white box inset in the left panel of Fig.\,\ref{W3main-zoom} represents a foreground region that is further zoomed in the right panel. Here we use a {\em Spitzer} three color composite background which uncovers a previously unknown mini-HFS with filaments represented by dark lanes converging to the central hub with a cluster of infrared point-sources, of which two are dominant sources. The greenish colour of this cluster implies an excess emission in the 4.5\,\mum band and therefore indicative of EGO's \citep{Cyganowski2008} and outflow activity. This HFS appears to extend up to 1\,pc in size and the B-field lines display a rough spiral pattern centered on the cluster of point sources or the hub. In Fig.\,\ref{W3Main-I-Bfield-Spitzer}, this foreground HFS appears as a concentrated peak of circular contours to the south-east of the main-hub. In the \herschel\, study by \citet{Rivera-Ingraham2013}, this mini-HFS is identified as the coldest clump W3\,Main\,SE.

\subsection{S\,106}

In S\,106 the brightest peak of Stokes I corresponds to the IRS\,4 source that is the junction of at least four filaments to the east ( marked in Fig.\,\ref{S106-I-Bfield-Spitzer}). A second peak to the west is connected by three filaments (marked as western filaments). The southern long filament connected to this peak, also appears to trace the walls of the bipolar cavity. Indeed, the central region of this target is significantly shaped by the bipolar radiation bubble/HII region. There are two important features traced by the POS B-field morphology in this region. First, the B-field pattern is aligned along the dense edge-on torus-like structure within which the driving source of the radiation bubble/outflow, namely IRS\,4, is located. However, at the outer regions of the filament, at lower Stokes {\it I} intensities, the B-field lines appear relatively more scattered, and close to perpendicular to the filament axis at the extreme north-eastern filament. Both on the eastern and western filament groups, the scattered B-fields with large $PF$ values appear to smoothly stream towards the central regions, especially close to the IRS\,4 peak as $PF$ decreases. Second, the POS B-field appears to be aligned along the arc-shaped walls of the bipolar cavity, both to the north and the south. The southern dense filament appears to be mostly curved along the edge of the bipolar cavity and even in this filament the B-field lines align along the filament axis. Overall, it suggests that the B-field morphology is significantly re-arranged by the bipolar outflow/radiation bubble.

\begin{figure*}[!h]
   \centering
  \resizebox{6.cm}{!}{\includegraphics[angle=0]{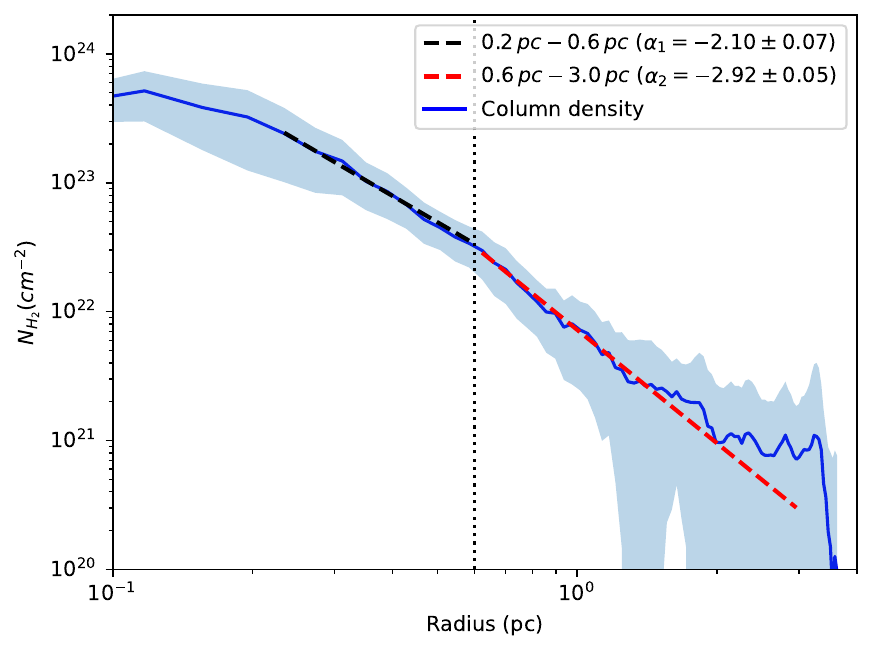}}
  \resizebox{6.cm}{!}{\includegraphics[angle=0]{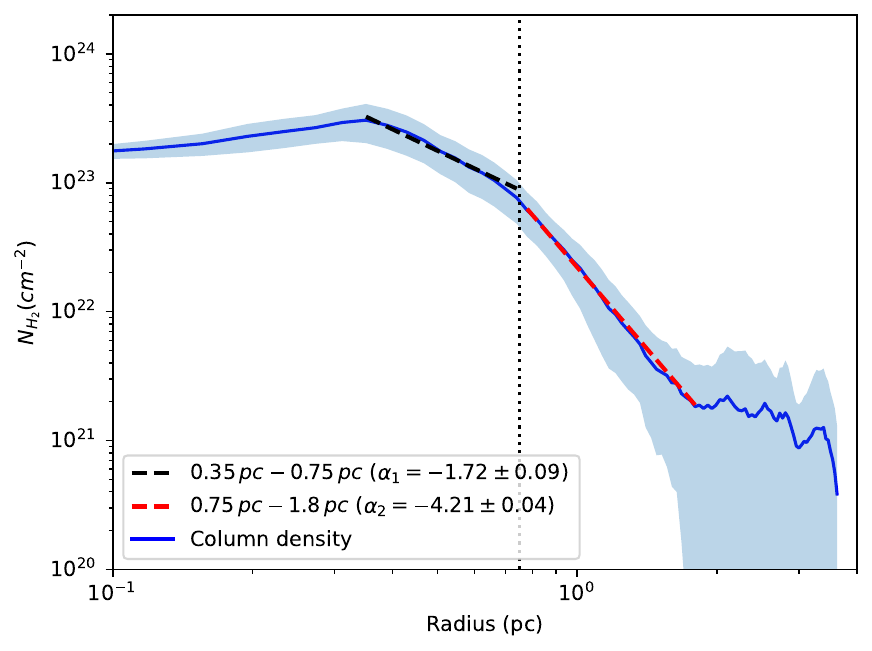}}
  \resizebox{6.cm}{!}{\includegraphics[angle=0]{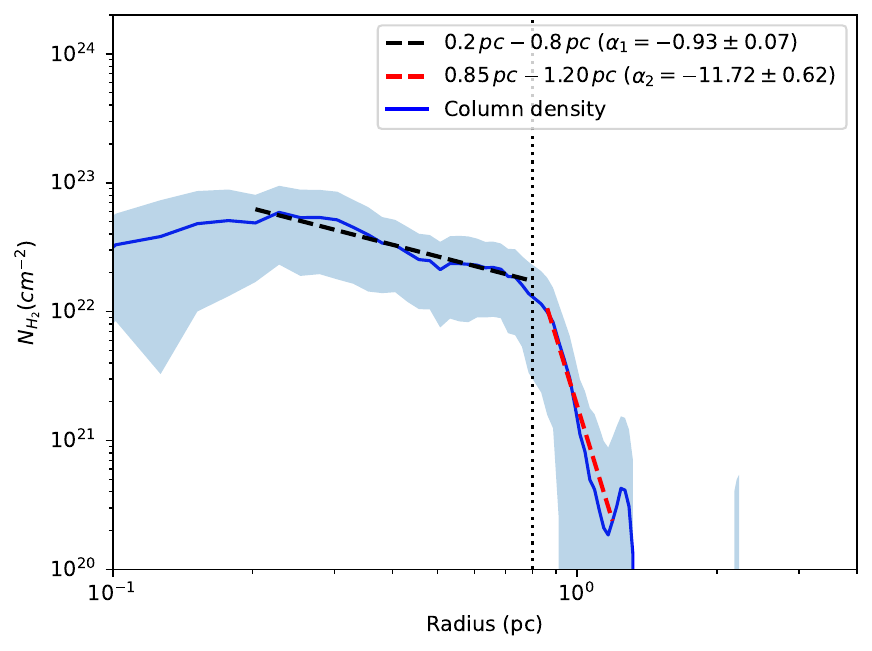}}
  \caption{Left to right: Circularly averaged radial profiles of column densities and associated power-law fits displayed for targets W3(OH), W3 Main and S106. The shaded region indicates the standard deviation of the  values in each annulus. The vertical dotted lines are marked at 0.6, 0.75 and 0.8\,pc respectively for the left to right panels, representing the approximate transition point between power laws. In W3-Main and S\,106 the center of the hub is located in between the "two nodes" (see Sec.\ref{hubdef} for details and Fig.\,\ref{hubzoom}).}          
  \label{radialProf}
    \end{figure*}

\begin{table*}
 \caption[]{Properties of the HFS  and their hubs.} 
    \label{table1}
\tabcolsep 4pt
 {\centering \begin{tabular}{l|c|cc|cccc|c|c|c}
    \toprule
    \multirow{2}{*}{Target} &
     Evolutionary &
     \multicolumn{2}{|c|}{Hub central position }&
        \multicolumn{4}{c}{Hub properties$^{a}$} &
     \multicolumn{1}{|c}{Total Mass} &
      \multicolumn{1}{|c}{Luminosity$^{d}$} &
      \multicolumn{1}{|c}{Distance}  \\
    & stage& RA (J2000) & DEC (J2000) & maj (pc) & min (pc) & PA (deg)$^b$ & size (pc)$^c$& \msun & 10$^4$\,L$_{\odot}$  & kpc \\
      \midrule
    W3(OH) &Early & 02:27:06.32 & +61:52:29.4 &1.2 & 0.68 & 30 & 0.94 & 2700 & 2.3$\pm$1.0& 2.0  \\
    W3-Main &Intermediate & 02:25:36.77 & +62:06:05.8 &1.5 & 0.89 & 93& 1.2 & 5700& 13.8& 2.0 \\
    S\,106 &Evolved & 20:27:24.89 & +37:22:45.6 & 1.8 & 1.11 & 92 & 1.5& 940& 9$\pm$1& 1.3 \\
    \hline
     W3\,SE &Early& 02:25:53.78 & +62:04:11.9 &0.32 & 0.26 & 30 &0.31 & - & - & 2.0 \\
    \bottomrule
  \end{tabular} 
  \begin{list}{}{}
 \item[]{{\bf Notes:} 
 We compute the total mass within the observed field of view of our targets for  $\nhh\ge10^{21}\NHUNIT$. The $\nhh$ maps are derived for $T=20$\,K as explained in Sect.\,\ref{Coldens}.$^a$ The major axes are set from the radial profile fitting, and using {\it SAOImageDS9} the ellipses are fit by hand to the equal density contour. $^b$PA of the ellipse. $^{c}$size = (maj$+$min)/2.  $^{d}$See references in Sect.\,\ref{intro}.
}
\end{list}      
   \par} 
\end{table*}

\section{Properties of the hub and filament regions towards HFSs}\label{obsprop}

\subsection{Defining and identifying the hub center}\label{hubdef}

The total intensity (Stokes {\it I}) maps discussed in the previous section were used to derive the column density maps (Sect.\,\ref{Coldens}) and identify filamentary structures in each target. The method to identify filamentary structures is presented in Appendix\,\ref{App1b}. \citet{Myers2009} defined hubs as high column density low-aspect ratio objects compared to filaments that are low column density, high-aspect ratio structures. However, given the two central peaks of Stokes {\it I} emission, especially in W3\,Main and S\,106, it is not straightforward to define the boundaries of a hub. To aid the identification of a hub in the same way in all three targets, we have used further deliberations made by the F2C paradigm of \citet{Kumar2020}. The F2C paradigm advocates the presence of two embedded peaks within the hub as a result of the evolution of the hub from stage II to stage III as defined by \citet{Kumar2020}. Considering the existence of two peaks (for evolved sources)
is necessary to first define the {\em center} of the HFS, and therefore it is the first constraining parameter of the hub. 

For this purpose, we examined the literature for all known sign-posts of star formation in the central regions of the targets and display a zoom-in view of the central regions in Fig.\,\ref{hubzoom}. Even though W3(OH) and S\,106 do not possess two major peaks in the column density maps unlike W3\,Main, it turns out that both targets have well-cataloged compact embedded infrared clusters adjacent to the sub-mm peak. In W3(OH), this embedded cluster \citep{Carpenter2000} reveals itself as the bluish group of stars with a bright blue-white star. W3(OH) itself is surrounded by young deeply embedded infrared stars, but the focus of several detailed studies, mostly using (sub)mm interferometry, has been on the young, luminous pair of objects W3(OH) and W3(H$_2$O), which are marked in Fig.\,\ref{hubzoom}. In W3\,Main, the "two peaks" or the "two nodes" of the hub region are well-identified at both sub(mm) and infrared wavelengths.  In S\,106, the embedded cluster is identified as a loose group of infrared excess young stellar objects \citep{Saito2009}. 

We define the central position of the hub as the projected midpoint of the two peaks. Thus the central positions of the hubs are the midpoints between the W3(OH) sub(mm) peak and the bluish-white star of the embedded cluster for W3(OH),  the two sub(mm) peaks for W3\,Main, and IRS\,4 and the center of the infrared cluster in S\,106. The center of the hubs are marked using a black box-circle symbol in Fig.\,\ref{hubzoom} and listed as the central position in Table.\,\ref{table1}. 

Free-free emission at centimeter wavelengths observed with the Very Large Array (VLA) is an important indicator of the evolutionary state of the ionized region, and used as a standard tool to study compact (ultra and hyper) HII regions. In Fig.\,\ref{hubzoom}, we over-plot the best available centimeter continuum data from the VLA data archive to visualize all the compact HII regions in each of the targets. As evident from the figure, they are the most compact in W3(OH), and the most extended in S\,106, further justifying our choice of the evolutionary sequence from the youngest to the oldest. Owing to the well-known multiple episodes of star formation in W3\,Main, other HII regions from OB stars surround the present episode. These are marked as W\,3k and BIRS30 in Fig.\,\ref{hubzoom}.

\subsection{Measuring the hub size from the HFS column density radial profile}\label{profiles}

\begin{figure*}[!h]
   \centering
  \resizebox{6.cm}{!}{\includegraphics[angle=0]{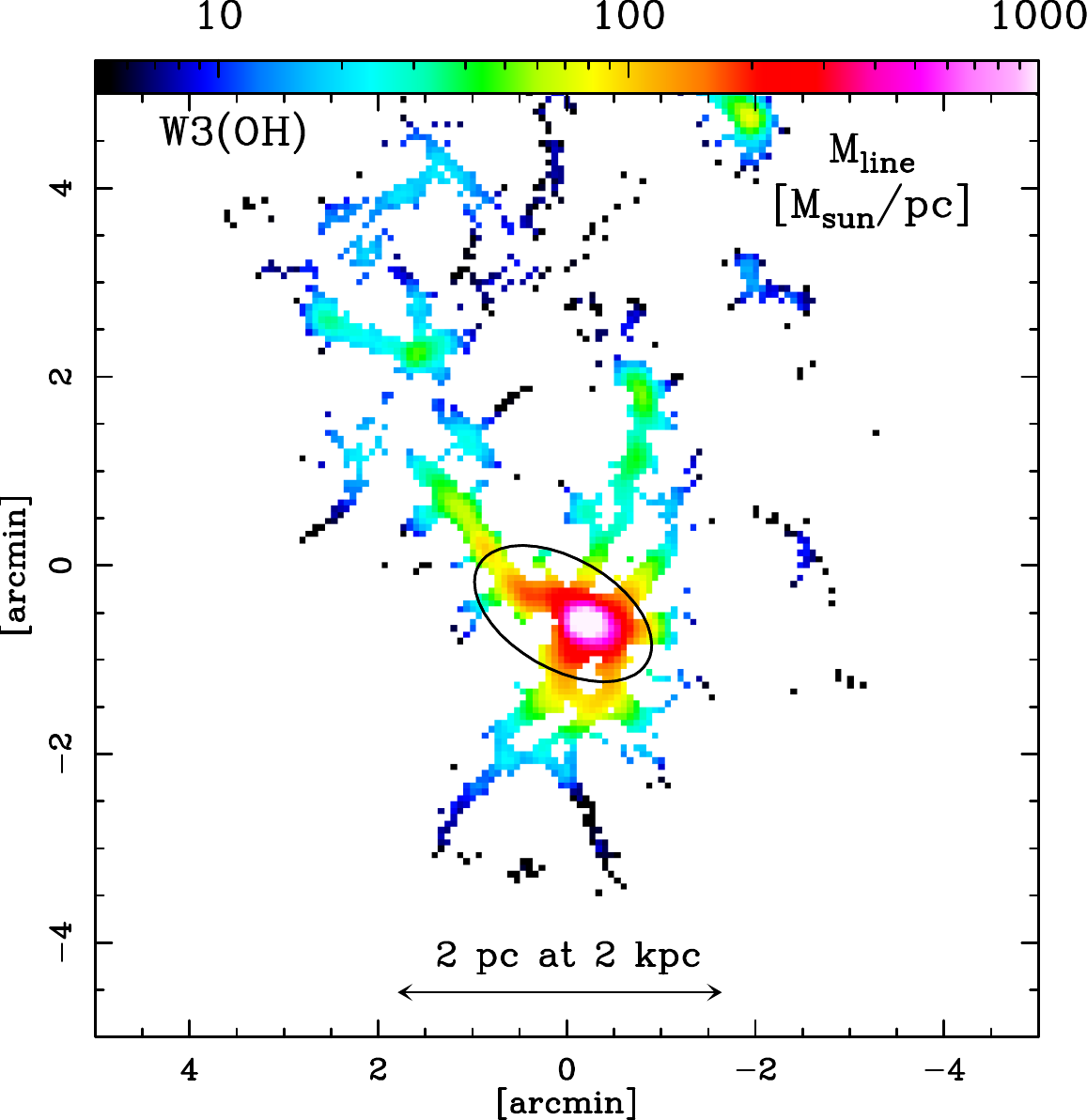}}
  \resizebox{6.cm}{!}{\includegraphics[angle=0]{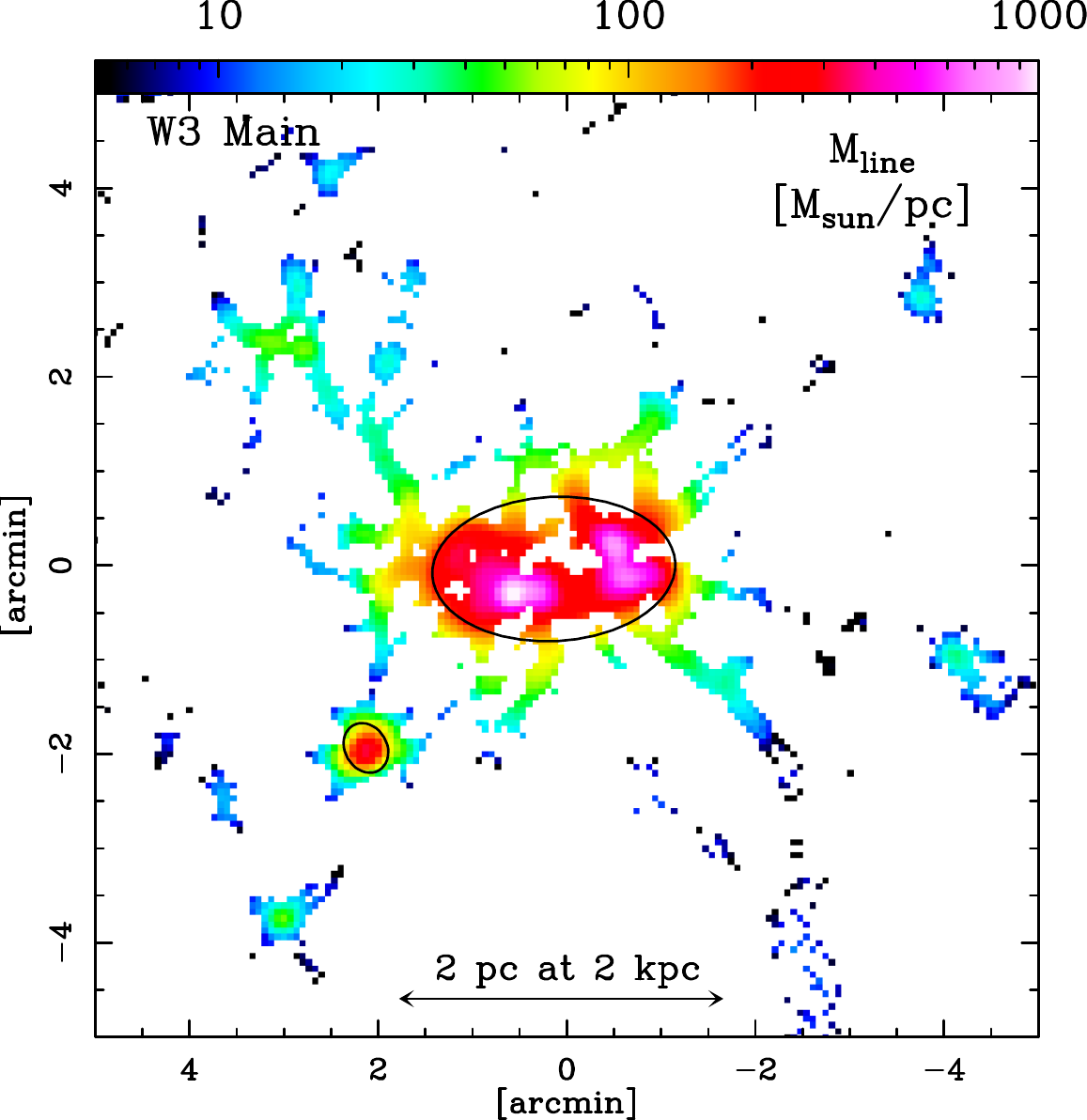}}
  \resizebox{6.cm}{!}{\includegraphics[angle=0]{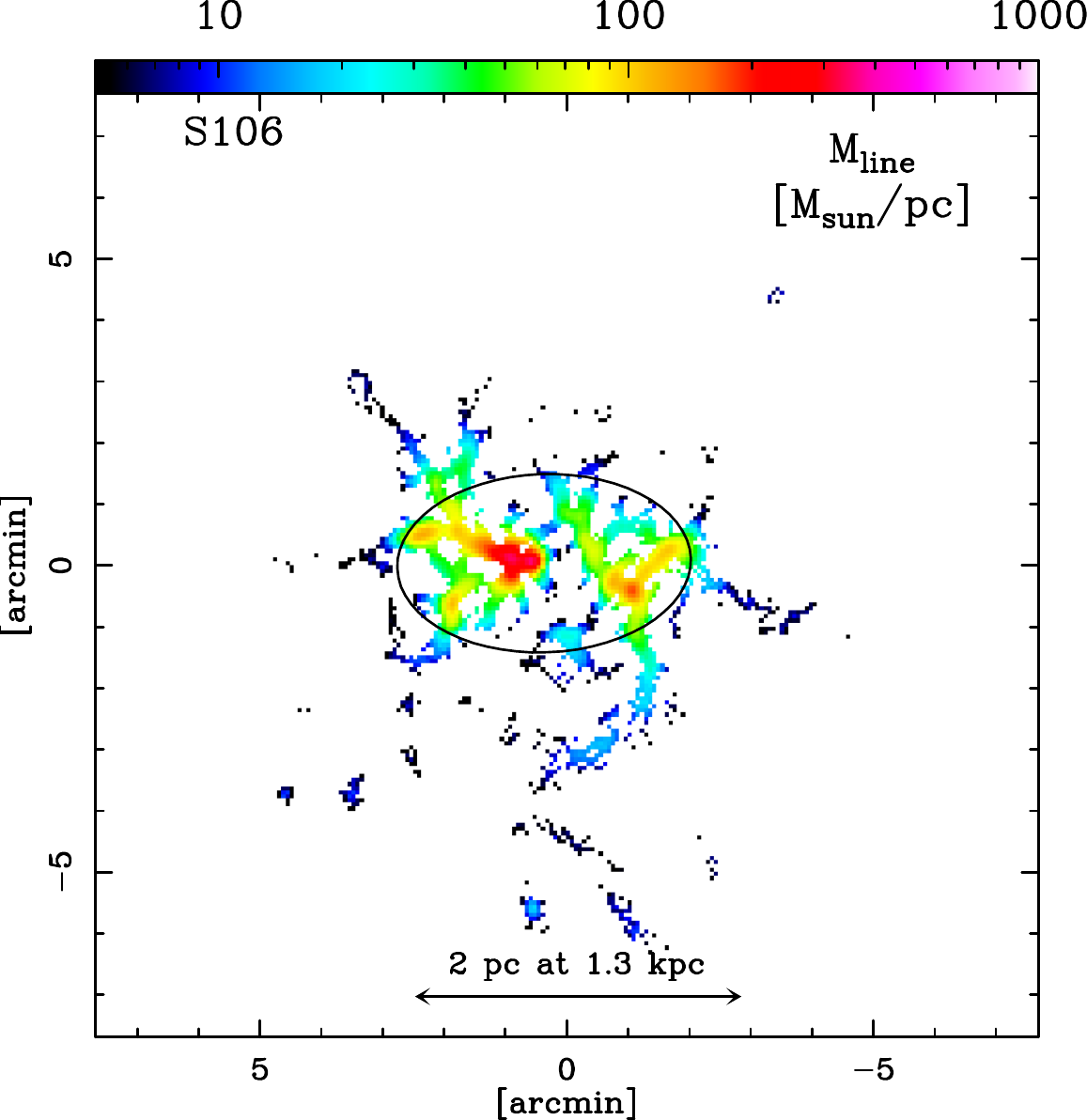}}
  \caption{Maps of filament line mass, $M_{\rm line}$  (in  \msun/pc)  
  for W3(OH), W3 Main, and S106 from left to right. All three panels have the same physical scale of 5.8\,pc$\,\times\,$5.8\,pc and the same color scale. The black ellipses indicate the hub sizes as given in Table\,\ref{table1}.}          
  \label{MlineEllipse}
    \end{figure*}

Having identified the center of the hub, it is necessary to define a radius or boundary of the hub. To aid this purpose, we produced radial plots of azimuthally averaged column density for each target as shown in Fig.\,\ref{radialProf}. The observed data was fit with between one and three power-laws while allowing fitting ranges to obtain least fitting errors. In all three cases, attempts to fit the data with a single power-law clearly failed. For W3(OH) and W3\,Main, three power-laws could be fit, but the third one represented the very outer regions with poor signal-to-noise, which we therefore omitted. Also, in W3\,Main and S\,106 we deliberately truncated the inner-limit, as the dip within the 0.1-0.2\,pc is created by the lower density mid-point region between the two nodes which is chosen as the hub-center. As evident from the figure, the radial plots are best fit with two distinct power-law slopes, which separate the entire HFS into two distinct regions. The turn-over point is a natural outcome of minimizing the errors on the two power-law fits. We define this turn-over point in the power-law slopes as the boundary of the hub. With the turnover point as reference, we will call the inner and outer regions as hub and filaments, respectively. Using the center we fitted an ellipse to equal column density contours by setting the semi-major axis of the ellipse approximately equal to the scale defined by the turnover point in the radial profiles. Only the minor-axis and PA of the ellipse were fit. The projected major and minor axes and the position angles of the fitted ellipses are listed in Table\,\ref{table1} as hub properties, and are shown as black ellipses in Figs.\,\ref{hubzoom} \& \ref{MlineEllipse}. Apart from the notions inspired by the F2C paradigm, we also note that the 2D projected geometry in Stokes {\it I }emission is approximately elliptical which prompted the choice of an ellipse against other contours or geometries. For the foreground mini-HFS W3\,Main SE however, we have not conducted a similar analysis owing to the lack of sufficient pixels encompassed within the JCMT 14\arcsec beam. Only an equal density contour was used to fit an ellipse, the properties of which are listed in Table\,\ref{table1} and shown in Fig.\,\ref{MlineEllipse}. We notice an increasing size of the hub for the early to evolved HFS evolutionary stage.

The column density radial profiles reveal several important features: 

\begin{itemize}
\item
a) The central region, not fitted with power-laws, shows a flattening in W3(OH) ($\le 0.2$\,pc) and a dip with positive slopes in W3\,Main ($\le 0.35$\,pc) and S\,106 ($\le 0.2$\,pc). This dip is due to the low density valley between the "two peaks" of star formation that are represented by the pairs of sub(mm) peaks and infrared clusters.\

\item
b) The regions separated by the turnover points (between the two power-laws) are smoothly connected in the early evolutionary stage target W3(OH), and become increasingly distinct as a function of evolutionary time, in W3\,Main and S\,106. The difference in the slopes of power-laws on either side of the turnover point increases from the early stage W3(OH) target to the oldest region S\,106, indicating how the two regions, the hub and filaments, evolve as star formation progresses. \

\item
c) If a power-law slope of -2 or lower were to be interpreted as suggestive of gravitating gas \citep{Ward-Thompson1994}, and slopes higher than -2 as non-gravitating gas, the hub in W3(OH) is undergoing gravitation collapse, while that in W3\,Main and S\,106 are not, indeed they may be even relaxing. On the contrary the filament regions in all three targets are gravitationally collapsing to the hub, albeit at different rates. These features are similar to the density profiles discussed for the Mon\,R2 region \citep{Kumar2021}. Even though these observed variations of the power-law slopes contain important physics, we refrain from a detailed interpretation at this point.\

\end{itemize}

    \begin{center}
\begin{table*}
 \caption[]{Line mass properties of hub and filaments}
    \label{table2}
\tabcolsep 4pt
 {\centering \begin{tabular}{l||c|cccc|c||c|cccc|c}
    \toprule
    \multirow{3}{*}{Target} &
       \multicolumn{6}{c|}{Hub} & \multicolumn{6}{|c}{Filaments} \\
     & pixel number & \multicolumn{4}{|c|}{Line mass [\msun\,pc$^{-1}$]} & slope &  pixel number & \multicolumn{4}{|c|}{Line mass [\msun\,pc$^{-1}$]} & slope  \\
          &  &  mean  & median & SD & max && &mean & median & SD & max& \\
      \midrule
    W3(OH) & 282 & 281 & 163 & 332 & 1924  & $-1.2$  & 1583  & 24 & 17 & 22 & 139  & $-1.4$ \\
    W3-Main & 529 & 333 & 285 & 177 & 1087 & $-1.0$ & 1899 & 36 & 23 & 36 & 211 & $-2.2$\\
   W3\,SE &  &  181 & 89 & 314 & 192 & & & & & & \\
   S\,106 & 1188  & 67 & 53 & 59 & 350 &$-0.85$& 1059& 14 & 9 & 12 & 59 &$-1.4$\\
    \bottomrule
  \end{tabular} 
  \begin{list}{}{}
  \vspace{.2cm}
 \item[]{{\bf Notes:} 
 The SD columns give the standard deviation of the line mass values for the hub and filament regions indicating roughly the width of the distributions. The max columns give the maximum line mass values. The slope columns give the indices of the power-law fits. 
    }
 \end{list}      
   \par} 
\end{table*}
\end{center}
    
\begin{figure*}[!h]
   \centering
  \resizebox{18.cm}{!}{\includegraphics[angle=0]{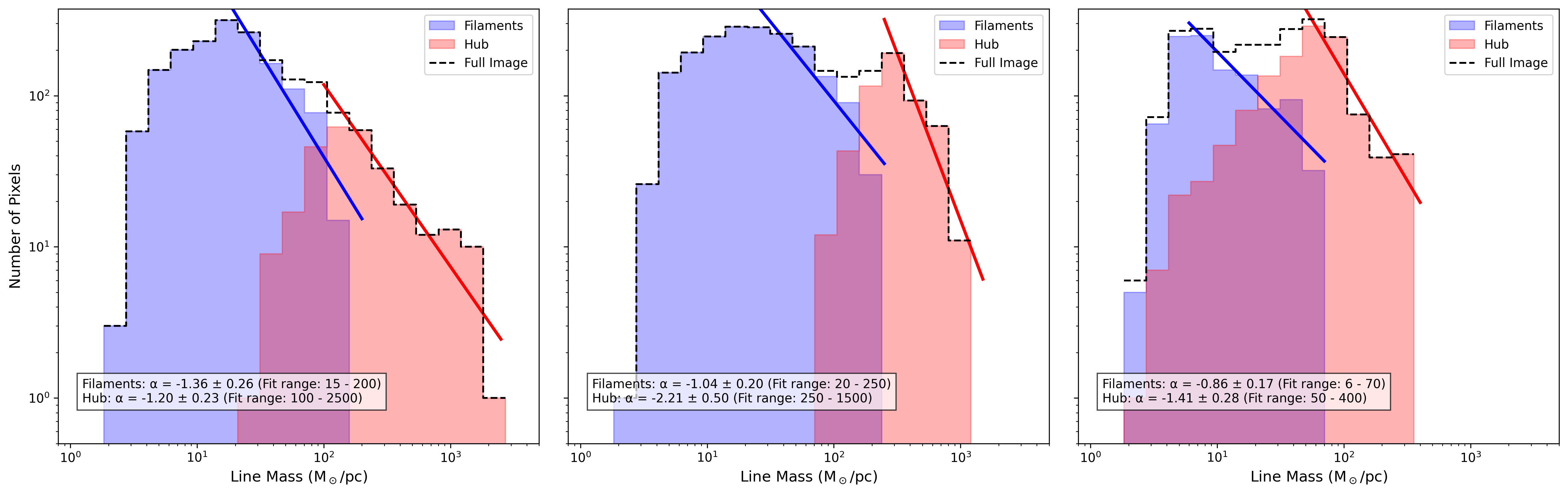}}
  \caption{Line mass histograms for the full image (black dashed line), filament (blue), and hub (red) regions for W3(OH), W3-Main, and S 106 from left to right. The fitted power-laws are shown by blue (filament) and red (hub) lines, and the fitted slopes are indicated in each panel.}       
  \label{Mlinehisto}
    \end{figure*}
    
\subsection{Filament line mass maps of HFS}\label{Sec.Mline}
    
The column density maps were used to produce line mass maps. The line mass   ($M_{\rm line}=\mu_{\rm H_2}m_{\rm H}  \nhh \, W_{\rm fil}$) is  derived from the  \nhh map (Sect.\,\ref{Coldens}) and assuming a constant filament width of $W_{\rm fil}=0.1$\,pc \citep{Arzoumanian2011,Arzoumanian2019}.
 In each region, the emission from extended structures has been masked to select only the filamentary structures using
curvature maps as described in  Appendix\,\ref{App1b}. The filament line mass maps are displayed for all the three targets in Fig.\,\ref{MlineEllipse}.

As evidenced in Fig.\,\ref{MlineEllipse}, the depth of our observations are such that the faintest detected structures in all the three targets represent filaments 
$\gtrsim10\sunpc$. Theoretically, filaments with line masses  larger than the critical equilibrium value for isothermal cylinders \citep[e.g.][]{Inutsuka1997},  $M_{\rm line, crit}^T = 2\, c_{\rm s}^2/G\sim27\,$M$_\odot$/pc \citep{Ostriker1964}, where  $c_{\rm s}\sim0.24$\,\kms\ is the sound speed at $T\,\sim\,20$\,K, are gravitationally unstable and fragment to form stars.
In addition, recent observational studies proposed that the typical mass of the cores/stars formed along a given filament is proportional to the filament line mass \citep{Andre2019,Shimajiri2023,Arzoumanian2023}.  
Consequently, higher line mass filaments would form preferentially higher mass stars.
It can be seen that the central hub regions (represented by the brightest emission and therefore high line mass filaments in Fig.\,\ref{MlineEllipse}) correspond to the center of (high mass) star formation activity as evidenced by Fig.\,\ref{hubzoom}. These findings are in accordance with the formation of massive stars only in hubs postulated by the F2C paradigm \citep{Kumar2020}. Note that the only other region which includes line mass of 100\sunpc or more in the studied target fields is the W3\,Main SE mini-HFS.

The hub region of W3(OH) displays an elliptical condensation enclosing a few hundred \sunpc with a clear peak representing nearly a 1000\sunpc,  which coincides with the well-studied OH and H$_2$O maser sources \citep{TurnerWelch1984,Giese2024}, and multiple outflows \citep{Zapata2011}. In W3-Main, the "two nodes" of the hub are distinctly visible, with peaks representing \ml$\sim$1000\sunpc. The left node coincides with the massive young source IRS\,5. The right node includes two distinct peaks aligned north-south, the northern peak coinciding with the hyper-compact HII region C (see Fig.\,\ref{hubzoom}), while the southern peak may be a younger core/massive stellar object, that can be resolved only with higher angular resolution studies. The hub region of S\,106 is relatively more dispersed including mostly \ml$\sim$50-100\sunpc structures, but the peak region corresponding to IRS\,4 has two roundish condensations of \ml$\sim$300\sunpc (red color in Fig.\,\ref{MlineEllipse}) which also drives the bipolar radiation bubble (Fig.\,\ref{S106-I-Bfield-Spitzer}). 

In order to quantify and compare the line mass properties of the hub and filament regions, we used, respectively, all the pixels inside and outside the black ellipses marked in Fig.\,\ref{MlineEllipse}. For each target, the number of pixels enclosed in these regions are listed in Table\,\ref{table2}. For each target, Fig.\,\ref{Mlinehisto} displays the pixel-wise histograms of line masses for the hub and filament regions compared to that of the total field-of-view. These histograms represent an equivalent Filament Line Mass Function (FLMF) \citep[cf.][for observational and simulation results, respectively]{Andre2019,Abe2021} in these targets\footnote{Note that the FLMFs in \citet{Andre2019} and \citet{Abe2021} correspond to the line mass distributions of individual filaments, while the FLMF in Fig.\,\ref{Mlinehisto} are the line mass distributions of the pixels tracing the filaments.}. Also, in light of the finding that hub is a network of short and very high-density filaments \citep{Kumar2021}, the peaks in Stokes I and therefore column-density actually represent filaments of high line mass that facilitate massive star formation. These filaments within the hub region have been identified using curvature maps as described in Appendix.\,\ref{App1b} It can be seen that only in the youngest source W3(OH), the FLMF shows a relatively smooth power-law (at high line mass end) that includes both the filament and hub FLMFs, starting roughly from the 20\sunpc up to the limit of 2000\sunpc. There is however a small departure from the continuous function around 100\sunpc corresponding to the peak of the hub FLMF. The power-law indices of the filament and hub regions are -1.36$\pm$0.26 and -1.2$\pm$0.23, respectively. In W3\,Main, the histogram shows two distinct peaks, one at \ml $\sim$20\sunpc with a broad distribution, and another narrower at \ml $\sim$200\sunpc that correspond to the filament and hub regions, and are fitted with slopes of -1.04$\pm$0.2 and -2.21$\pm$0.5, respectively. Similarly in S\,106, the histogram has two peaks each corresponding to the peak of filament and hub histograms with power-law indices -0.86$\pm$0.17 and -1.41$\pm$0.28, respectively. Comparing the FLMFs between these regions it appears that at the earliest evolutionary stage represented by W3(OH), a Salpeter-like FLMF is witnessed in the hub which gradually weakens as star formation progresses represented by the -0.86 power-law for S\,106. In contrast, the FLMF in the filament region appears to be similar in the earliest evolutionary stage and the most evolved stage as represented by W3(OH) and S\,106. However, the FLMF has 
a steeper power-law in W3\,Main when the star formation activity is highest, as evidenced by the numerous compact HII regions and the young massive stars in the hub.

In summary, similarly to the hub size and the radial column density profile, we find distinct signatures of the filament line mass distribution for each of the HFSs reflecting their different evolutionary stages. Next, we examine how polarization properties change as a function of the evolution of the HFS and their star formation activity.

\section{Polarization properties}\label{Sec.PolProp}

In this work, because we have assembled targets at 
different evolutionary stages with a variety of star formation sign-posts from stars to outflows, we can compare the polarization properties as a function of evolutionary stage to better understand the impact of the star formation activity and feedback on the polarization properties of the  dense gas.

\subsection{Distributions of $PF$ and $PI$ in hubs and filaments}\label{PFPIhisto}

\begin{figure*}[!h]
   \centering
  \resizebox{6.cm}{!}{\includegraphics[angle=0]{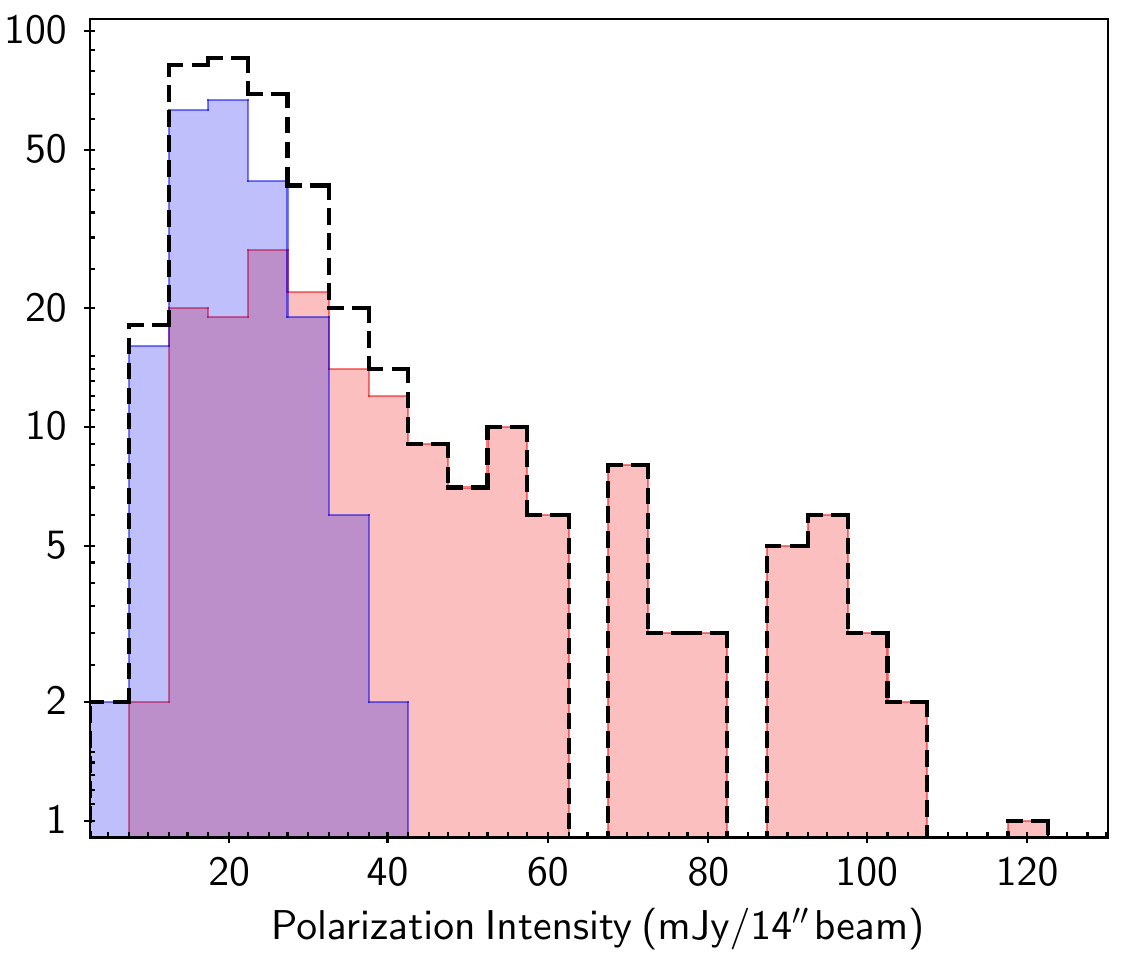}}
  \resizebox{6.cm}{!}{\includegraphics[angle=0]{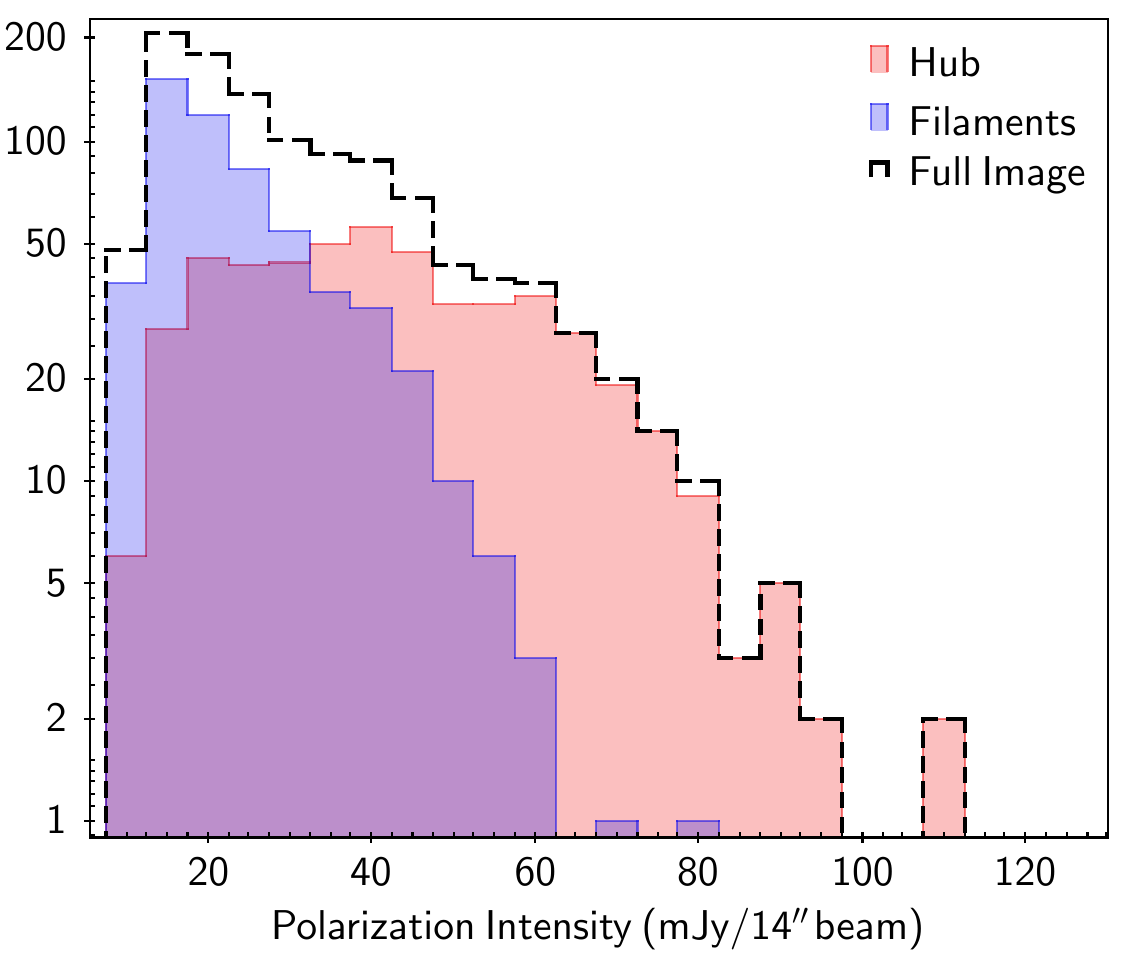}}
  \resizebox{6.cm}{!}{\includegraphics[angle=0]{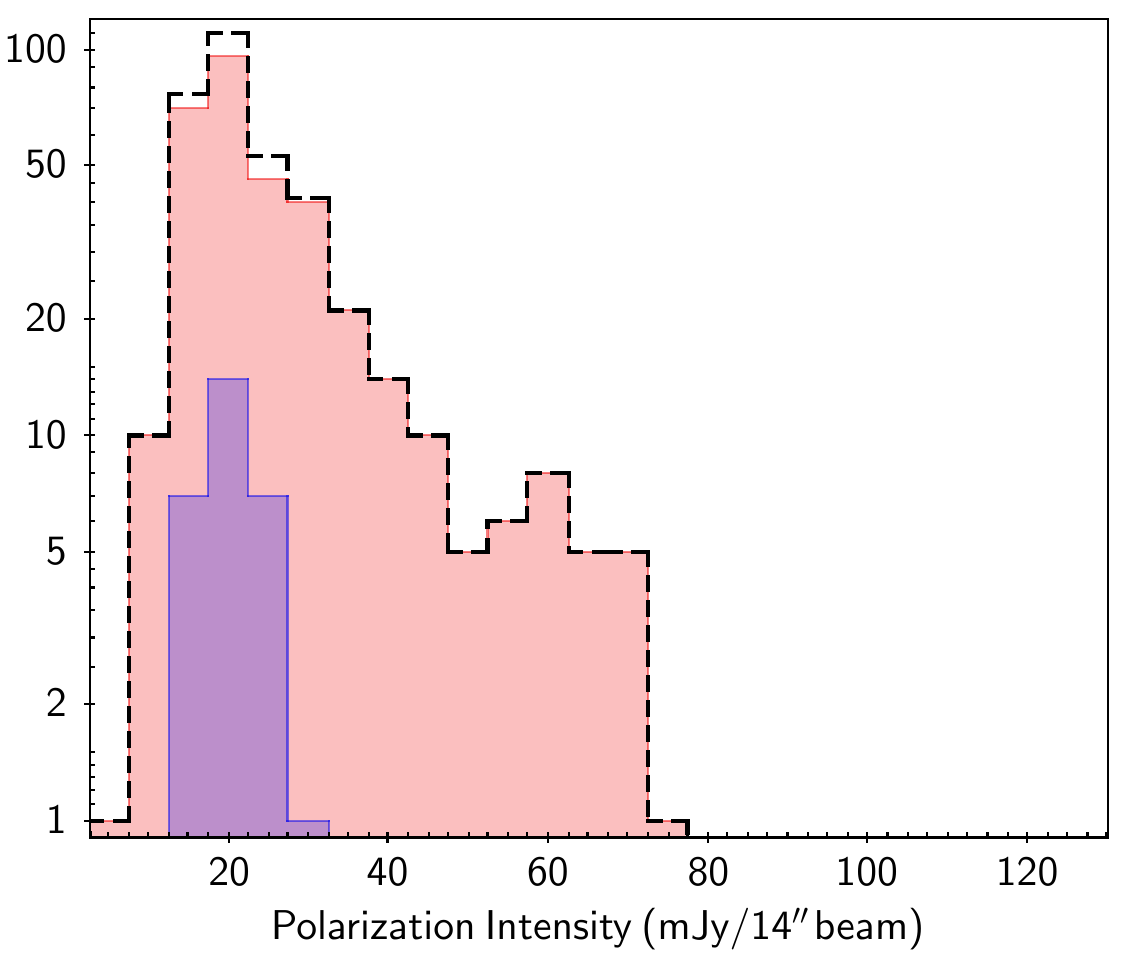}}
  \caption{Polarized intensity ($PI$) histograms for the total  (black), hub (red), and filament (blue) regions for W3(OH), W3-Main, and S 106 from left to right. The vertical axes indicate the number of pixels per bin.}
  \label{PIhisto}
    \end{figure*}

\begin{figure*}[!h]
   \centering
  \resizebox{6.cm}{!}{\includegraphics[angle=0]{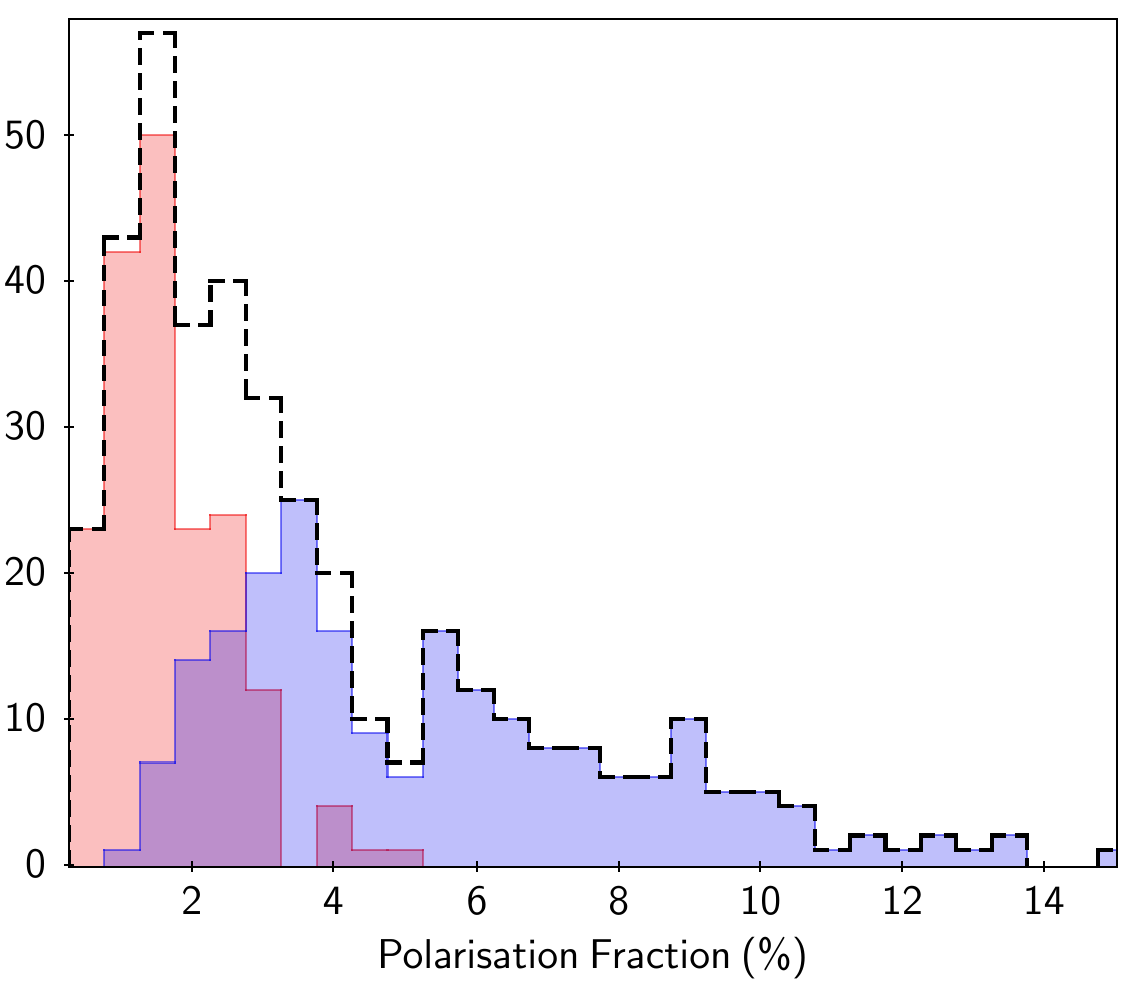}}
  \resizebox{6.cm}{!}{\includegraphics[angle=0]{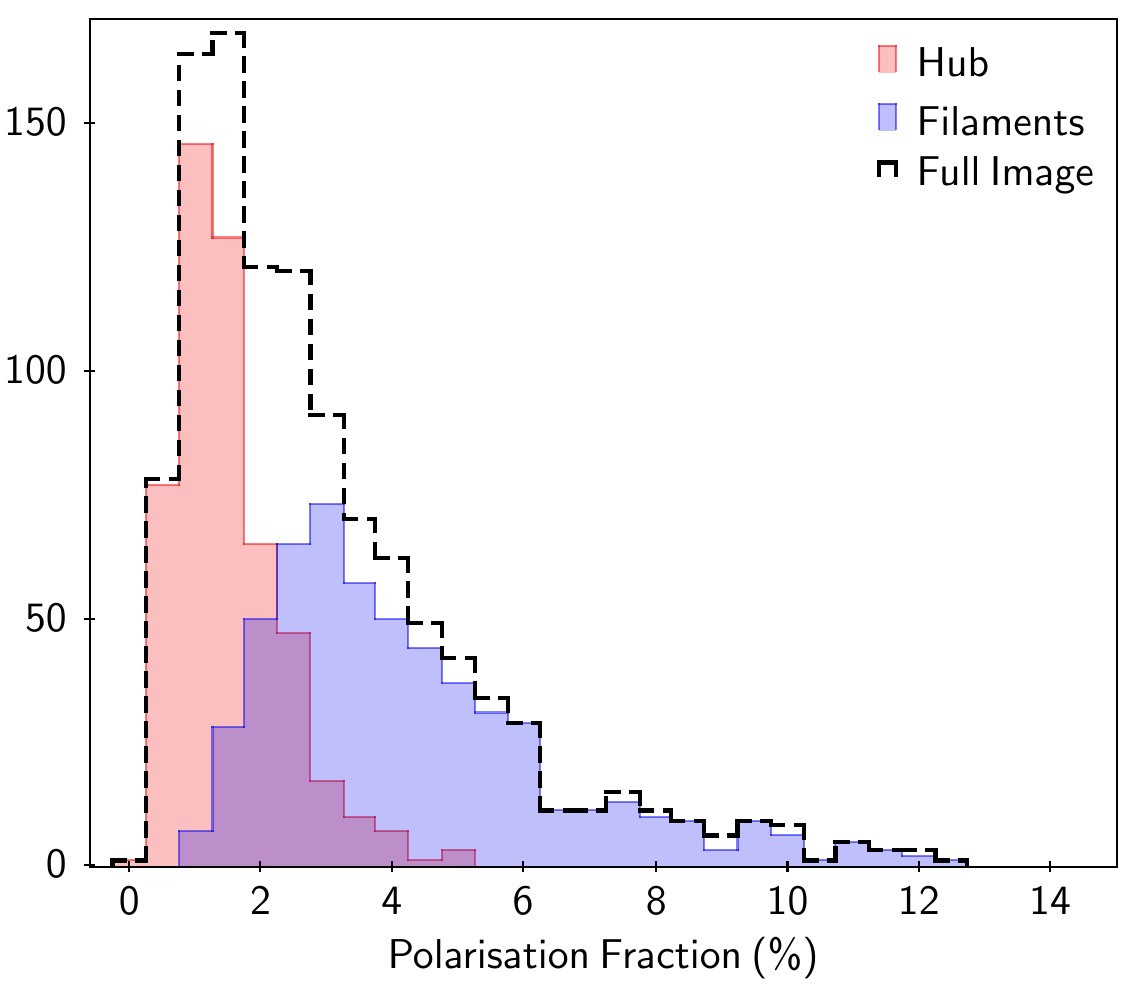}}
  \resizebox{6.cm}{!}{\includegraphics[angle=0]{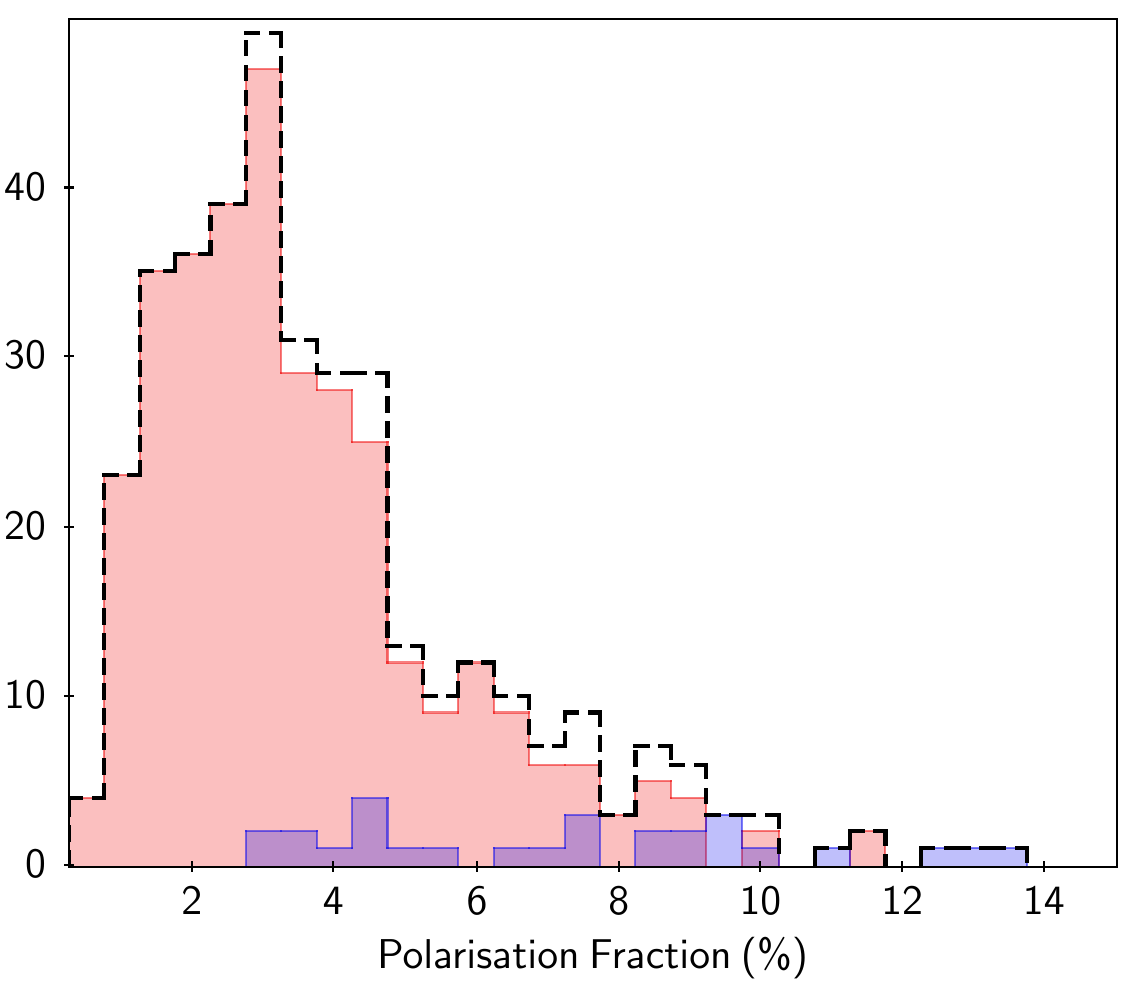}}
  \caption{Polarization fraction ($PF$) histograms for the total  (black), hub (red), and filament (blue) regions for W3(OH), W3-Main, and S 106 from left to right. The vertical axes indicate the number of pixels per bin.} 
  \label{PFhisto}
    \end{figure*}

Having defined and identified the hub and filament regions of the HFS in each target, we now examine the behavior of the observed polarization properties in these two regions and compare with that of the region as a whole for each HFS and among the targets. In Figs.\,\ref{PIhisto} \&\,\ref{PFhisto} respectively, we plot the histograms of polarized intensity ($PI$) and polarization fraction ($PF$). In producing these histograms we used maps with 4\arcsec size pixels, and have applied the combined criteria that the plotted data points have 
$PI/\delta PI > 5$ and  error in $PF$, $\delta PF,<3\%$. We have, however, examined the histograms with a more relaxed constraint of error in $PF$ with $\delta PF<3\%$, which only adds a few pixels at high $PF$ values arising at the edges of filaments. 
On the contrary the plotted data points are robust and represent the crest regions of filaments and within the hub. 

It can be seen from Fig.\,\ref{PIhisto}, that the highest and lowest values of polarized intensity are respectively found in the youngest and oldest targets namely W3(OH) and S\,106. Even though W3\,Main is the most massive HFS of all the three targets, the hub region of W3(OH) has larger number of pixels at $PI>$80 mJy/14\arcsec beam.  The $PI$ histograms of S\,106, due to its late stage of evolution, stand out from the other two targets, owing to a spatially resolved bipolar radiation bubble. 

For W3(OH) and W3-Main, the $PF$ histograms show a broader distribution for the filaments with larger values (up to 14\%) reached for W3(OH) compared to the narrower distributions for the hub region peaking at values around 1\% (Fig.\,\ref{PFhisto}).  
The $PF$ histograms of S\,106 in the hub region shows a broader distribution peaking at 3\% with values reaching up to 8-10\%.

\subsection{Spatial variation of $PF$ and $PI$}\label{PFPI}

\begin{figure}[htbp]
   \centering
  \resizebox{8.cm}{!}{\includegraphics[angle=0]{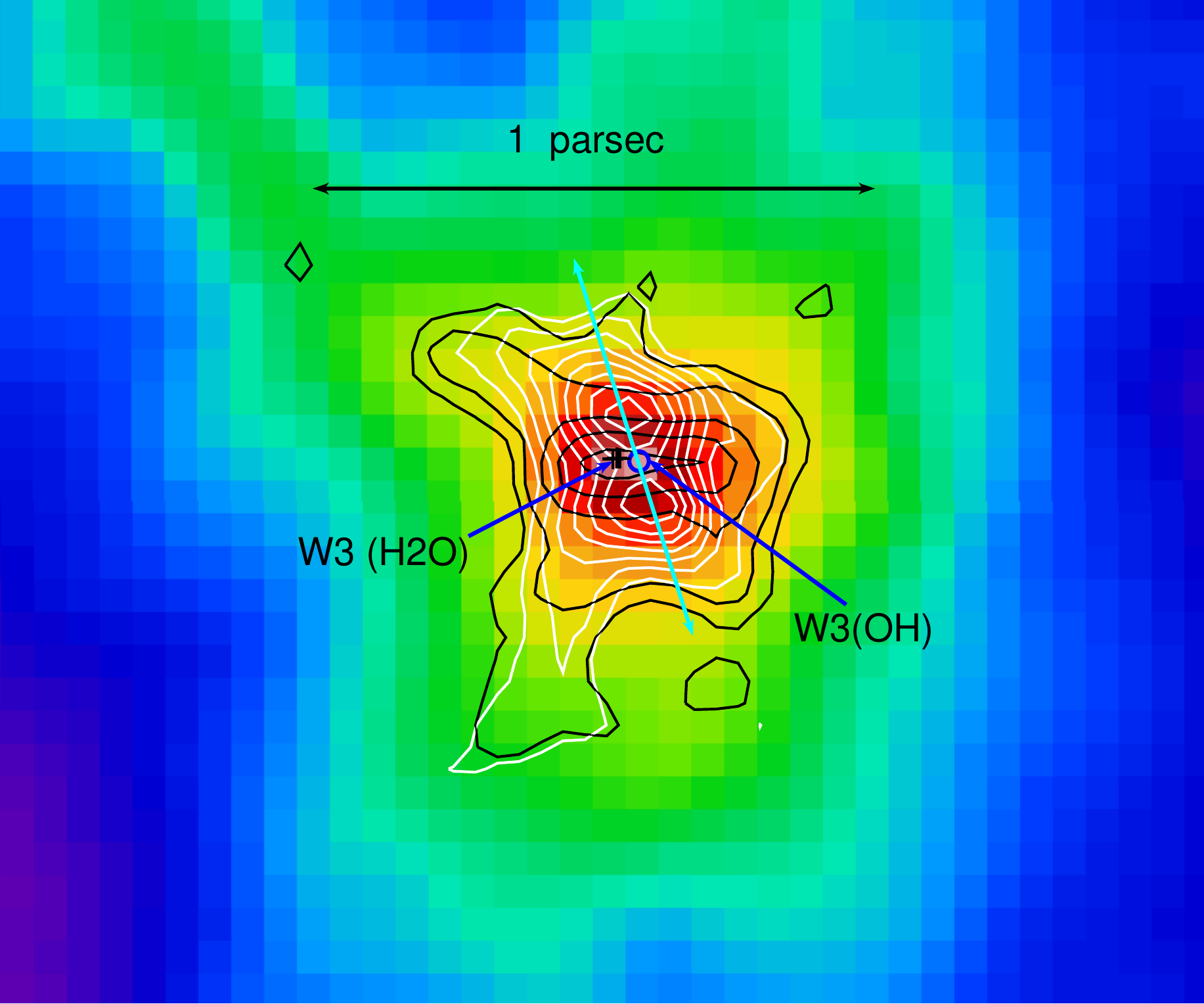}}
  \resizebox{8.cm}{!}{\includegraphics[angle=0]{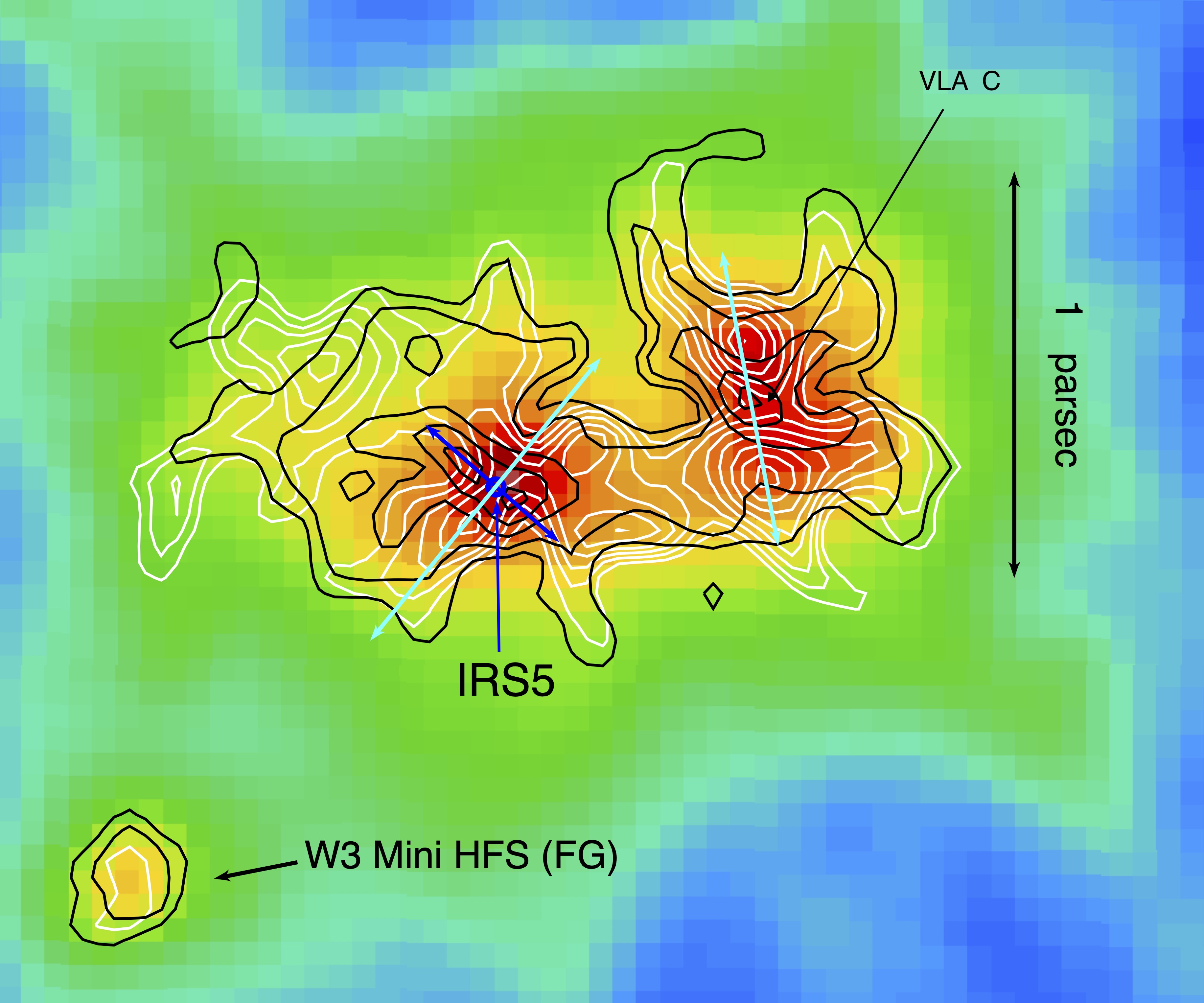}}
  \resizebox{8.cm}{!}{\includegraphics[angle=0]{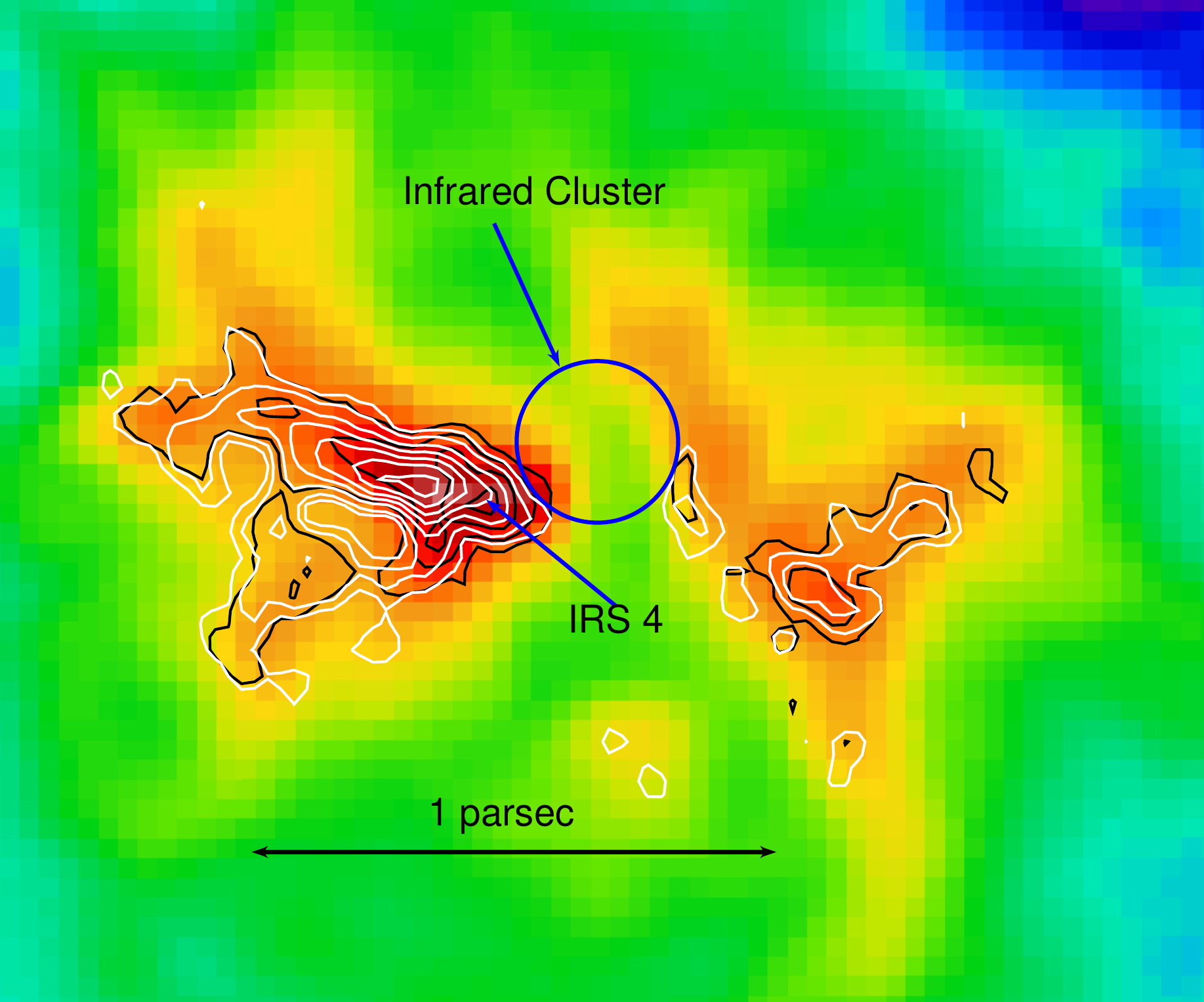}}
  \caption{Clockwise from top left: Spatial correlation between total intensity, $PF$, and $PI$ in W3(OH), W3\,Main and S\,106 displayed over \herschel\, SPIRE 250\,\mum image. Black contours represent $1/PF$ with levels 0.25, 0.5, 1, 1.5 \& 2 (for W3(OH) and W3\,Main, top) and 0.1, 0.3, 0.6 \& 0.9 (for S\,106, bottom). White contours display $PI$ with levels from 20 to 90 mJy/beam in intervals of 10 mJy/beam for W3(OH) and W3\,Main (top), and levels 5, 10, 20, 30, 40, 50 \& 60 mJy/beam for S\,106 (bottom). The cyan vectors in the top two panels mark bipolar patterns in $PI$ that correspond to outflows (see Sec.\,\ref{PFPI}). The blue vector in W3\,Main identifies the low $PF$ elongated structure centered on IRS\,5. This is similar to the elongated disk structure in W3(OH) traced by $1/PF$ contours.
       }          
  \label{PolContours}
    \end{figure}

The spatial variation of $PF$ and $PI$ in all three targets were examined and compared with the location of dominant massive young stellar sources and the sub-mm peaks in each target. Because $PF$ diminishes inside the highest density regions, we computed $1/PF$ and plotted it, along with $PI$, on the \herschel\, SPIRE 250\,\mum images as shown in Fig.\,\ref{PolContours}. The most luminous (massive) young stellar objects and known massive outflows are marked. 

In W3(OH) the sources OH and W3(H$_2$O) are embedded in a disk like structure traced by the $1/PF$ contours (black) while $PI$ (white contours) reveals a distinct bipolar shaped structure centered on this disk. Using SMA observations, \citet{Qin2016} uncover two outflows originating from W3(H$_2$O) which is roughly aligned north-south but tilted more towards north-east compared to the cyan vector shown in Fig.\,\ref{PolContours}. Understandably, the disk shaped structure traced by the black contours hosts several massive young stellar sources at very early evolutionary stage, including the OH source and the three cores associated with the H$_2$O maser Taylor-Welch object. The bipolar pattern from our 14\arcsec resolution maps indicate the net result of the multiple outflows in the region, all of which are effectively in the north-south direction \citep[see Fig.\,4 of][]{Qin2016} The disk like structure is well centered on the peak of 250\,\mum emission. The OH maser source is associated with a hyper-compact HII region as uncovered by the VLA observations \citep{Dzib2014}, it is a relatively evolved source with no outflow detection from \citet{Qin2016}.

The $PI$ and $1/PF$ contours are more complex in W3\,Main owing to a number of massive ionizing and young stellar sources crowded in the two nodes of the hub. In the left node, the $1/PF$ contours reveal an elongated structure (marked with a blue vector) on either side of IRS\,5 (coinciding with the 250\,\mum peak) perpendicular to which lies a massive outflow mapped in CO by \citet{Li2019} (see also Sec.\,\ref{W3main}). The $PI$ contours are not striking in tracing this bipolar outflow (cyan vector), however, two lobes can be visualized that roughly match with the axis of the outflow. In the right node of the hub, the $PI$ and $1/PF$ contours reveal complementary structures; $PI$ contours trace a clear bipolar structure, at the center of which $1/PF$ forms a condensed peak, coinciding with the VLA source C. The $1/PF$ contours can also be seen to trace the walls of the northern bipolar lobe. The 250\,\mum emission here is split between the condensation traced by $1/PF$ and also with the northern lobe of the bipolar pattern traced by $PI$. These two peaks of 250\,\mum may correspond to the two small peaks observed in the Stokes {\it I} and consequently the line mass maps (see Sec.\,\ref{profiles}). SMA and IRAM\,30m observations of the full hub-region of W3\,Main in both continuum and spectral lines \citep{Wang2012} reveal that the region contain several massive sources and multiple outflows that render this region complex. Our 14\arcsec maps here are tracing the major outflow cavity walls and associated cores where the driving sources are located. 

In S\,106, the peak traced by $1/PF$ coincides well with the source IRS\,4, and while both $1/PF$ and $PI$ trace similar structure, the peak traced by $PI$ is shifted to the left of IRS\,4. The region to the north and south, covered by the well-studied S\,106 bipolar radiation bubble/outflow lacks significant 850\,\mum emission thus no comparison can be made.

\subsection{POS B-field angle (\bang) vs filament orientation (\fang)}

\begin{figure*}[]
   \centering
  \resizebox{6.cm}{!}{\includegraphics[angle=0]{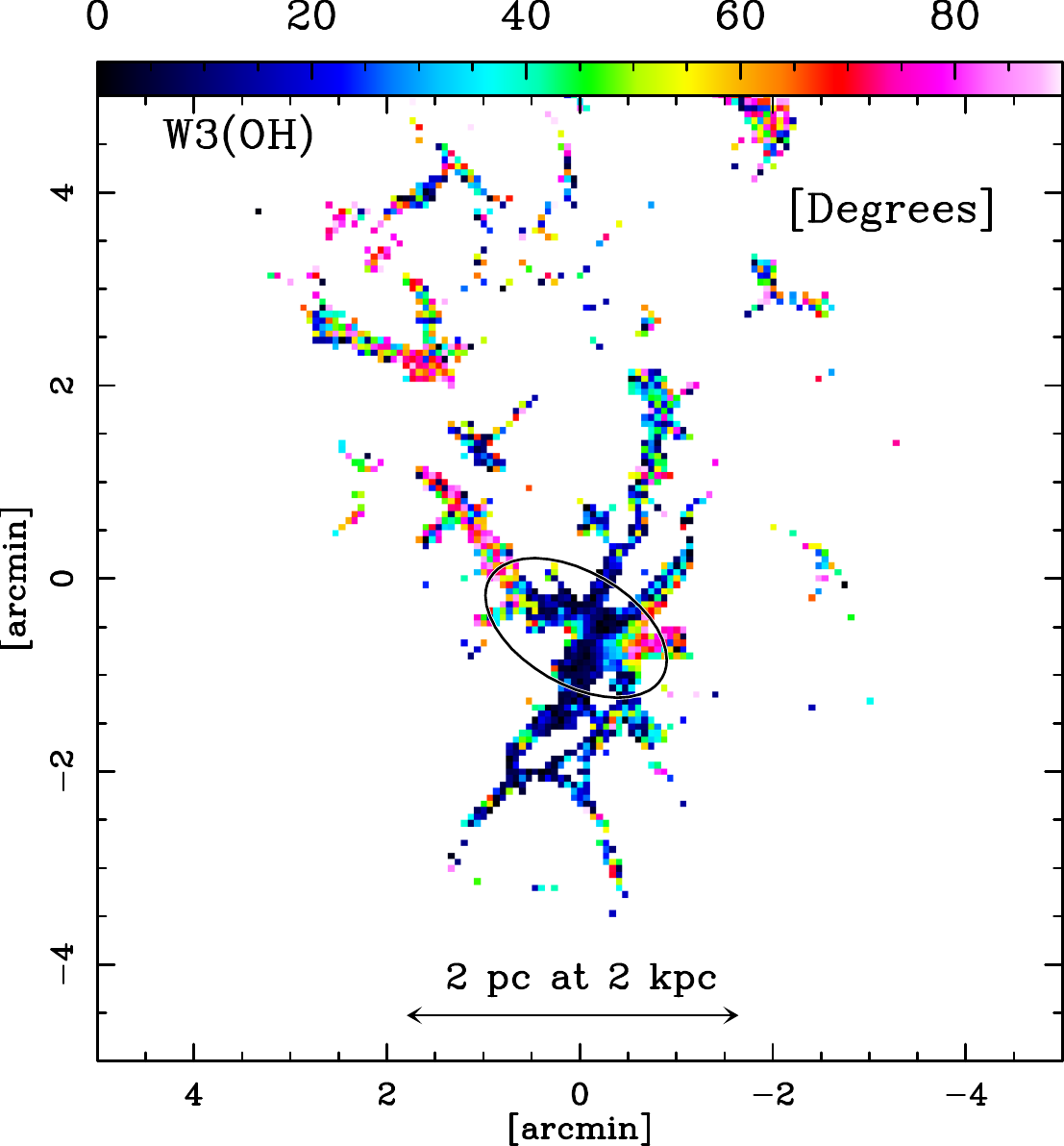}}
  \resizebox{6.cm}{!}{\includegraphics[angle=0]{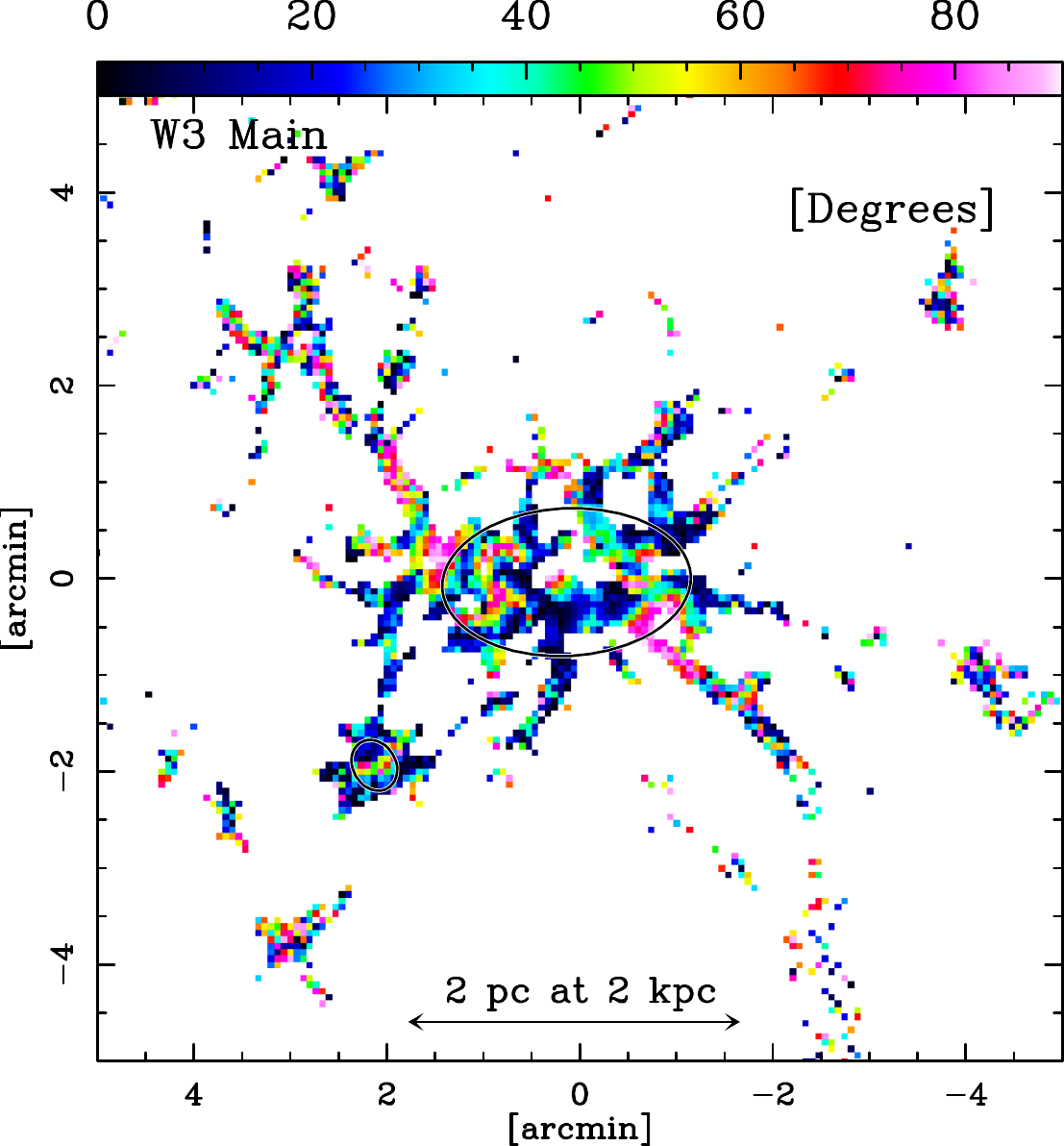}}
  \resizebox{6.cm}{!}{\includegraphics[angle=0]{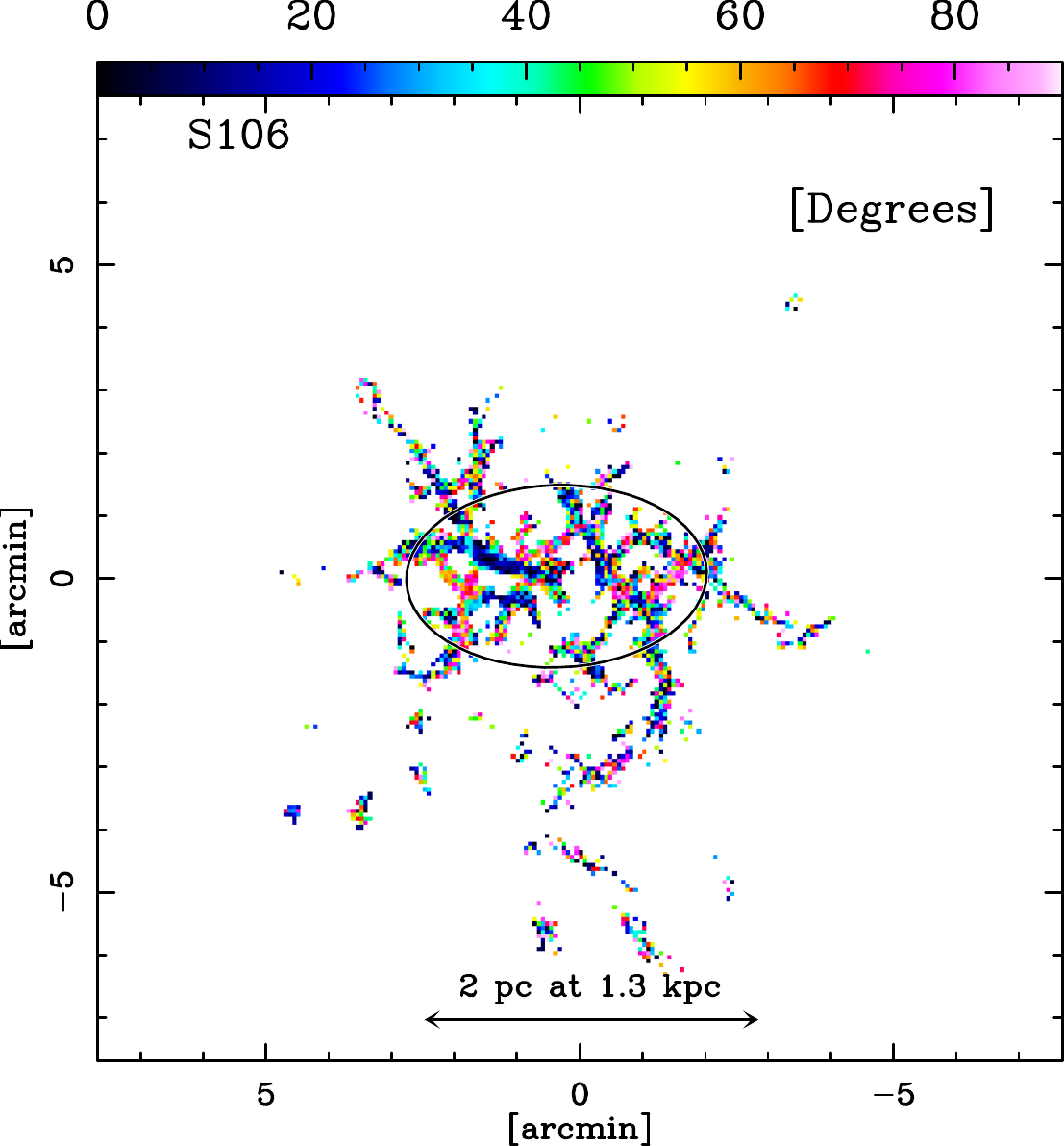}}
\vspace{-.1cm}
  \caption{ Maps of the difference between the position angle of the filaments \fang\ and the $\chi_{B_{\rm POS}}$ angle for W3(OH), W3-Main, and S\,106 from left to right. The black ellipses indicate the hub sizes as given in Table\,\ref{table1} and shown in Fig.\,\ref{MlineEllipse}.
}          
  \label{Diff_angle}
    \end{figure*}

In order to examine how  \bang\ behaves with respect to the  different filamentary structures over the entire region, we computed the difference between  \bang\ and the filament orientation \fang\ derived from our filament map (cf. Appendix\,\ref{App1b}). These maps are shown in Fig.\,\ref{Diff_angle} for all the three targets. It can be seen that the densest filaments, especially those close to the central regions of the HFS where the hub is located, appear in dark blue in Fig.\,\ref{Diff_angle} representing a small difference between  \bang\ and \fang. This suggests that the \bang\ is more aligned with the axis of the filaments towards the hubs. In contrast, the large difference between \fang\ and \bang\ (represented by pink and redder colors) is found in the lower density areas of the filaments. In W3(OH) the dark bluer colors are concentrated towards the hub with little or no redder colors. In W3\,Main the small angle differences are prominent both in the central hub and also in the foreground mini-HFS SE source. However, there is also a significant mixture of larger angle differences in the W3\,Main hub, unlike W3(OH). This is likely due to the presence of multiple compact HII regions and young bipolar outflows in the central hub of W3\,Main that is in the process of altering the density structure of the hub region. This mixed features of both large and small angle differences in the central hub region is most prominent in S\,106 where a well-resolved bipolar outflow/radiation bubble driven from the IRS\,4 has significantly reshaped the surrounding material.

\begin{figure*}[!h]
   \centering
  \resizebox{6.cm}{!}{\includegraphics[angle=0]{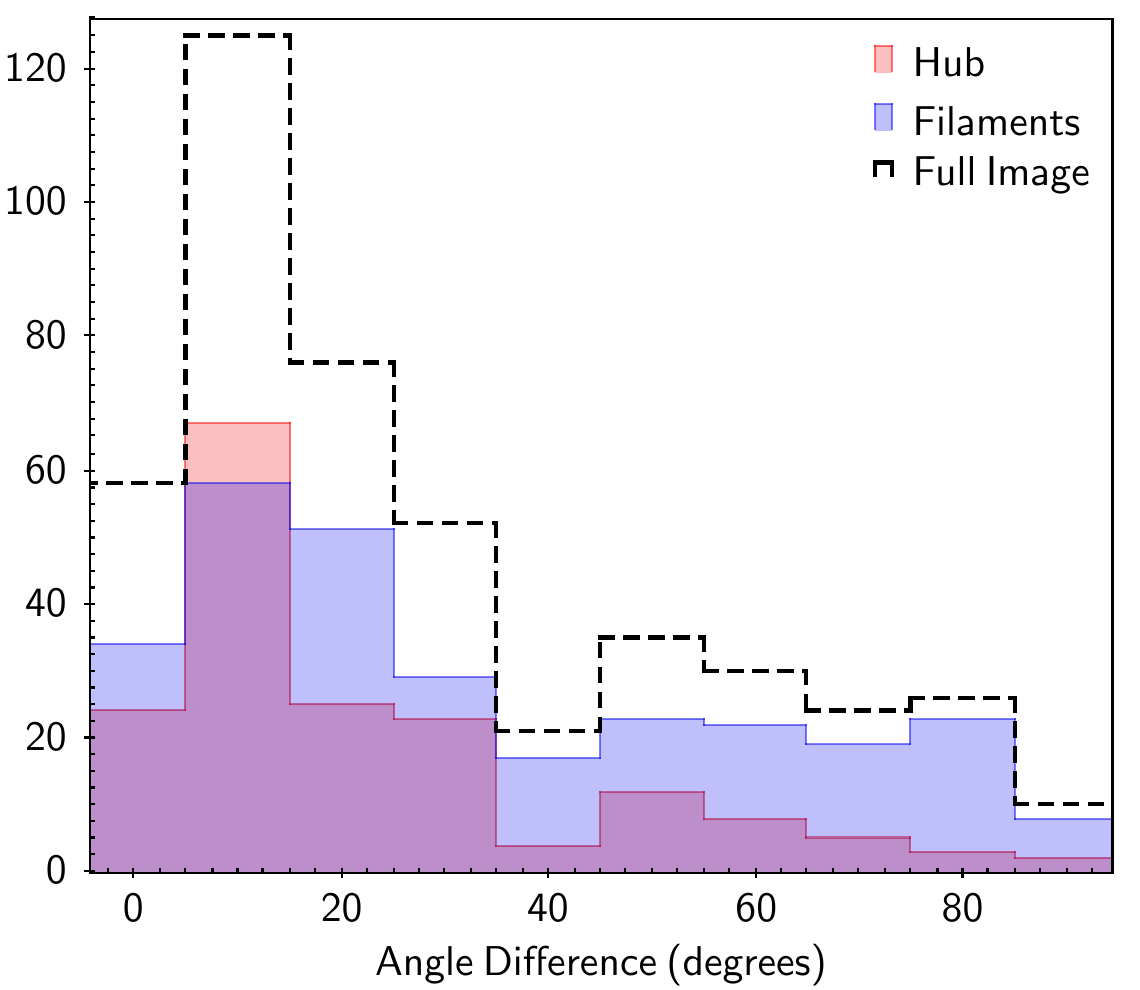}}
  \resizebox{6.cm}{!}{\includegraphics[angle=0]{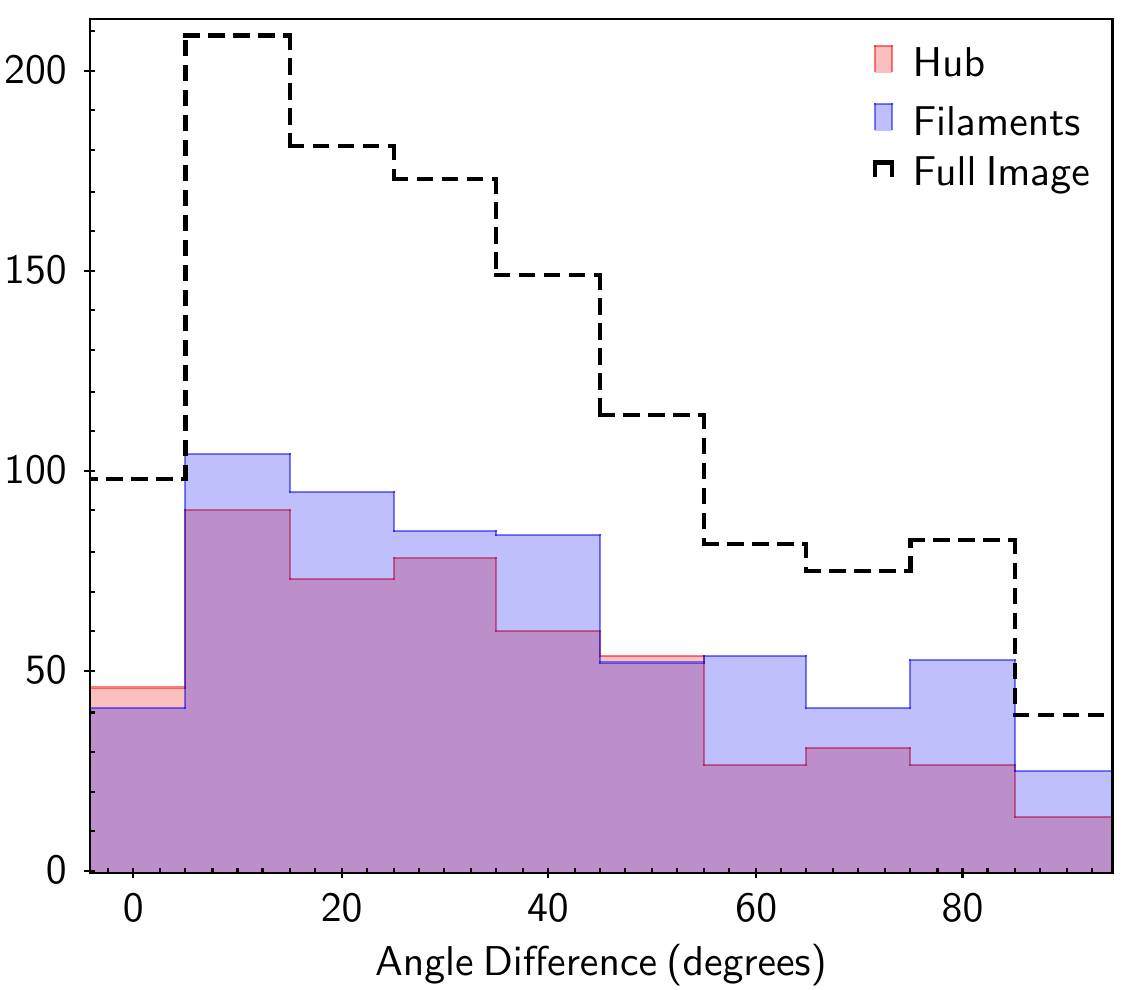}}
  \resizebox{6.cm}{!}{\includegraphics[angle=0]{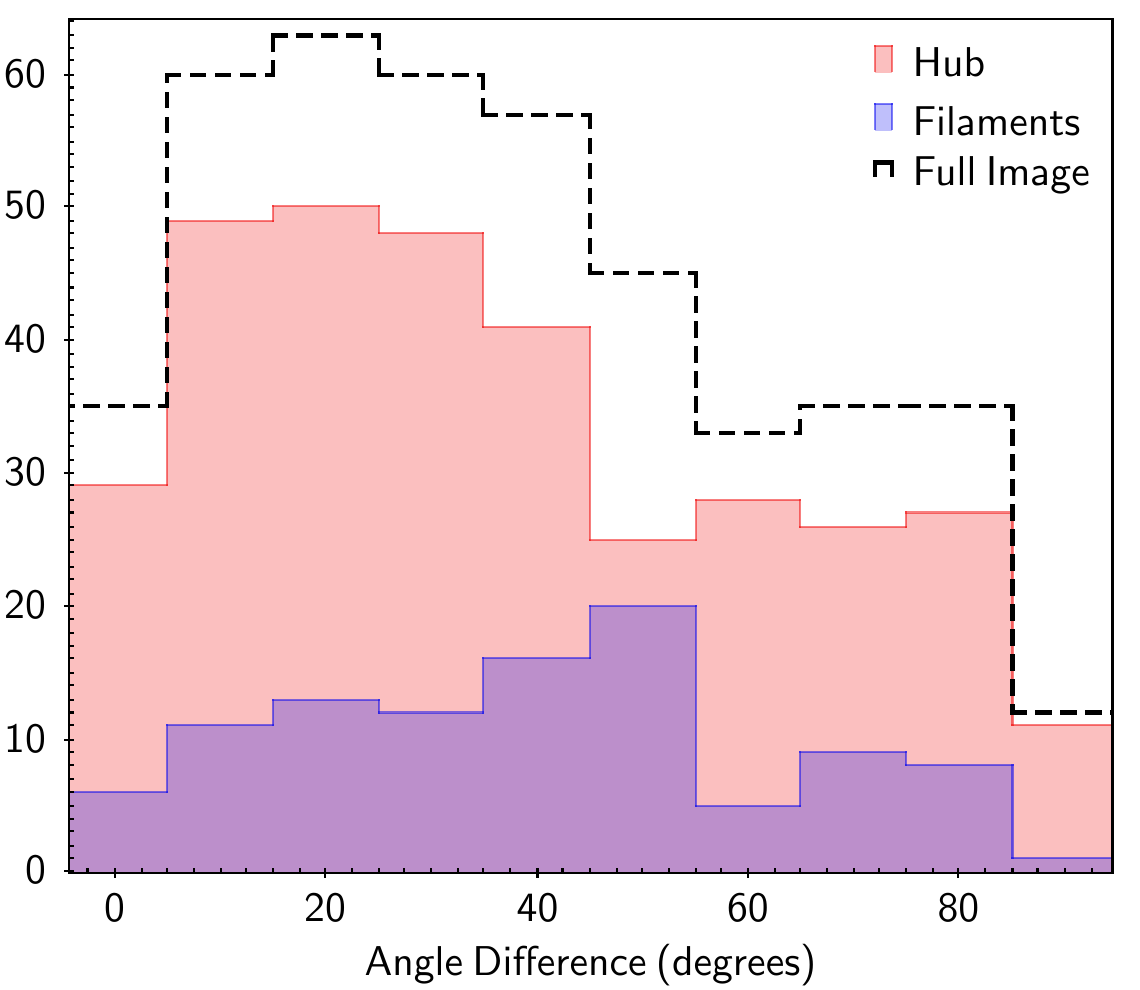}}
  \caption{ Histograms 
  of the difference between the filament orientation and  $\chi_{B_{\rm POS}}$  for the total  (black), hub (red), and filament (blue) regions, 
  for W3(OH), W3-Main, and S\,106 from left to right. 
   The vertical axes indicate the pixel number. Here pixels with the error on $\chi_{B_{\rm POS}}$ smaller than 20 degrees are chosen. 
       }          
  \label{Banglehisto}
    \end{figure*}

To further explore these maps, we also plot the histograms of the difference angle, \fdiff=|\bang-\fang|,  for pixels located in the hub, filament, and the full field as shown in Fig.\,\ref{Banglehisto}. In producing these histograms only pixels with  errors on \bang smaller than 20\degree\ are selected. The choice of \bang error was made by comparing the distribution of $PI/\delta PI>1$ and $>5$ with the corresponding \bang error which showed that the major concentration of data points was within an error of $< 20\degree$ with no particular correlation to the $PI/\delta PI$ cuts. In W3(OH), the  histogram of \fdiff\ for the entire region shows a significant peak around 10\degree\ indicating that the POS B-field is mostly parallel to the filaments, both inside and outside the hub, albeit with a larger number of pixels with perpendicular or random orientation along the filaments outside the hub. The histogram of the full region of  W3-Main also peaks at $\sim10-15$\degree\ however with a broad power-law like distribution up to $\sim80-90$\degree\ indicating larger mis-alignments and perpendicularity between the POS B-field and the filaments. This trend (peak towards parallel orientations and power-law distribution of large \fdiff\ values) is observed both for the hub and the filament regions. Finally, in the most evolved S\,106 HFS, the results are different with the presence of two peaks, the first largest peak at angles  $<45$\degree\ and the second smaller peak for  $>45$\degree. This distribution is mostly dominated by the distribution in the hub, while the angle difference for the filaments shows a random distribution. These histograms show noticeable  differences between our three targets possibly tracing the  evolution stages of the HFSs, as discussed further below. 

\section{Discussion}\label{Sec.disc}

In this work, we examined the properties of  three star cluster-forming HFSs in different evolutionary stages in an attempt to understand their evolution. We have combined observations of star-forming dense gas and young stellar population to obtain our results. The total intensity images at 850\,\mum trace the dense gas in the HFS, 
identifying the filament and hub regions in each target, while allowing us to quantify column densities and line masses. The changes in polarized intensity and polarization fraction with the progress of star formation and evolution of the HFS have been examined. The HII region emission from VLA and the information on outflows from the literature have served to assess the influence of feedback during the evolution.

\subsection{Polarization and magnetic field properties of  hub-filament systems}\label{Sec.disc.pol}

We analyzed JCMT POL-2 data of polarized dust emission at 850\,\mum for three selected HFSs in consecutive evolutionary stages, from early to evolved, to investigate the global evolution of the magnetic field structure. The histograms of relative orientation, \fdiff=|\bang-\fang|, between the B-field angle (\bang) and the filament orientation (\fang) shows a variation; 1) between the hub and the outer filamentary parts of the HFSs, and also 2) as a function of the evolutionary stage of the HFS (Fig.\,\ref{Banglehisto}). At early stages (for W3(OH)),  \bang\ and \fang\ are mostly aligned with a small dispersion of the \fdiff\ histogram. At the intermediate stage (for W3-Main), the \fdiff\ distribution broadens with a peak at small angles of parallel orientation and a power-law distribution up to \fdiff$\sim90\degree$ for perpendicular relative orientation. At the most evolved stage of S\,106, the \fdiff\ histogram towards the hub shows two peaks, with a stronger peak for the parallel configuration  and a weaker peak for the perpendicular configuration. 

The current theoretical models  propose the formation of high line mass filaments through accumulation of matter through flows along the B-field lines resulting in filaments perpendicular to the B-field \citep[e.g.,][]{Inoue2018,Pineda2023,Abe2024}. This theoretical prediction is compatible with the observations of high line mass filaments not connected to hubs \citep[e.g.,][to refer to a few examples]{Palmeirim2013,Cox2016,Ching2022,Pattle2023}. In HFS however, observations show that the relative orientation between the B-field and the filaments changes from mostly perpendicular to mostly parallel as the density increases towards the hub region \citep[e.g.,][]{Wang2020,Pillai2020,Arzoumanian2021,Wang2024}. The reorientation of the B-field from perpendicular to parallel along the high-line mass filaments connected to massive hubs has been interpreted as resulting from dynamical flows along the filaments carrying the matter and the frozen-in B-field towards the hub \citep[as suggested by MHD simulations, e.g.,][]{Gomez2018,Suin2025}. Consequently, we interpret the parallel configuration of filaments and B-field in our three targets as the influence of the gravity of the hub onto the surrounding filaments and B-field. In addition, the parallel relative orientation between the filaments and the B-field could result from projection effects as discussed for example in \citet{Doi2020}.

Previous observational studies of star forming clouds have shown that the POS B-fields are reorganized tangentially along the swept-up material in the shells formed by the expanding HII regions \citep{Tang2009HII,planck2016-XXXIV,Arzoumanian2021, Cortes2021, Tahhani2023}. We observe a similar effect in the northeastern region of W3(OH) (Fig.\,\ref{W3OH-I-Bfield-Spitzer}), though it is not directly associated with a known HII region in this case. In contrast, W3\,Main exhibits a B-field geometry modulation that can be partially attributed to the compact HII regions present within its hub (Fig.\,\ref{W3Main-I-Bfield-Spitzer}). Moreover, an hourglass-shaped pattern embedded within the hub (Fig.\,\ref{W3main-zoom} left) aligns with the most massive molecular outflow identified in the region \citep{Li2019}.

We also find that the POS B-fields can also be reshaped by massive outflows such as in \citet{Lyo+2021} and their associated radiation bubbles, as evident in S\,106 and the central region of W3\,Main (Fig.\,\ref{W3main-zoom} left). In S\,106 (Fig.\,\ref{S106-I-Bfield-Spitzer}), the B-field lines trace the cavity walls of the well-documented massive outflow and radiation bubble driven by IRS\,4. This reordering of the B-field along the outflow cavity walls may likely explain the high $PF$ values observed in the hub of S\,106 (cf. Fig.\ref{PFhisto}).The increase of $PF$ in this context may result from a combination of the increase of grain alignment efficiency from the anisotropic radiation from the massive central star \citep[according to the RAT theory][]{LazarianHoang2007,Hoang2020} and the reduced depolarization due to the reordering of the B-field structure. Outflows reshaping the B-fields have been previously seen in the study of numerous other targets \citep{Hull2017,Fernandez-Lopez2021, Pattle2022, Karoly2023, Encalada2024}.

Across all observed HFS targets, on average lower $PF$ values are consistently found towards the hubs as opposed to the lower column density emission outside the hubs. Such a decrease of $PF$ with increase in column density is compatible with previous observations \citep[cf.][for a comprehensive review]{Pattle2023} interpreted as resulting from the combined effects of loss of grain alignment, variation of grain properties, change of the 3D B-field configuration, and depolarization due to averaging along the line-of-sight and within the beam. 

A particularly interesting feature in our data is the correlation of low $PF$ regions, tracing the dense cores and/or disks/toroids, with the geometric centers of bipolar outflow-shaped patterns traced by $PI$, most prominently seen in W3(OH) and W3\,Main near ionizing sources (Fig.\,\ref{PolContours}). The edge-on disk-like morphology of the $1/PF$ contours in W3(OH) likely represents magnetized, self-gravitating toroidal structures, analogous to those observed in the G31.41 HFS \citep{Beltran2019, Beltran2024}. The G31.41 HFS remains one of the most thoroughly studied example of massive star-forming HFS, with detailed observations across multiple scales, angular resolutions, wavelengths, and tracers. \citet{Beltran2024} report that $PF$ values are approximately 2\% around four spatially resolved individual massive sources and increase up to $10\%$ in the broader hub. 
The observations of G31.41 are modeled as a magnetized toroidal structure hosting  massive sources, with an hourglass-shaped B-field pattern emerging along the polar direction. The $PF$ and $PI$ patterns observed in Fig.\,\ref{PolContours} align with the scenarios depicted by \citet{Beltran2024}. These findings underscore the complex interplay between gravity, magnetic fields, massive outflows, and the radiation environment within HFS, offering new perspectives on the role of B-fields in massive star formation and cluster evolution.

\subsection{Star formation in Hubs}\label{Sec.2eyes}

\subsubsection{Hub definition and its relation to cluster formation}
In this study, for the first time, we define the "hub" based on column density radial profiles of multiple targets. We do so by demonstrating that the column density distributions in our three targets can be approximated as elliptical structures, with the radial density profiles showing distinct transitions at the boundaries of these ellipses (cf. Fig.\,\ref{radialProf}). Similar definition has been previously used for Mon\,R2 by \citet{Trevino-Morales2019} and \citet{Kumar2021} resulting in a radius of 0.8--1.2\,pc (see Table.\,\ref{table1}). The systematic turnover of column-density slopes at the hub boundaries in different targets suggest a deeper implication that may shed light into the physics of star formation.

One intriguing feature of young stellar clusters (YSCs) in the Milky Way is that the cluster radii are highly constricted (median R$_c$=0.58$^{+0.32}_{-0.20}$\,pc for d$\,<\,$2\,kpc, \citet{Kharchenko2013}) which is larger for massive ($>10^4$\,\msun) YSCs (R$_c=1.5^{+2.35}_{-0.42}$\,pc) \citep{Zwart2010}. In addition, the cluster radii have a weak dependence \citep{Krumholz2019} on mass (R$_c \propto$ M$^{1/3}$, up to $10^5$\,\msun). We note that the observed cluster sizes are in close agreement with the hub sizes ($\sim$1\,pc), derived both in this work and other regions \citep[example Mon\,R2][]{Trevino-Morales2019,Kumar2021}. This correlation is intriguing, given that star formation takes place in the filaments of the HFS which are several times larger, ranging from 3-10\,pc \citep{Kumar2020}. 

\subsubsection{Double-nodes or two peaks of star formation in Hubs}
In \citet{Kumar2020}, we first highlighted that nearby star-forming regions such as Orion and NGC 2264 exhibit two distinct centers of star formation within their central hub regions, each demonstrating a clear evolutionary difference: one center being younger than the other. These nodes within the hub represent the densest and most massive gas concentrations, displaying characteristics indicative of ongoing star cluster formation. In this study, we extend the initial identification of the two star formation peaks by \citet{Kumar2020} through a clear detection of the corresponding nodes in all three targets analyzed. We have shown that the two nodes are consistently associated with discernible concentrations of young stars at distinctly different evolutionary stages (Sect.\,\ref{hubdef}). Remarkably, these features remain evident across all three targets studied here, regardless of their evolutionary state.

\begin{table*}
 \caption[]{Examples of well studied HFSs displaying the "two peaks" or  "double nodes" of star formation.}
    \label{table4}
\tabcolsep 4pt
 {\centering \begin{tabular}{l|c|cc|cc|c}
    \toprule
    \multirow{2}{*}{Target} &
     \multirow{2}{*}{Evolutionary Stage} &
      \multicolumn{2}{|c|}{Nodes} &
      \multicolumn{2}{|c|}{Separation} &
      \multicolumn{1}{|c}{Distance} \\
    & & young & old & angle (\arcsec) & proj (pc) & pc \\
      \midrule
    NGC\,1333 & early & SSV13/IRAS3 & BD+30$\degree$549 & 426\arcsec\, & 0.5 & 238 \\
    W3(OH) & early & Taylor-Welch & IR Cluster & 37\arcsec\, & 0.36 & 2000\\
    G31.41 & early & P2 & P1 & 41\arcsec & 0.75 & 3750\\
    NGC\,6334 & intermediate & NGC6334I(N) & NGC6334I & 112\arcsec\, & 0.7 & 1300 \\
    Orion & intermediate& BN/KL & Trapezium & 70\arcsec\, & 0.14 & 412 \\
    NGC\,2264 & intermediate & IRS2 (NGC2264D) & IRS1 (NGC2264C) & 410\arcsec\,  & 1.6 & 800 \\
    Aquila Rift & intermediate & Serpens South& W40  & 20.6\arcmin\, & 1.46 & 260 \\
    W3\,Main & intermediate & IRS\,5 & VLA-C & 70\arcsec\, & 0.68 & 2000\\
    S\,106 & evolved & IRS\,4 & IR cluster & 55\arcsec\, & 0.35 & 1300 \\
    \bottomrule
  \end{tabular} \par}
\end{table*}

In Table\,\ref{table4}, we compile a list of well-studied star-forming regions that we propose as examples of these “two star formation nodes.” The separation between these nodes is generally less than 1\,pc, with exceptions such as NGC 2264 \citep{Peretto2006} and the Aquila Rift HFS \citep{Bontemps2010}. While there is some debate and discrepancy in distance estimates for regions like W40 and Serpens South \citep[see][for discussion]{Bontemps2010}, it is noteworthy that both sources are centrally located in the large-scale \herschel\ maps presented by these authors. The smallest node separation is observed between Orion’s BN/KL and the Trapezium cluster, at 0.14 pc.

The double node of star formation has not yet been systematically explored, and its prevalence as a common feature of HFS remains unclear. However, in the well-studied regions listed in Table\,\ref{table4}, the presence of both young and older nodes appears systematic. These clumps are distinguishable not only through clear indicators of youth and evolved states but also by notable temperature differences, even when located in close proximity. For instance: NGC 6334 I(N) is colder than NGC 6334 I \citep[see Fig. 1]{Arzoumanian2022}. In NGC 1333, the SSV13 cluster is embedded within a colder clump compared to the SVS3/BD+30$\degree$549 region \citep[see Fig. 2]{Hacar2017}.
Despite these observations, the physical origin of such well-defined dual centers of star formation activity remains largely unexplored and poorly understood. 

We hypothesize that the two observed star-forming nodes may represent different stages in a sequential star formation process within the theoretical expectation that star formation continue (and accelerate) over many million years \citep{Inutsuka2015,Inutsuka2017}. Given the timescale over which stellar feedback disperses gas, it is plausible that only two such nodes — one relatively younger than the other — are visible within a single dense hub at any given time.

We further suggest that earlier star formation episodes, now traced by young star clusters, are located near — but no longer physically associated with — the currently active star-forming hub, having lost their connection to dense gas. A notable example is the Orion Trapezium cluster, located near BN/KL, which is no longer associated with dense gas, likely due to its dispersal by stellar feedback \citep[see Fig.1\, of][]{Hacar2018}. Yet, it represents a major star formation node from the previous episode. Conversely, the Orion South dense clump, situated adjacent to BN/KL, could represent the next or emerging star formation node \citep[also found in Fig.1\, of][]{Hacar2018}.

Extending this idea, we propose that dense hubs in very early evolutionary stages may host only a single, active star-forming node alongside a second, pre-stellar one. A potential example is the Mon R2 hub: it contains a cluster of five IRS sources forming one node, and a cold, dense pre-stellar clump or filament located $\sim$0.3\,pc southeast of the IRS cluster, observed in sub-millimeter emission \citep[the elongated highest density (white color) feature in Fig.\,F.\,1 of][]{Didelon2015}, forming the second.

Based on these observations, we propose that the "double-node" configuration of star formation represents a continuous evolutionary sequence. The spatial configuration of dense gas (the hub) and associated star clusters changes over time. At a given epoch, one node appears young while the other is older. In the previous generation, the younger node may have been pre-stellar; over time, as stellar feedback clears out dense gas or accretion reorganizes the hub’s structure (e.g., through shock compression or filament collisions), the older node evolves into an exposed star cluster located outside the current dense hub.

Future theoretical modeling and more systematic observational studies of star-forming nodes — both within and around dense hubs — will be crucial to better understand the nature and implications of this proposed "double-node" configuration in hub-filament systems (HFSs).

\subsection{Evolution of magnetized hub-filament systems}\label{sec.disc.evolution}

HFS are now well established as progenitors of star cluster formation. The initial mass function (IMF) reliably characterizes the mass spectrum of stars within these clusters, where low-mass stars dominate not only in number but also in mass, because the IMF power-law exponent is smaller than -2. The relatively few massive stars are known to form inside hubs, possibly ubiquitously \citep{Kumar2020}. The goal of this study has been to identify indicators of the evolution of the HFS and in particular hubs, as star formation proceeds.

\subsubsection{Line mass evolution} As elaborated in Sec.\,\ref{Sec.Mline}, the central intensity peaks that constitute the hubs naturally enclose the highest line mass filaments and the overall slopes of the FLMF (represented by line-mass histograms in Fig.\,8) display discernible variations as the HFS evolves. In particular a smooth Salpeter-like slope enclosing both the hub and non-hub regions in W3(OH) breaks out into two different slopes in the intermediate and evolved targets W3\,Main and S\,106 respectively. This likely indicates that star formation proceeds significantly in the hub-region turning dense gas into stars. As the highest line mass filaments in the hub-region result in massive star formation and fragmentation producing other low- intermediate-mass stars, one can expect the FLMF slope turning steeper as evidenced in W3\,Main and S\,106 histograms. In contrast, the external filament regions do not display significant alterations in their line-mass distributions. At very evolved stages nearing the end of star formation and destruction of the HFS, even the filament regions may change their mass distributions.

\subsubsection{Hub sizes} Next, as evident from Table.\,1 (see Sec.\,4.2), there is a noticeable increase in the size of the hub from the early to evolved HFS evolutionary stage, in-spite of the significant differences in their total mass and luminosities. Even though this effect can not be generalized given the limited sample size, the trend is important because the "hub size" effectively reflects the final agglomeration of the dense gas that is responsible for the formation of the young stellar cluster. As the HFS evolve, the process of accumulating dense gas from the outermost parts of the HFS to the central hub is nearly complete, effectively representing the maximum size of the hub. At the same time, the feedback effects from massive star formation in the hub would have its impact on the lowest density matter, especially at the outer regions of the HFS. This is because feedback propagates most easily in the low density inter-filamentary cavities, eroding the low density material. The highest density structures in the hub are the last structures to wane at the end of star formation.

One interesting feature from Fig.\,13 (see Sec.\,6.1) is the distinct peaks in B-field orientation angles with respect to the filament in the most evolved stage. In the hub, the orientation is mostly parallel. Considering again the limited sample, and projection effects, we cautiously suggest that this observation may be suggesting that the dragging of B-fields along the highest density filaments is mostly complete. While this process of gravity dragging the B-fields begins at the earliest stages of the HFS, finds it maximum effect at the most evolved stage.

\subsubsection{Evolution of PF and PI}

The distributions of $PF$ for the hub region is found to broaden with the evolutionary state of the target (see Sec.\,\ref{PFPIhisto} and Fig.\,\ref{PFhisto}). In comparison the outer filament regions maintains similar widths for the $PF$ distribution in the case of W3(OH) and W3\,Main but display a large scatter in S\,106.  Given that low {\it PF} values at high column density is a product of several possible effects such as loss of grain alignment, variation of grain properties, and depolarization due to line-of-the-sight averaging, we can only speculate possible reasons. In the early stages such as W3(OH), the presence of nearly edge-on toroidal structures (Fig.\,\ref{PolContours}) can result in depolarization effect from line-of-sight averaging. As star formation proceeds, and massive young stars emerge, shedding much uv radiation on the surroundings, increased $PF$ in S\,106 can be the result of re-alignment of dust grains due to radiation. Also, the highest and lowest $PI$ values in the hub are found in the youngest and oldest targets (Sec.\,\ref{PFPIhisto} and Fig.\,\ref{PIhisto}). In essence, at early evolutionary stages, the high density hub structure is more ordered resulting in less depolarization in W3(OH) compared to the more evolved W3\,Main. At later stages, the emergence of young stars and associated radiation and waning dense structures can result in higher $PF$ values with broader distributions and decreasing $PI$ values, as evident from S\,106.

\subsubsection{B-fields in HFS and their evolution}

Massive stars preferentially form in hubs \cite{Kumar2020}. While magnetic fields initially aid the formation of massive stars, eventually they get reshaped from their stellar feedback. There is mounting observational evidence \citep{Beltran2019} and theoretical considerations \citep[see, for example,][]{Krumholz2015} pointing to the critical role of magnetic fields in the assembly of massive stars. Inside the hubs, magnetic fields become especially significant by creating the ideal conditions for regulated mass accumulation and providing support against gravitational collapse at the highest densities. An increasing number of multi-scale observations of magnetic fields \citep{JunhaoLiu2023,Beltran2024,Koch2022} have revealed their structure from HFS to core scales. The few-parsec scale observations encompassing the three HFS presented in this study, combined with detailed B-field analyses of G31.41 HFS \citep{Beltran2024,Law2025} and NGC\,6334 \citep{JunhaoLiu2023} at sub-parsec to circumstellar scales, offer some of the most comprehensive insights to date on the role of magnetic fields in massive star formation.

From the roughly orthogonal orientation in the outer filament regions, the B-fields reorient to become parallel at the highest densities inside the hub. This shift in orientation underscores the importance of magnetic-pressure support —even more than turbulent support— in massive star-forming hubs. An increasing number of observational studies corroborate the idea that magnetic support plays a more dominant role than turbulence in these environments \citep{Beltran2019, Beltran2024, Rawat2024}. The unique configuration of HFS is critical in this context, as they naturally assemble the highest density network of filaments within their hubs, a prerequisite for high-mass star formation \citep[see, e.g.,][]{Kumar2021}. As higher densities are assembled within the hubs of HFS — via mechanisms such as filament coalescence — B-fields undergo modulations in both strength and morphology. These changes appear to provide the magnetic support necessary \citep{Girart2009,Krumholz2015} to restrict fragmentation \citep{Chen2019} and sustain continued mass accretion onto massive stellar seeds. Observations of magnetically supported toroidal structures in G31.41 HFS \citep{Beltran2019} and disk-like magnetic structures around W3(OH) offer compelling evidence supporting this proposition.

Once an HFS begins forming young stellar clusters, especially with massive stars, feedback mechanisms —such as outflows, radiation bubbles, and HII regions — reshape the surrounding B-fields \citep{Tahhani2023, Khan2024}. For instance, \citet{JunhaoLiu2023} find that in NGC 6334, at the highest densities, B-fields exhibit no preferred direction or show slightly perpendicular orientations relative to the dense gas. This is largely attributed to feedback-driven outflows. Our observations further suggest that this magnetic field reshaping predominantly occurs in the lower-density media surrounding the dense filaments. Within the densest structures themselves, B-fields remain largely unaltered, continuing to provide critical support for mass accumulation and regulating the star formation process. These results reinforce the notion that magnetic fields are integral to both the initiation and evolution of massive star formation within hub-filament systems, offering new insights into the multi-scale dynamics of these complex environments.

\section{Summary and conclusions}\label{Summary}

In this study, we investigated the characteristics of three star cluster-forming hub-filament systems — W3(OH), W3\,Main, and S\,106 — representing early, intermediate, and advanced evolutionary stages, respectively. We presented new observations of dust-polarized emission with JCMT/SCUBA-2/POL-2 and reanalyzed archival near-infrared and radio data from the VLA, WIRCAM, \spitzer, and \herschel. Our main findings are summarized below:

\begin{itemize}
\item
Column density and line mass maps derived from the 850\,\mum\ dust emission reveal the hub-filament nature of these regions, with central hubs showing the highest column densities and line masses ranging from several hundred to a thousand \msun\,pc$^{-1}$. The plane-of-sky magnetic field angles ($\chi_{B_{\rm POS}}$) were compared to Stokes I maps, showing that B-fields are predominantly perpendicular or randomly orientated with respect to lower-density filaments in outer regions, allowing the possibility that B-fields are mostly perpendicular to the filaments in 3D \citep{Doi2020}. However, they become aligned with higher-density structures within the hubs.\

\item
Archival VLA and near-infrared data, together with column density maps, reveal two primary star formation nodes within the hubs. These nodes exhibit clear evolutionary differences, such as young stellar clusters visible in near-infrared adjacent to deeply embedded massive YSOs or compact HII regions. \

\item 
Ellipses were fit enclosing the two nodes of star formation. Azimuthally averaged radial column density profiles, centered at the midpoint between these two nodes, were used to define the "hub." These profiles exhibit a two-power-law structure, transitioning at r$\sim$0.6-0.8\,pc, with power-law indices 
 $>-2$ inside and $<-2$  outside the hub boundary (Fig.\,\ref{radialProf}). As the HFS evolve, we notice an increase in the hub-size, and a steeper FLMF slope inside the hub.\

\item
Histograms of polarization fraction ($PF$) and polarized intensity ($PI$) for both "hub" and "filament" regions illustrate an evolutionary trend from the youngest system, W3(OH), to the most evolved, S\,106, indicating the influence of star formation feedback. Contour maps of $1/PF$ and $PI$ reveal disk- and bipolar outflow-shaped patterns centered on the most luminous sources in each region (Fig.\,\ref{PolContours}). We observed that massive bipolar outflows and radiation bubbles significantly reshape the magnetic fields, which align along the outflow cavity walls. This feature is consistent across all three targets. The sources driving these outflows correspond to regions with the lowest $PF$ values, likely due to large grain destruction or increased grain alignment from enhanced anisotropic radiation of young stars.\

\item
Histograms comparing the difference between ($\chi_{B_{\rm POS}}$) and filament orientation angles were generated for both hub and filament regions. The results align with observed morphologies, indicating B-fields are generally parallel to dense hub structures and perpendicular in lower-density filament areas. In the evolved system S\,106, these angle differences display both parallel and perpendicular alignments within the hub, suggesting modifications driven by outflow activity. \

\item
In W3\,Main, we identified a previously unreported spiral-shaped mini-HFS in the foreground, which had been recognized as the W3\,Main SE clump in prior \herschel\ studies.\

\item
Alongside our current targets, we compiled a list of nearby star-forming regions where a "double-node" pattern of star formation is evident, consistently showing evolutionary differences between adjacent nodes. This pattern may be a common feature in many HFS and could reflect the physical conditions governing their formation and evolution.
\end{itemize}

This is the first study to examine the HFS properties as a function of evolutionary stage using only three targets. To better understand the change of the properties of HFS as they evolve in time and as star formation proceeds, a statistical analysis of a large sample is required. Such a study should examine the distribution of hub sizes, line-mass functions and magnetic field orientations as a function of evolution. High angular resolution data are needed to resolve the structure of the hubs (e.g., ALMA, JWST) and molecular line observations are needed to describe the kinematics. Quantitative comparisons of results form MHD numerical simulations at different epoch of the simulation will also be important to understand the physical processes responsible in the formation and evolution of star-cluster forming hub-filament systems.

 \begin{acknowledgements}
We thank the anonymous referee for a very thoughtful, encouraging and hawk-eyed review that greatly enhanced the clarity, bibliographic completeness and quality of this manuscript. This research is partially supported by Grants-in-Aid for Scientific Researches from the Japan Society for Promotion of Science (KAKENHI 18H05437, 19H01938 and 21H00045). MSNK was supported as a visiting scientist at the Theoretical Astrophysics Laboratory, Nagoya University through the aid of KAKENHI 18H05437. RSF was supported by the Visiting Scholars Program provided by the NAOJ Research Coordination Committee, NINS (NAOJ-RCC-23DS-050).
Data analysis was in part carried out on the Multi-wavelength Data Analysis System operated by the Astronomy Data Center (ADC), National Astronomical Observatory of Japan.
The James Clerk Maxwell Telescope is operated by the East Asian Observatory on behalf of The National Astronomical Observatory of Japan; Academia Sinica Institute of Astronomy and Astrophysics; the Korea Astronomy and Space Science Institute; the National Astronomical Research Institute of Thailand; Center for Astronomical Mega-Science (as well as the National Key R\&D Program of China with No.\,2017YFA0402700). Additional funding support is provided by the Science and Technology Facilities Council of the United Kingdom and participating universities and organizations in the United Kingdom, Canada, and Ireland. 
Additional funds for the construction of SCUBA-2 were provided by the Canada Foundation for Innovation.
The authors wish to recognize and acknowledge the very significant cultural role and reverence that the summit of Mauna Kea has always had within the indigenous Hawaiian community. We are most fortunate to have the opportunity to conduct observations from this mountain.
  \end{acknowledgements}

\bibliographystyle{aa}
\bibliography{aa54901-25} 

\begin{appendix}

\section{Stokes I maps and selection of the filamentary emission}\label{App1}
\subsection{Stokes I maps}\label{App1a}
\begin{figure*}[!b]
   \centering
  \resizebox{6.4cm}{!}{\includegraphics[angle=0]{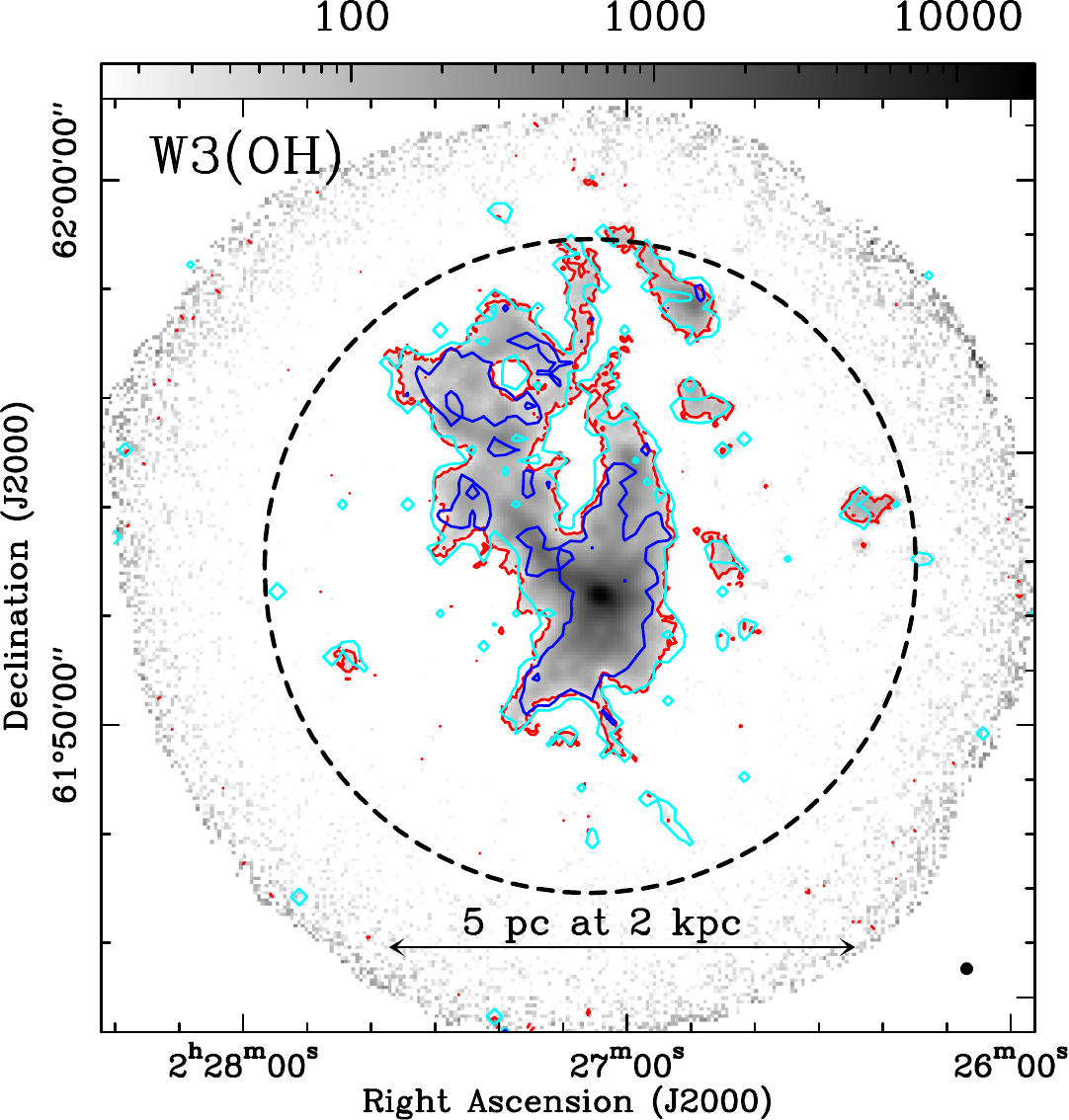}}
  \resizebox{5.8cm}{!}{\includegraphics[angle=0]{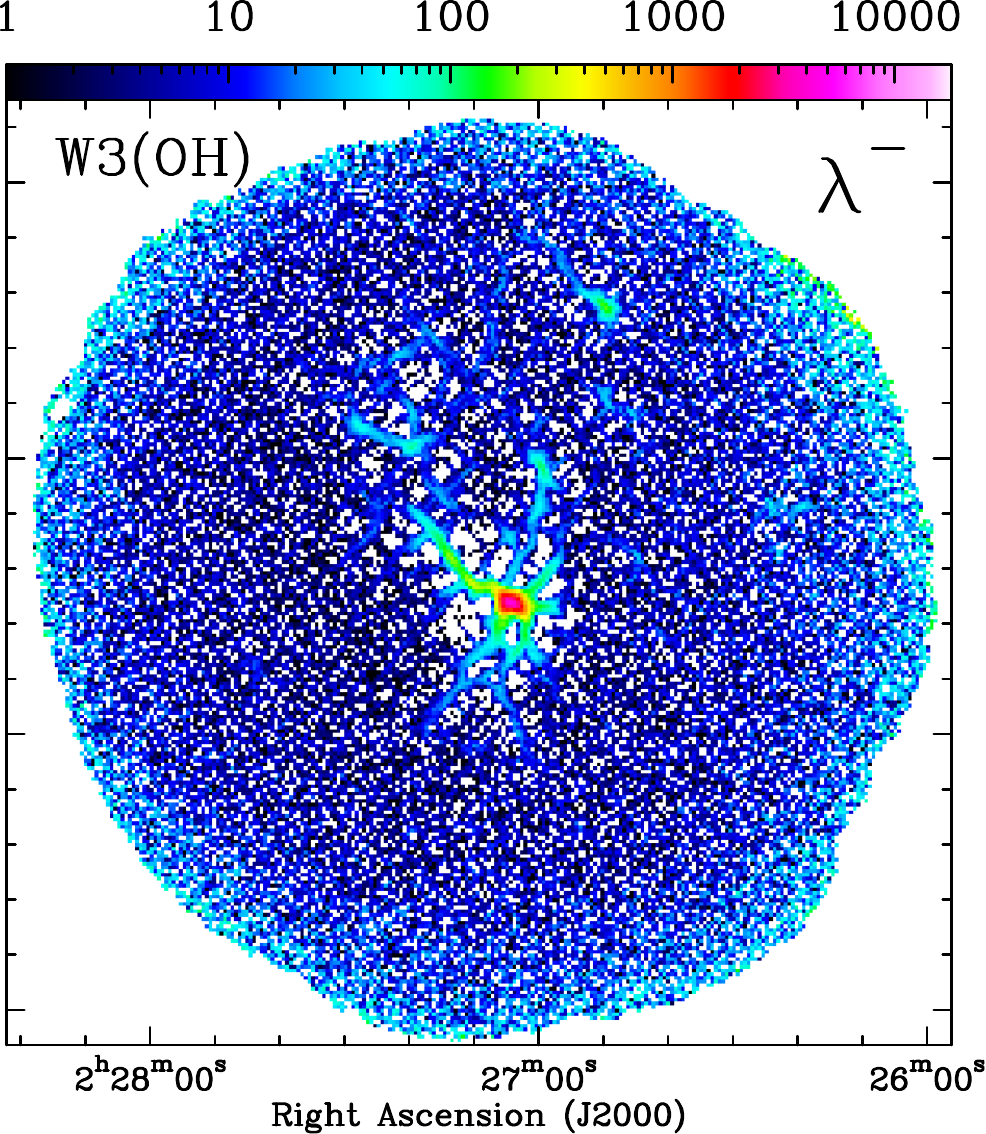}}
    \resizebox{5.8cm}{!}{\includegraphics[angle=0]{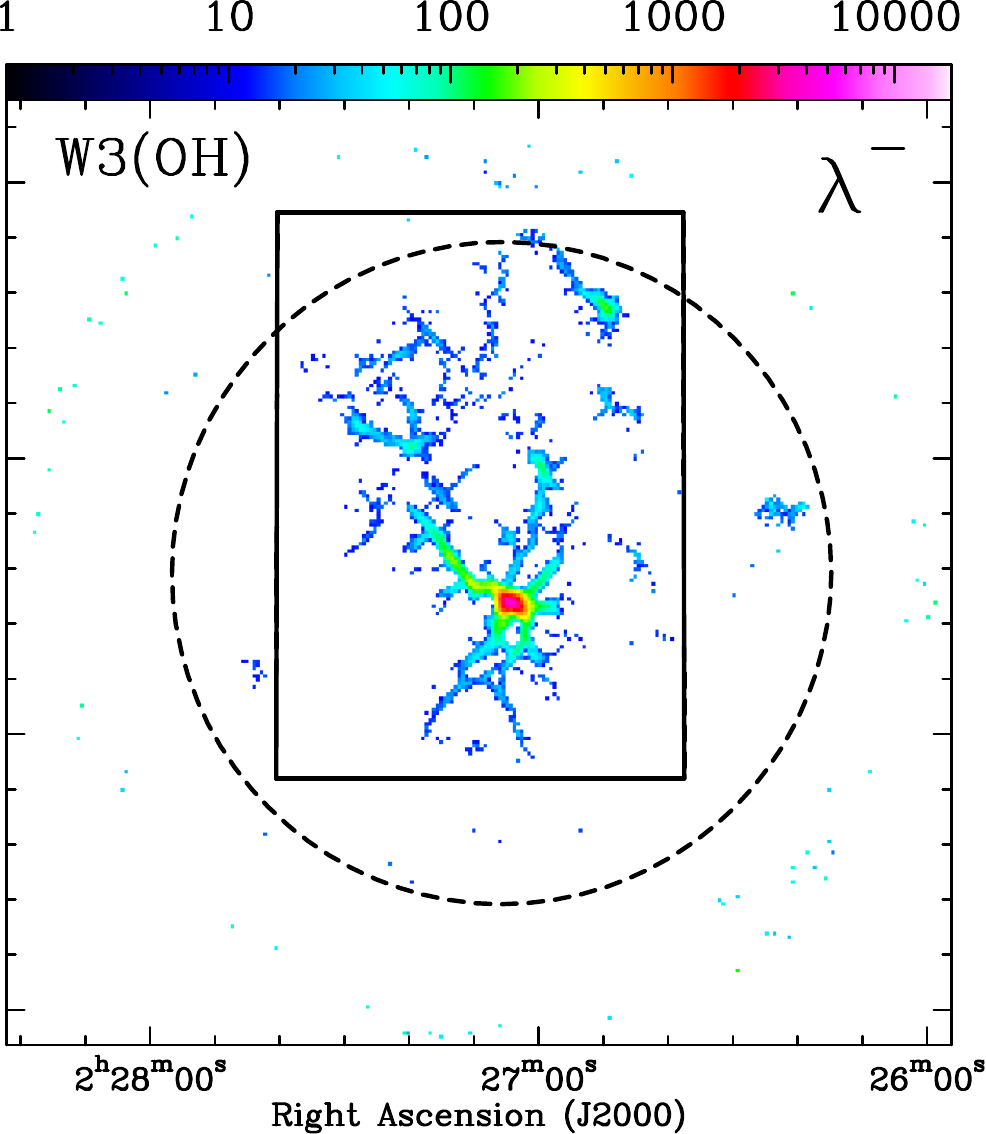}}
\vspace{-.1cm}
  \caption{ {\it Left:} Stokes  $I$ map at 850\,$\mu$m  observed with the JCMT POL-2  towards W3(OH) in unit of mJy\,beam$^{-1}$. 
  The HPBW resolution of the maps shown on this figure is $14\arcsec$ and the pixel size is $4\arcsec$. The red contours correspond to $I/\delta I>5$. The cyan and blue contours correspond to $PI/\delta PI>1$ and $PI/\delta PI>5$, respectively. 
 The black dashed circle have a diameter of $12\arcmin$. 
 {\it Middle:} The minimum curvature map (in unit of mJy\,beam$^{-1}$) of W3(OH) derived from the Stokes  $I$ map. 
 {\it Right:}  W3(OH) minimum curvature map (in unit of mJy\,beam$^{-1}$) of W3(OH) where pixels with  $I/\delta I<5$ are masked. The filamentary structures are enhanced by the minimum curvature map. The black rectangle is the field shown in Fig.\,\ref{W3OH-I-Bfield-Spitzer}. The dashed circle is the same as on the left panel.}          
  \label{Curv_W3OH}
    \end{figure*}

\begin{figure*}[!b]
   \centering
  \resizebox{6.4cm}{!}{\includegraphics[angle=0]{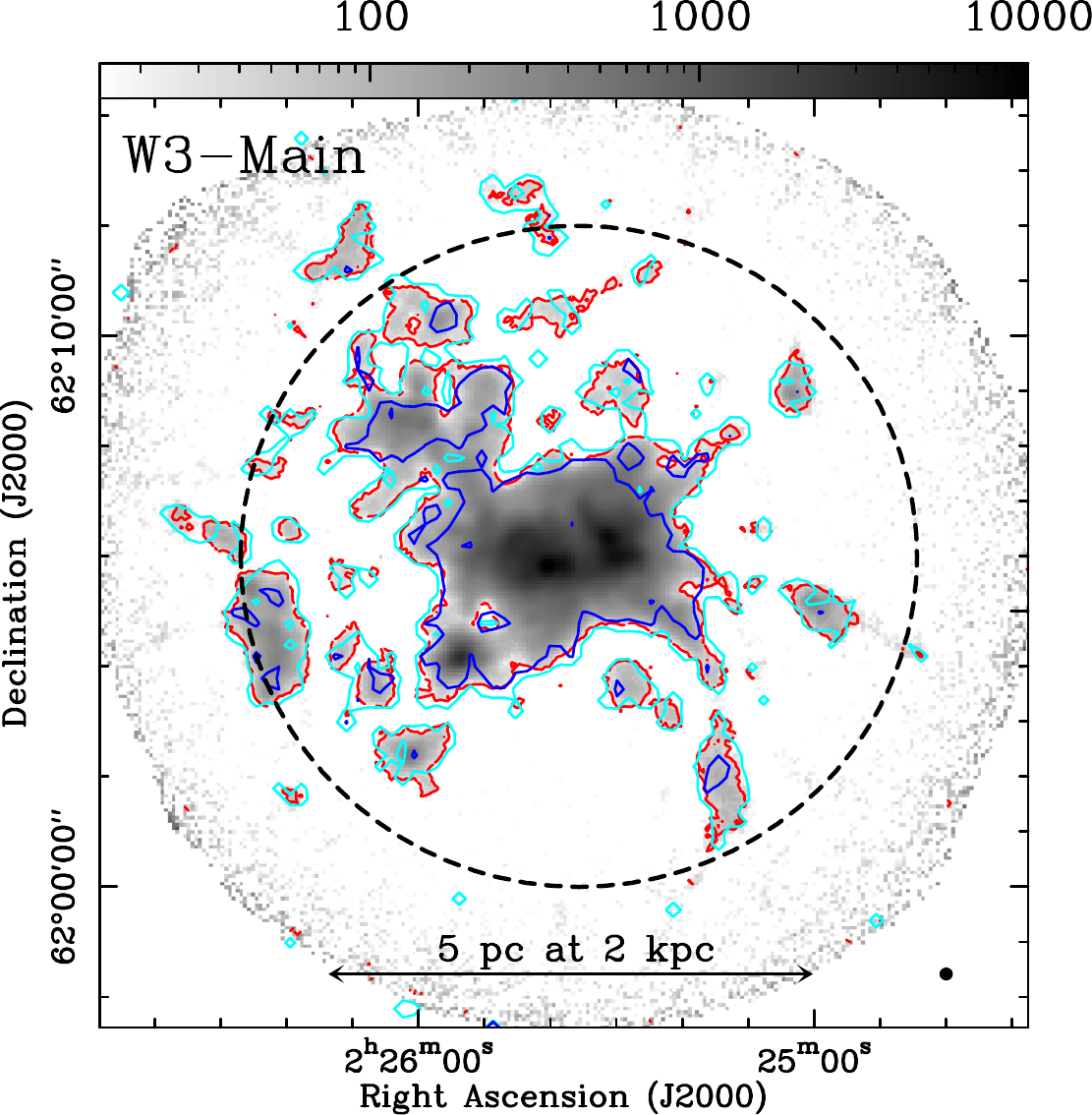}}
  \resizebox{5.8cm}{!}{\includegraphics[angle=0]{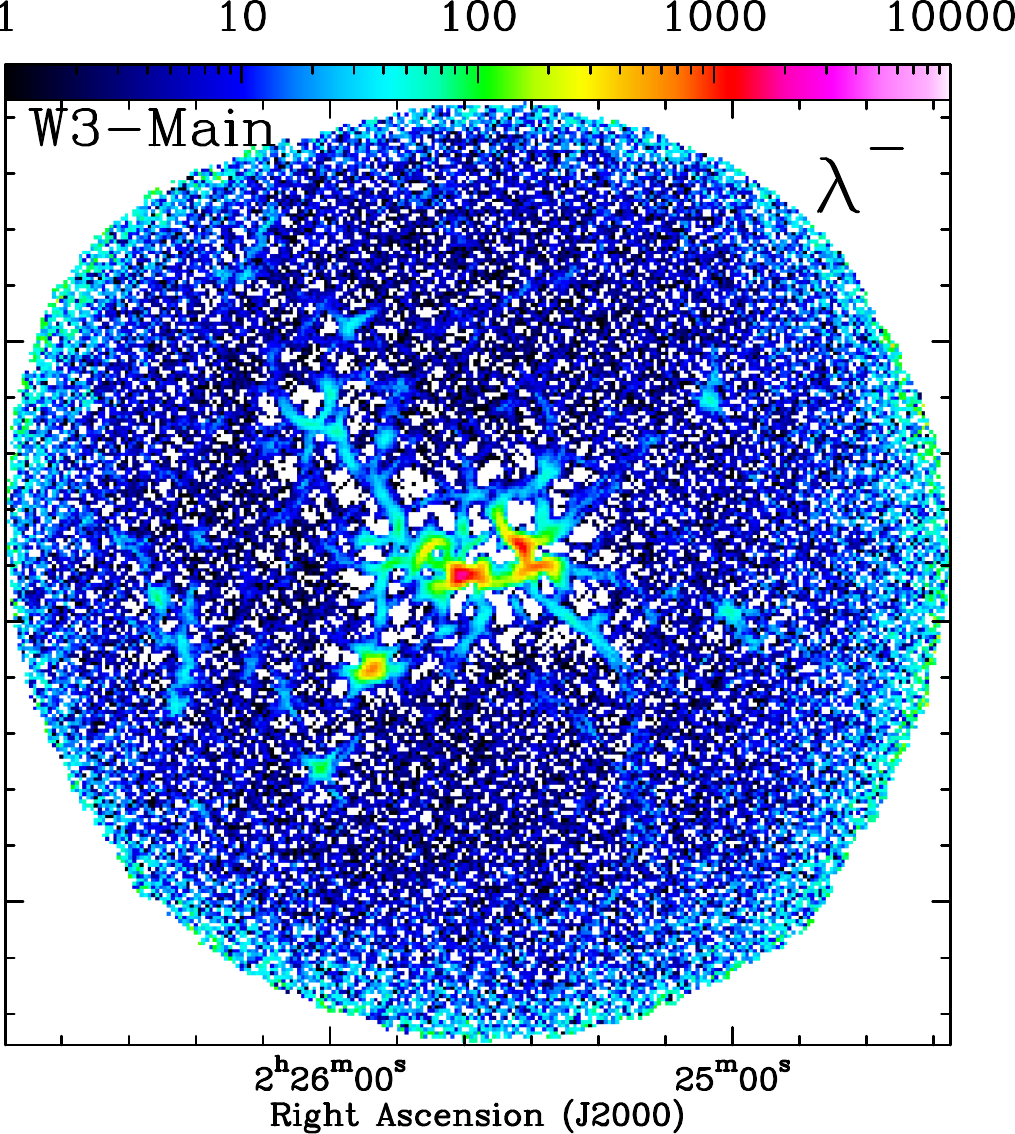}}
    \resizebox{5.8cm}{!}{\includegraphics[angle=0]{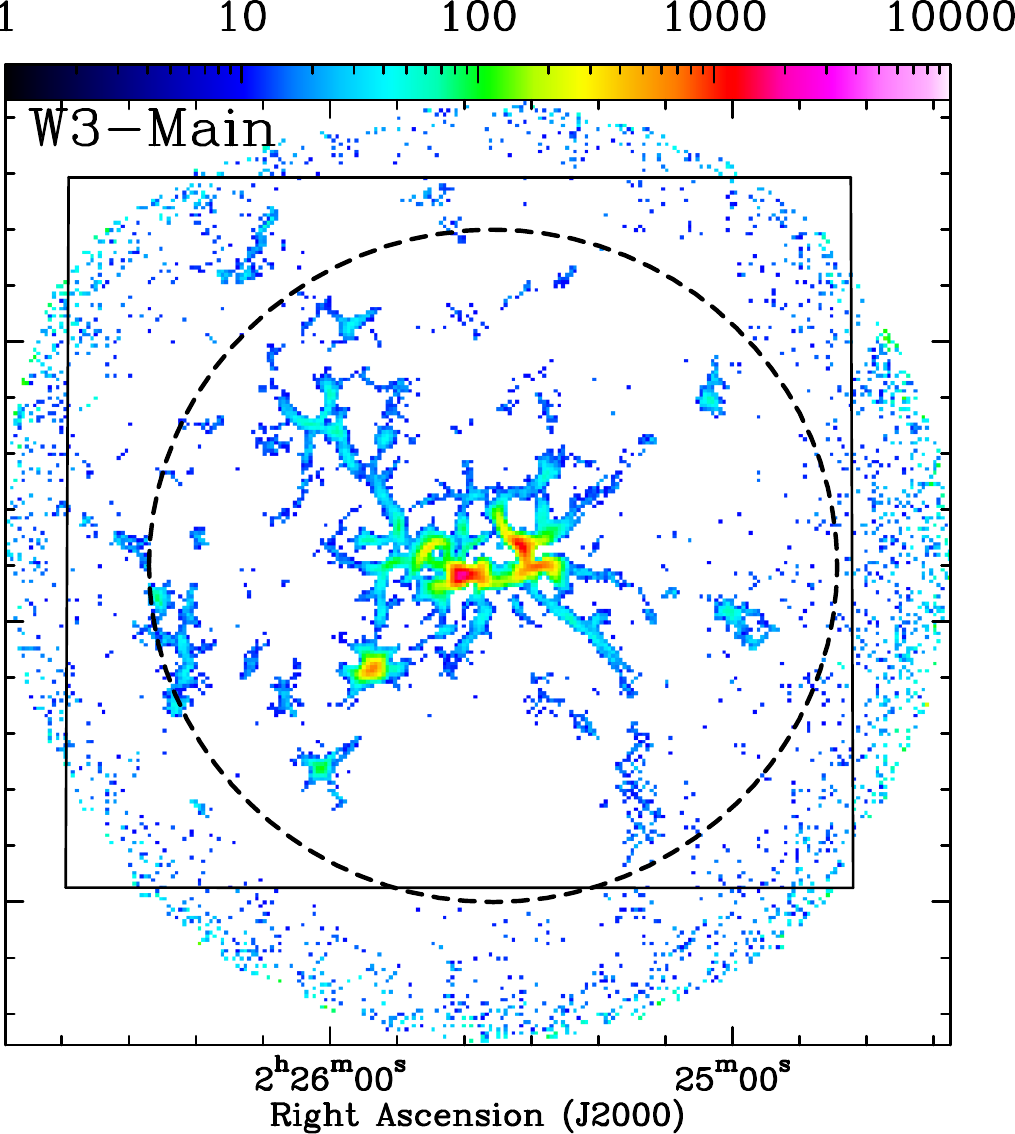}}
\vspace{-.1cm}
  \caption{Same as Fig.\,\ref{Curv_W3OH} for W3-Main. The black rectangle  on the right hand side panel corresponds to the area shown in Fig.\,\ref{W3Main-I-Bfield-Spitzer}. 
}          
  \label{Curv_W3Main}
    \end{figure*}

\begin{figure*}[]
   \centering
  \resizebox{6.3cm}{!}{\includegraphics[angle=0]{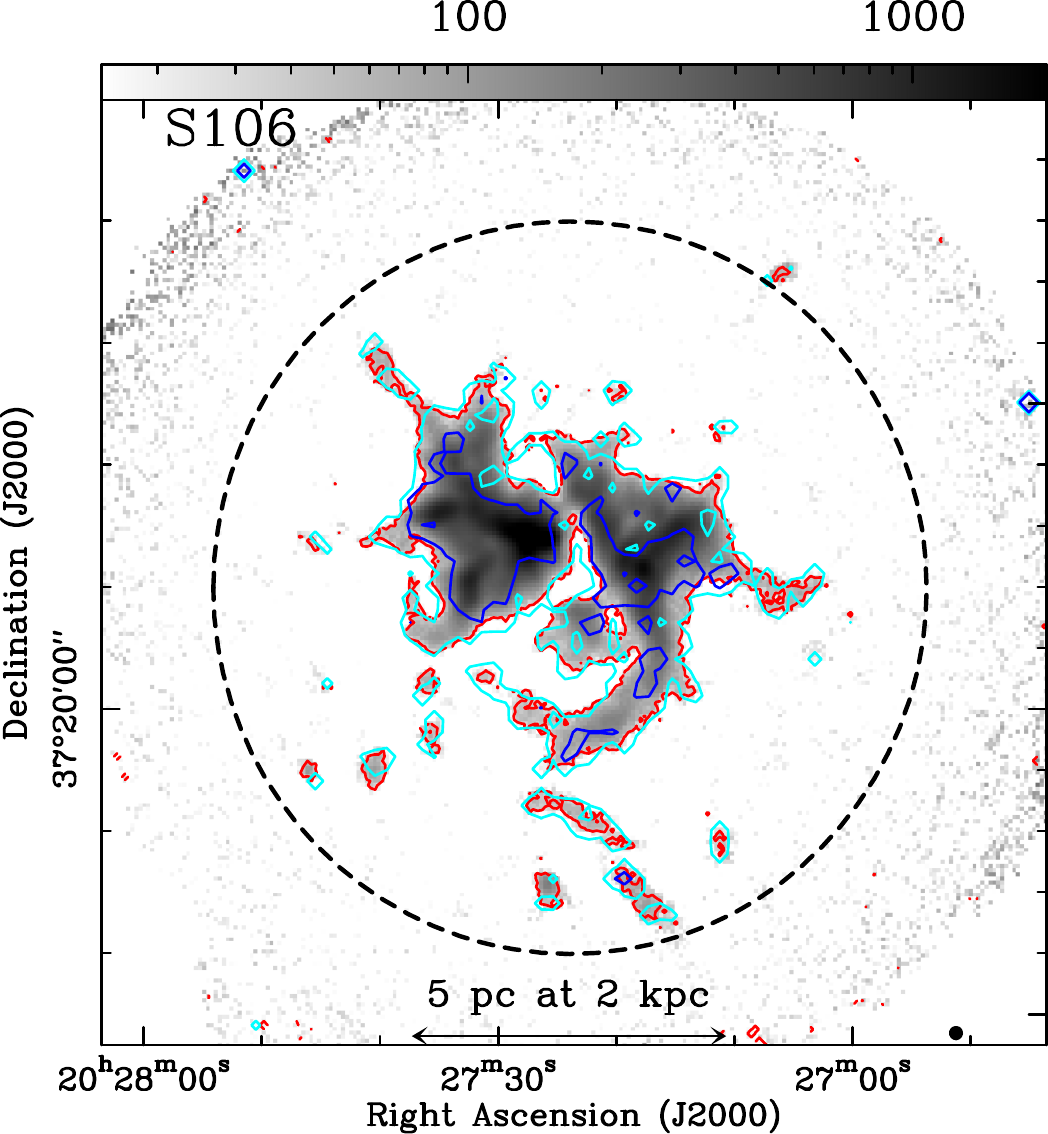}}
  \resizebox{5.95cm}{!}{\includegraphics[angle=0]{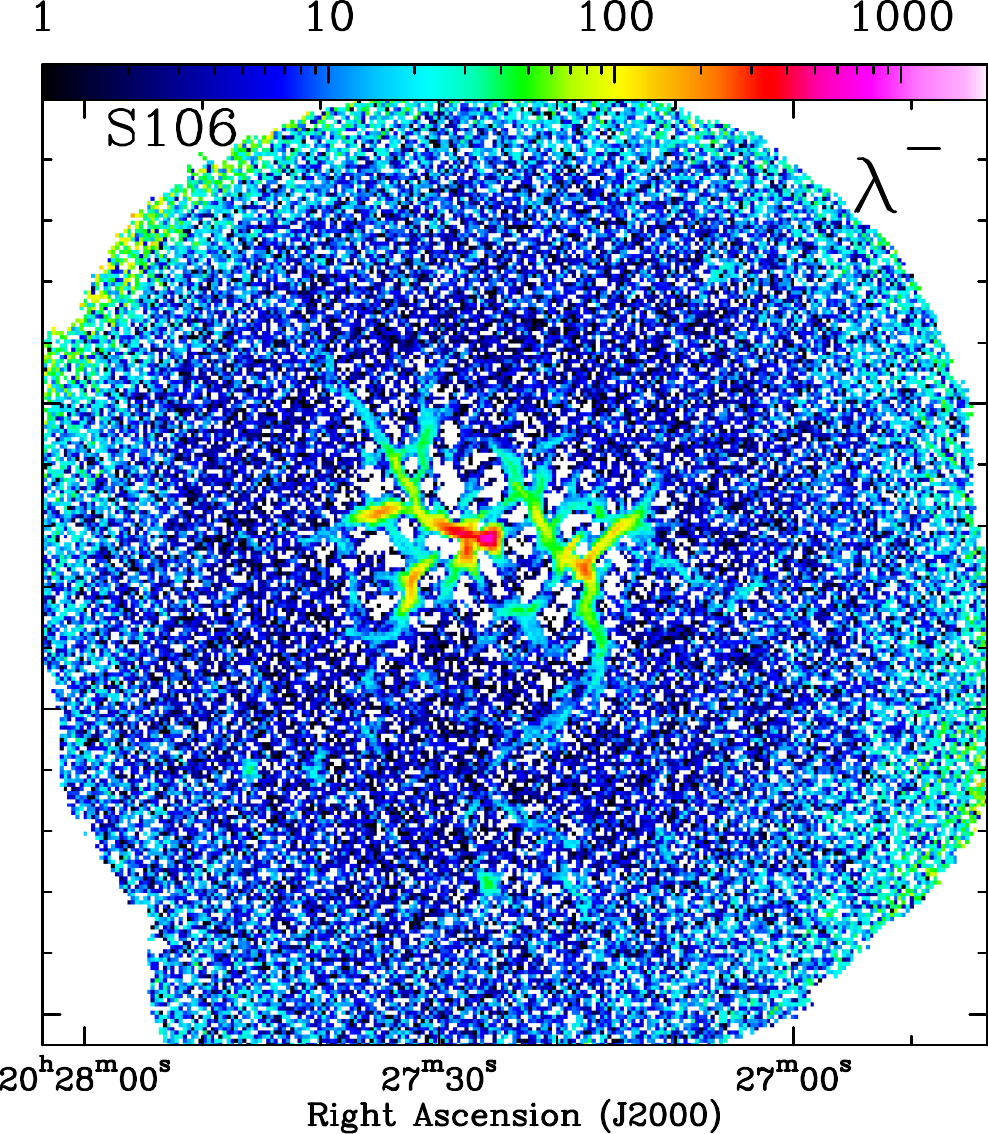}}
    \resizebox{5.95cm}{!}{\includegraphics[angle=0]{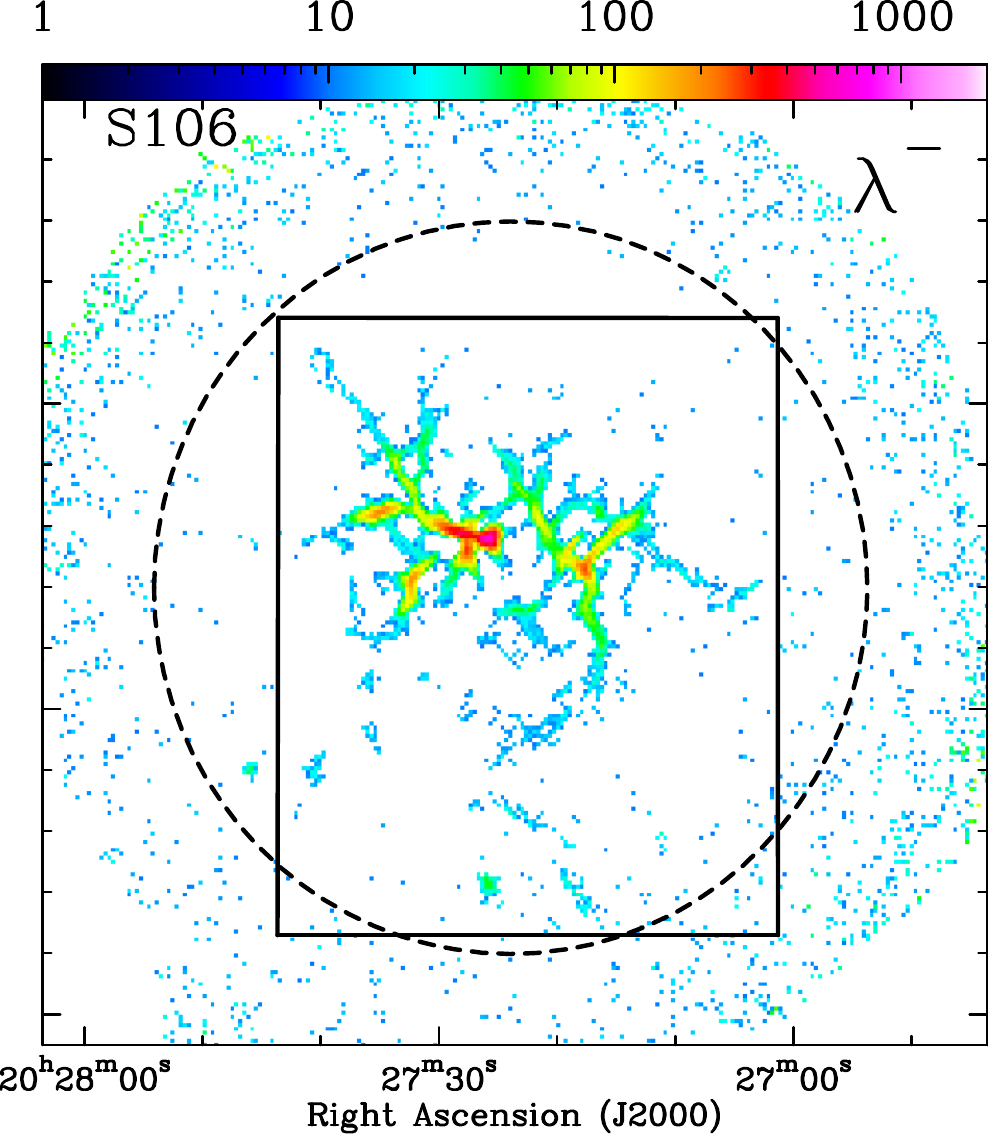}}
\vspace{-.2cm}
  \caption{Same as Fig.\,\ref{Curv_W3OH} for S106. The black rectangle  on the right hand side panel corresponds to the area shown in Fig.\,\ref{S106-I-Bfield-Spitzer}. 
}          
  \label{Curv_S106}
    \end{figure*}

\begin{figure*}[]
   \centering
  \resizebox{6.cm}{!}{\includegraphics[angle=0]{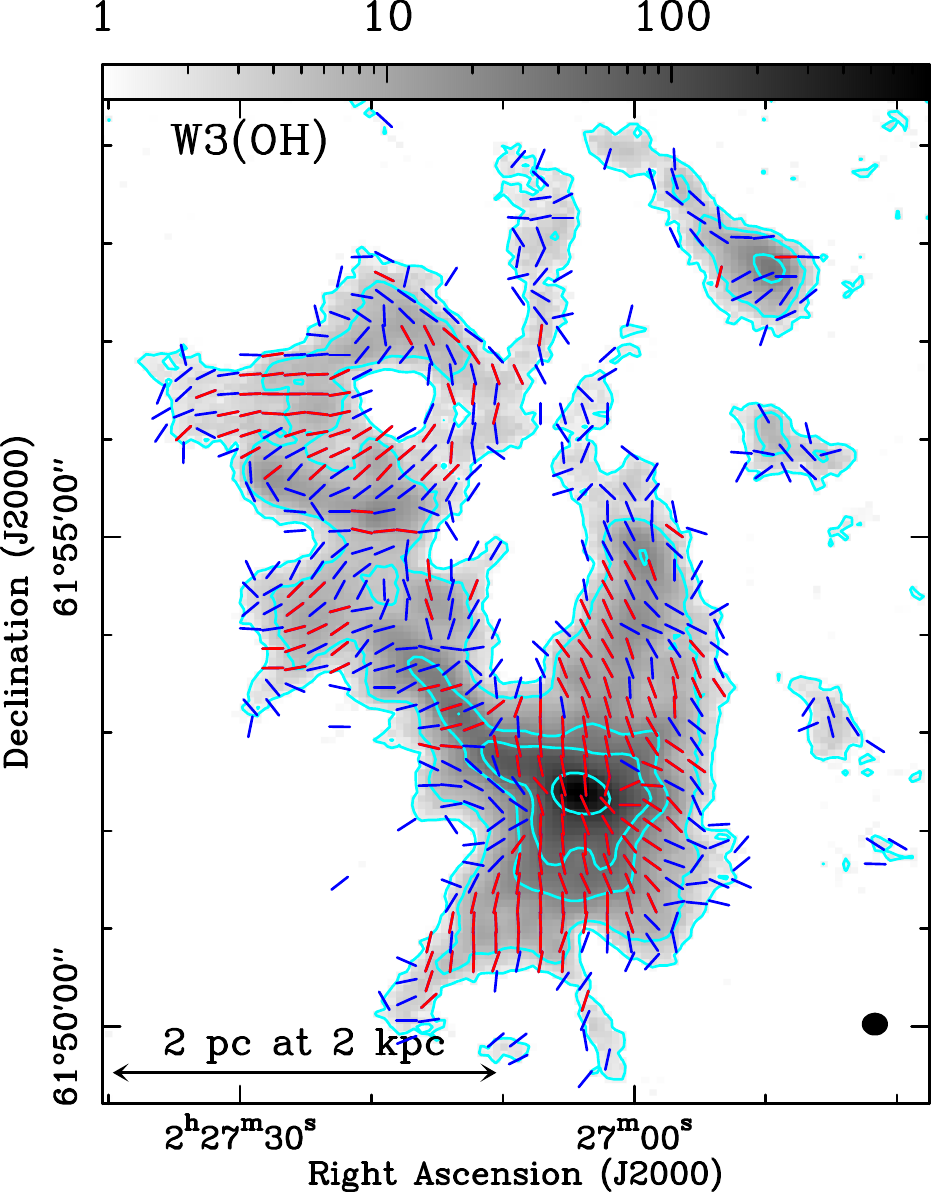}}
 \resizebox{6cm}{!}{\includegraphics[angle=0]{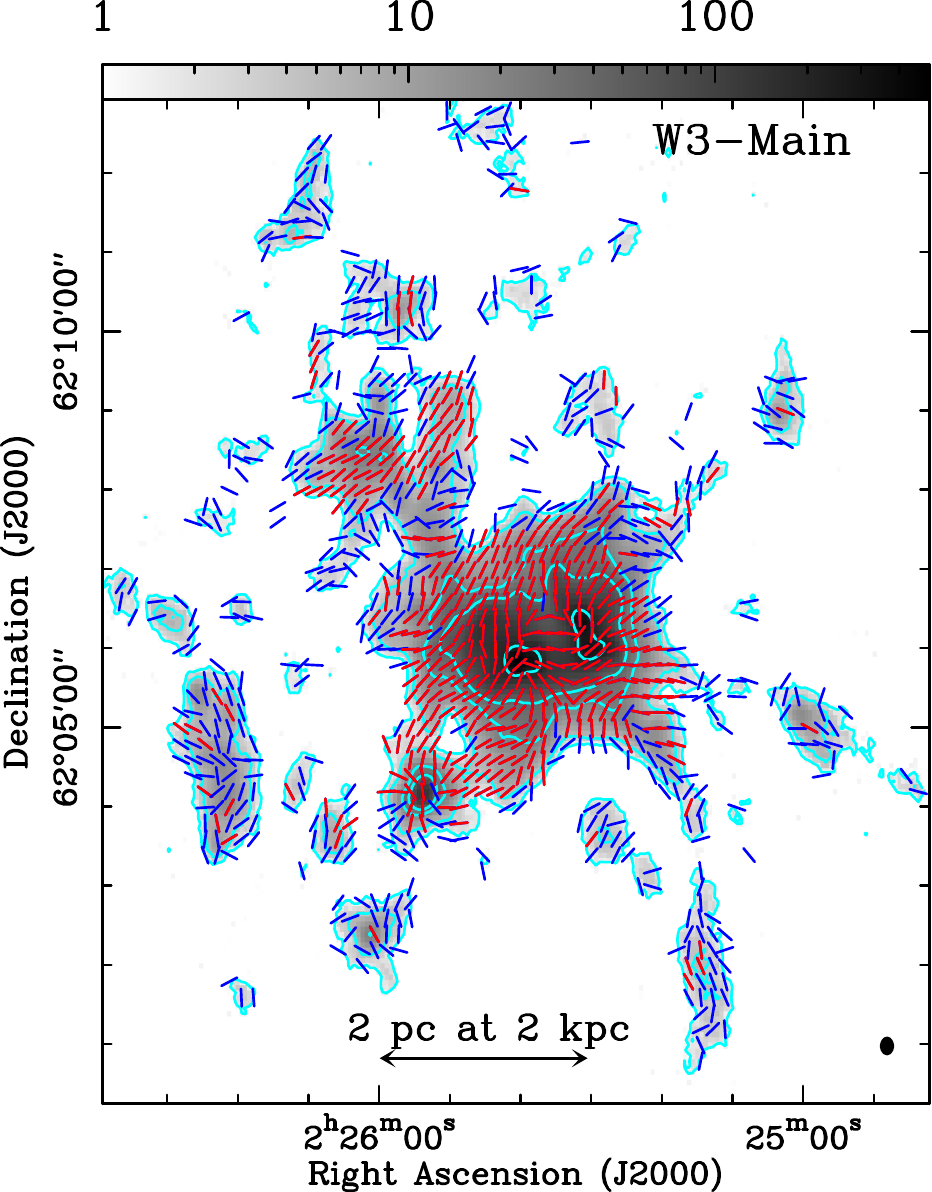}}
    \resizebox{6cm}{!}{\includegraphics[angle=0]{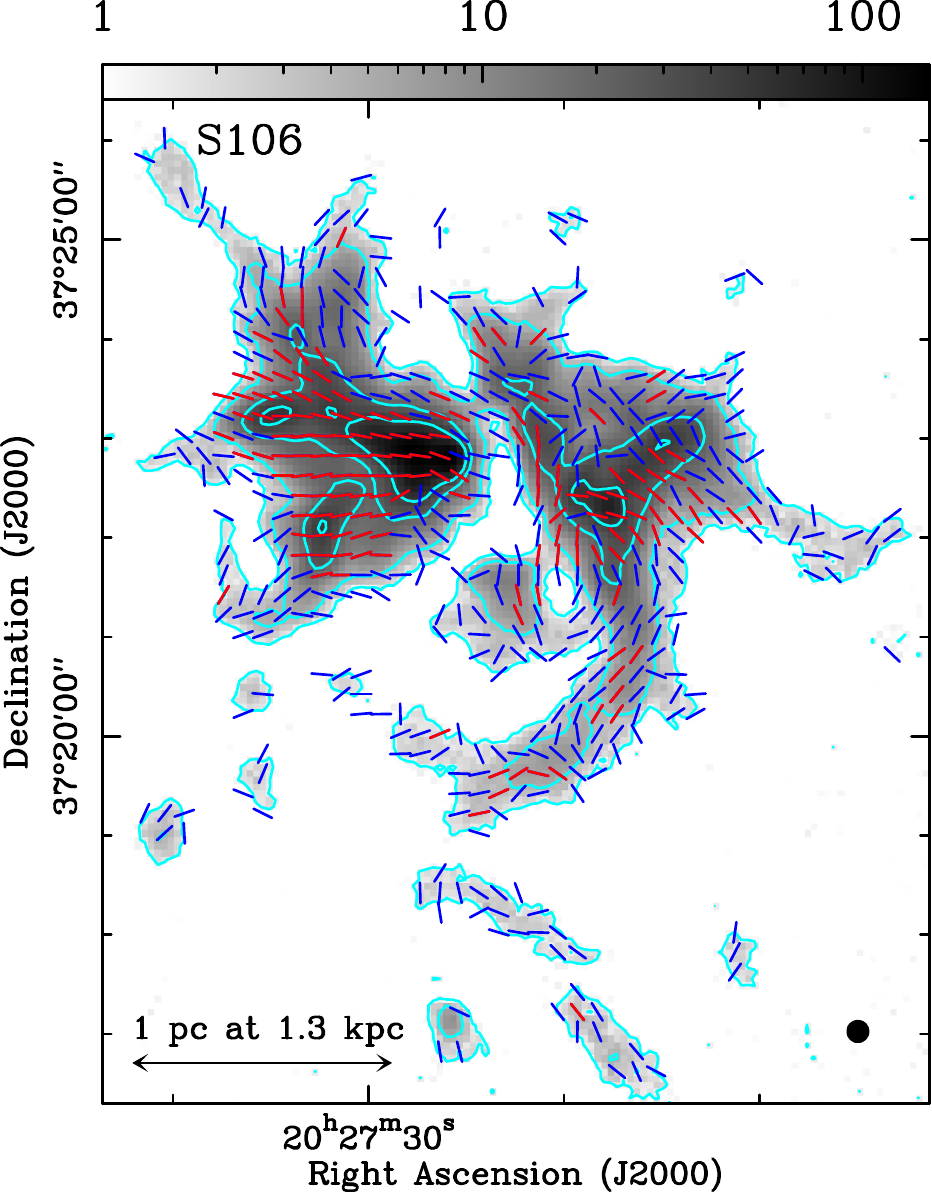}}
\vspace{-.1cm}
  \caption{
 Column density map, in units of $10^{21}\NHUNIT$, of W3(OH), W3\,Main and S\,106 from left to right.  The contours are 
  1.4, 4.6, 23, 46, and 231\,$\times10^{21}\NHUNIT$ equivalent to the Stokes I contours shown in Figs.\,\ref{W3OH-I-Bfield-Spitzer}, \,\ref{W3Main-I-Bfield-Spitzer}, and \,\ref{S106-I-Bfield-Spitzer}. The  blue and red  lines  show the $\chi_{B_{\rm POS}}$ angle
  for  $1<PI/\delta PI<5$ and $PI/\delta PI\ge5$, respectively, and $I/\delta I>5$.
 The lengths the lines are normalized to better represent the POS B-field orientation. 
 The data are at an angular resolution of $14\arcsec$. 
}          
  \label{coldensMaps_angle}
    \end{figure*}

The left panels of Figs.\,\ref{Curv_W3OH}, \,\ref{Curv_W3Main}, and \,\ref{Curv_S106} show the full Stokes  $I$ map at 850\,$\mu$m  observed with the JCMT SCUBA-2/POL-2  towards the three regions studied in this paper. 
The mean $\delta I$ uncertainties of the  Stokes  $I$ maps at 850\,$\mu$m are 2.4\,mJy/14\arcsec-beam for W3(OH), 2\,mJy/14\arcsec-beam for W3-Main, and  3.8\,mJy/14\arcsec-beam for S106. The mean $\delta PI$ uncertainties of the  polarized emission maps  are 1.5\,mJy/14\arcsec-beam for W3(OH), 1.2\,mJy/14\arcsec-beam for W3-Main, and  1.7\,mJy/14\arcsec-beam for S106. 
From the Stokes $I$ maps at 850\,$\mu$m and assuming a uniform temperature of $T=20\,$K,  we calculate the  column density  $\nhh$ maps (see details in \ref{Coldens}). Fig.\,\ref{coldensMaps_angle} show the column density maps for the three targets with the POS B-field angles. 


\subsection{Selection of the filamentary emission}\label{App1b}

To remove the extended emission of the clouds and select the filamentary structures, we compute the Hessian matrix  $H$  of  the second-order partial derivatives  for all pixels of the Stokes I maps with a pixel size of 4\arcsec. The solution of the characteristic equation of the Hessian matrix provides two eigenvalues. The smaller of the two eigenvalues, $\lambda^-$, corresponds to  the minimum curvature of the intensity map \citep{Arzoumanian2012,Schisano2014,planck2016-XXXII}. The minimum curvature maps derived from the Stokes I are shown in the middle panels of Figs.\,\ref{Curv_W3OH}, \ref{Curv_W3Main}, and \,\ref{Curv_S106}, for W3(OH), W3-Main, and S106, respectively. Filaments are elongated structure which corresponds to a relatively small curvature along their main axis. Consequently,  filamentary structures are enhanced in the minimum curvature maps with respect to the surrounding extended emission. To derive "filament maps", we further mask the pixels of the $\lambda^-$ maps with  $I/\delta I<5$ 
(cf. the right hand side panels of Figs.\,\ref{Curv_W3OH}, \ref{Curv_W3Main}, and \,\ref{Curv_S106}). 

Using the  Hessian matrix  $H$, we also compute $\theta_{\rm fil}$ the local POS angle of the filamentary structures  for all the selected pixels of the filament maps as follow:

 \begin{equation}
 \theta_{\rm fil} = \frac{1}{2}\, \arctan\left( \frac{2H_{xy}}{H_{x^{2}} - H_{y^{2}}}\right) +90^\circ, \label{Eq:Angle}
\end{equation}

where the $x$ and $y$ axes are the horizontal and vertical axes  of the Stokes $I$ map in  Cartesian coordinates, 
and  $H_{x^{2}}=\partial ^{2}I/\partial x^{2}$,  $H_{y^{2}}=\partial ^{2}I/\partial y^{2}$ and  $H_{xy}=H_{yx}=\partial ^{2}I/\partial x\partial y$ \citep[see also][]{Arzoumanian2019}. 
    
    \end{appendix}

\end{document}